\documentclass[10pt]{article}
\usepackage{graphicx} 
\usepackage{amsmath} 
\usepackage{amsfonts}
\usepackage{amssymb}
\usepackage{subfigure}
\usepackage{multirow}

\title{\textbf{Realistic Tunneling States for the Magnetic Effects \\ in Non-Metallic Real Glasses}} 

\author{ Giancarlo Jug$^{a,d}$\cite{J}, Silvia Bonfanti$^{a,b}$ 
and Walter Kob$^c$ \\
$^a$Dipartimento di Scienza ed Alta Tecnologia and To.Sca.Lab \\
Universit\`a dell'Insubria,
Via Valleggio 11, 22100 Como, Italy \\
$^b$Laboratoire Charles Coulomb, Universit\'e de Montpellier \\
Place Eug\`ene Bataillon, F-34095 Montpellier Cedex 5 - France \\
$^c$Laboratoire Charles Coulomb, UMR CNRS 5221 \\ Universit\'e de Montpellier, 
34095 Montpellier, France \\
$^d$INFN -- Sezione di Pavia, Italy and IPCF -- Sezione di Roma, Italy}

\date{\today} 

\begin{document}

\maketitle 

\begin{abstract}
The discovery of magnetic and compositional effects in the low temperature
properties of multi-component glasses has prompted the need to extend the
standard two-level systems (2LSs) tunneling model. A possible extension
\cite{Jug2004} assumes that a subset of tunneling quasi-particles is moving in
a three-welled potential (TWP) associated with the ubiquitous inhomogeneities
of the disordered atomic structure of the glass. We show that within an
alternative, cellular description of the intermediate-range atomic structure
of glasses the tunneling TWP can be fully justified. We then review how the
experimentally discovered magnetic effects can be explained within the approach
where only localized atomistic tunneling 2LSs and quasi-particles tunneling in
TWPs are allowed. We discuss the origin of the magnetic effects in the heat
capacity, dielectric constant (real and imaginaryi parts), polarization echo
and SQUID magnetization in several glassy systems. We conclude by commenting on
a strategy to reveal the mentioned tunneling states (2LSs and TWPs) by means of
atomistic computer simulations and discuss the microscopic nature of the
tunneling states in the context of the potential energy landscape of
glass-forming systems.
\end{abstract}

\newpage

\section{Introduction}
The physics of glasses, especially at low temperatures, continues to attract
considerable interest. At low temperatures glasses are believed to be 
characterized by low-energy excitations going under the name of tunneling 
systems, or states (TSs) which are normally described in terms of double-welled 
potentials (DWPs) and two-level systems (2LSs) with energy asymmetry and 
tunneling barrier uniformly distributed in the amorphous solid \cite{Phi1981,
Esq1998}. Little is still known about the nature of the TSs, but the general 
consensus is still that the intermediate-range atomic structure of glasses is 
well described by Zachariasen's 1932 continuous random network model 
\cite{Zac1932,War1934} (thus, homogeneously disordered like for a liquid) and 
the 2LSs arise out of two slightly similar, localized atomic configurations. 
With this characterization the 2LSs have been employed in the 1970s and 80s to 
explain with some success the anomalies in the properties of glasses at low 
temperatures.
Recently, the field has been witnessing a renaissance since the discovery that 
there are unexpected magnetic effects in insulating glasses 
\cite{Str2000,Woh2001,Nag2004,Bra2004} and that the TSs are responsible for 
the decoherence of phase qubits made of superconducting Josephson-junctions 
\cite{Sim2004,Mar2005,Bur2006,Cla2008,You2005}.  
The 2LSs are thought to be ubiquitous in the junction's tunneling barrier 
which, though very thin, is considered to be amorphous and described by 
Zachariasen-Warren's model \cite{Zac1932,War1934}. On the other hand, 
the low-temperature glasses have also been studied as a paradigm where to 
conduct research on the physics of aging so common to many disordered systems 
out of equilibrium \cite{Ami2012,Osh2003}. In 
this context, the relevant degrees of freedom have also been described in terms 
of 2LSs, with some success \cite{Bur2013}.
     
Yet, deviations from the behaviour predicted by the standard tunneling model
(STM) have challenged the validity of the model in the case of multi-component
glasses with variable content of the good crystal-forming (GCF) component (e.g.
(SiO$_2$)$_{1-x}$(K$_2$O)$_{1-x}$ with changing $x$ \cite{Mac1985,Jug2010}) 
and especially in glasses of compositions like BaO-Al$_2$O$_3$-SiO$_2$ in the 
presence of a weak magnetic field \cite{Ens2002}. In such glasses (the mixed 
alkali-silicates have not yet been investigated in a field, but we predict 
important magnetic effects there, and $x$-dependent, too) a puzzling 
non-monotonous magnetic-field dependence has been revealed in most physical 
properties \cite{Str2000,Woh2001,Nag2004,Bra2004}. The magnetic effect is 
normally weak, but orders of magnitude larger than expected from basic 
thermodynamic considerations. The STM is unable to explain the compositional 
and magnetic effects, thus a suitable extension of the tunneling model for 
both situations has been proposed by one of us \cite{Jug2004}. This extended 
tunneling model (ETM) rests upon the existence in
the multi-component (or contaminated mono-component) glasses of regions of 
enhanced regularity (RERs) in the atomic structure of the incompletely-frozen 
(in fact) amorphous solid. A complete mathematical description and physical
justification of the ETM thus calls at the very least for a partial demise of 
the Zachariasen-Warren's vision of the intermediate structure of glasses.   

Research on the mechanism of the glass phase transition is also witnessing
intense activity \cite{Ber2011}, with almost all of the investigations 
conducted from the high-temperature (liquid or supercooled-liquid) side of the 
transition. Any attempt at a new look at (and demise of) the 
Zachariasen-Warren's continuous random network model description of the atomic 
structure of glasses from the 
low-temperature side is therefore likely to provide new insight also on the 
nature of the glass transition itself.   

In this paper we provide a new scenario for the atomic structure of glasses,
the cellular model, a description within which the phenomenological 
assumptions for the ETM's mathematical formulation become completely justified. 
The cellular model provides for a more realistic mathematical formulation
in terms of a tetrahedric four-welled tunneling potential, the triangular 
three-welled version employed being nothing but a poor-man's, but probably
very realistic, version affording a much 
simplified mathematical description. Within the cellular approch to the 
structure of glasses the most significant local tunneling potentials (also in 
terms of number density) turn out to be the DWPs, for a single (or very few) 
atomic particles, and the tetrahedric four-welled potential (TFWP) for a 
correlated cluster of $N\gg 1$ charged atomic particles. A reasonable and 
very useful simplification for the TFWP is then the replacement of the $N$ 
interacting and tunneling atomic particles with a single quasi-particle 
subject to a triangular tunneling potential (TWP) carrying renormalized 
parameters (charge, magnetic threaded area, energy asymmetry and tunneling 
probability) moving about one face of the tetrahedric potential. The 
renormalization is fully justified by the proximity 
of four similarly quasi-ordered, close-packed atomic cells (RERs) and the 
reasonable assumption that most of the charged particles will avoid the 
interstice's centre.
We present therefore the mathematical description of the TWP and its quantum 
mechanics in the appropriate limits for practical applications. We apply the 
results for the description of the density of states and the temperature and 
magnetic-field dependence for the specific heat of the glass, then for the 
dielectric constant (real and imaginary part) in the linear regime, and for 
the polarization echo - always in the presence of a magnetic field. Finally we
apply the model to the description of the temperature and magnetic-field 
dependence for the SQUID-magnetization, showing that the ETM keeps providing
a good description for this quantity up to room temperatures. As a check that
the tunneling states we use are realistic representations of the remnant 
degrees of freedom in the disordered solid phase we also outline a strategy 
for the study of these local tunneling states in the potential energy
landscape (PEL) with the help of classical numerical simulation of a large 
number of particles interacting with a BLJ potential.   

The paper is organized as follows. In Section 2 we introduce the cellular 
model for the structure of glasses, an important refinement of Zachariasen's
continuous random-network model, and examine the likely tunneling states that 
emerge from such cellular picture, concluding that only DWPs and TFWPs
should be relevant for the physics of glasses below the glass transition 
temperature $T_g$. In Section 3 we review the relevant quantum mechanics of
the three-welled, poor man's version of the TFWP, a version that has been 
used so far to obtain a reasonable single explanation for all the anomalies 
(and deviations from STM predictions) due to composition changes and to the 
magnetic field. We also show how to evaluate the magnetic density of states 
(DOS) $n(E,B)$ and, in Section 4, we examine the magnetic-field dependent 
heat capacity $C_p(T,B)$ comparing with some published data for the 
multi-silicates. In Section 5 we do the same for the dielectric constant, 
real part $\epsilon'$ and imaginary part $\epsilon''$, also showing some 
comparison with available data at low (kHz) frequency. In Section 6 we examine 
the application of the ETM to the explanation of some of the data for the 
polarization echoes in the silicates and in glycerol, also explaining the
so-called isotope effect. In Section 7 we examine the question of the 
paramagnetism of insulating glasses {\em per se}, that is that due to the 
tunneling states themselves and which we find to be stronger than the Langevin 
paramagnetism of the trace paramagnetic impurities contained in the 
multi-silicates. Section 8 contains a discussion about guiding principles on
how one should proceed, within a molecular dynamics simulation of a model 
glass former (the BLJ system in 3D, for example), in order to characterize
the relevant tunneling states from a numerical analysis of the PEL and thus
make contact with the original phenomenological modelization at low 
temperatures. Finally, Section 9 contains a Summary of the results obtained 
and our Conclusions.  

\section{The Cellular Model for the Atomic Structure of Glasses and the 
Three-Welled Tunneling Potential}

What is a glass? According to Shelby ``A glass can be defined as an amorphous 
solid completely lacking in long range, periodic atomic structure and 
exhibiting a region (of temperature) of glass transformation behaviour'' 
\cite{She2005}. This, however, does not mean that the atomic arrangement of a 
glass should be the same as that for a liquid, as is implicit in Zachariasen's 
1932 proposed continuous network picture \cite{Zac1932,War1934}, which has 
been widely adopted by scholars (at least in the West, see below). This 
picture differs from that of a liquid only in that a dynamical arrest has 
occurred, without specifying its ultimate origin. In a spin-glass the ultimate 
origin of dynamical arrest is magnetic frustration (with or without disorder), 
but for ordinary structural glasses it remains mysterious and an important 
open issue \cite{Ber2011,Ang1988,Lub2007,Sim2009}. Prior to Zachariasen's 
scheme, however, the Soviet scientist A.A. Lebedev had proposed, in 1921, the 
concept of ``crystallites" \cite{Leb1921}, small undefined crystal-like 
regions clumpt against each other in random orientation to contain altogether 
all atoms in the substance. Later, Randall proposed that these be real 
micro-crystals and explained the rounded-up X-ray spectra from glasses in this 
way \cite{Ran1930}. However, the density of 
glasses is some 10\% less than that of poly-crystalline aggregates and the 
thermal properties of glasses also cannot be explained through the 
Lebedev-Randall picture. Despite these observations, the West-Soviet 
controversy continues to these days. The Zachariasen-Warren model of glass
structure was in fact criticized by H\"agg \cite{Hag1935} in the West right in 
the early days of X-ray crystallography and a good review of the status quo of 
this controversy has been recently provided by Wright \cite{Wri2014} who 
concludes from a re-analysis of X-ray and neutron-scattering data from many 
covalent-bonded and network glasses that indeed {\em cybotactic groupings} 
(better-ordered regions) may well be present and frozen-in in most glasses, 
especially if multi-component. 
The formation of {\em polyclusters}, instead of crystallites, in most glasses 
is the latest claim by the Soviet school \cite{Bak1994,Bak2013}, based on 
observations 
and thermodynamic reasoning. As nicely set out by Bakai \cite{Bak1994}, the
incipient crystals forming at and below $T_m$ (melting point) are in constant
competition with kinetically swifter (for glass-forming liquids) polyclusters 
(or crystallites) that can win thermodynamically and kinetically over crystals 
during a rapid enough quench.  

For e.g. the oxyde glasses (e.g. window glass, of composition 
CaO-Na$_2$O-SiO$_2$) the dilemma is most simply put in these terms. Pure good 
glass-former (GGF) SiO$_2$ melts at around 1900 K and vitrifies at around 
1475 K, whilst pure GCFs CaO and Na$_2$O melt at 2886 K and 1405 K, 
respectively. Yet, the mixture CaO-Na$_2$O-SiO$_2$ vitrifies at around 
750-850 K (depending on \% composition) and that is way below the melting 
point of very bad glass-formers CaO and Na$_2$O. Since phase separation is 
commonplace in such ternary mixtures (the recent X-ray tomography imaging of 
the related molten system BaO-Na$_2$O-SiO$_2$ is compelling \cite{Bou2014}), 
what is then stopping the pockets and channels formed by CaO and Na$_2$O 
within the GGF SiO$_2$'s network to at least initiate some degree of 
crystalline ordering? In the mono-composed oxydes it will be up to the 
impurities to take the place of the GCF phase separation.

On the experimental side, the concept of de-vitrification is gradually taking 
shape with reports of metallic glasses \cite{Hwa2012} and also monocomponent 
high coordination covalent solids like Si forming {\em paracrystals} in the
amorphous solid \cite{Tre2012}. Therefore, the stance will be taken in this 
paper that only the purest mono-component glasses may abide to the 
Zachariasen-Warren continuously random-network model of a glass, whilst the 
vast majority of real glasses will be organized otherwise at the 
intermediate-range atomic structure.

\begin{figure}[!hbtp]
\centering
{
   \subfigure[]{\includegraphics[scale=0.3] {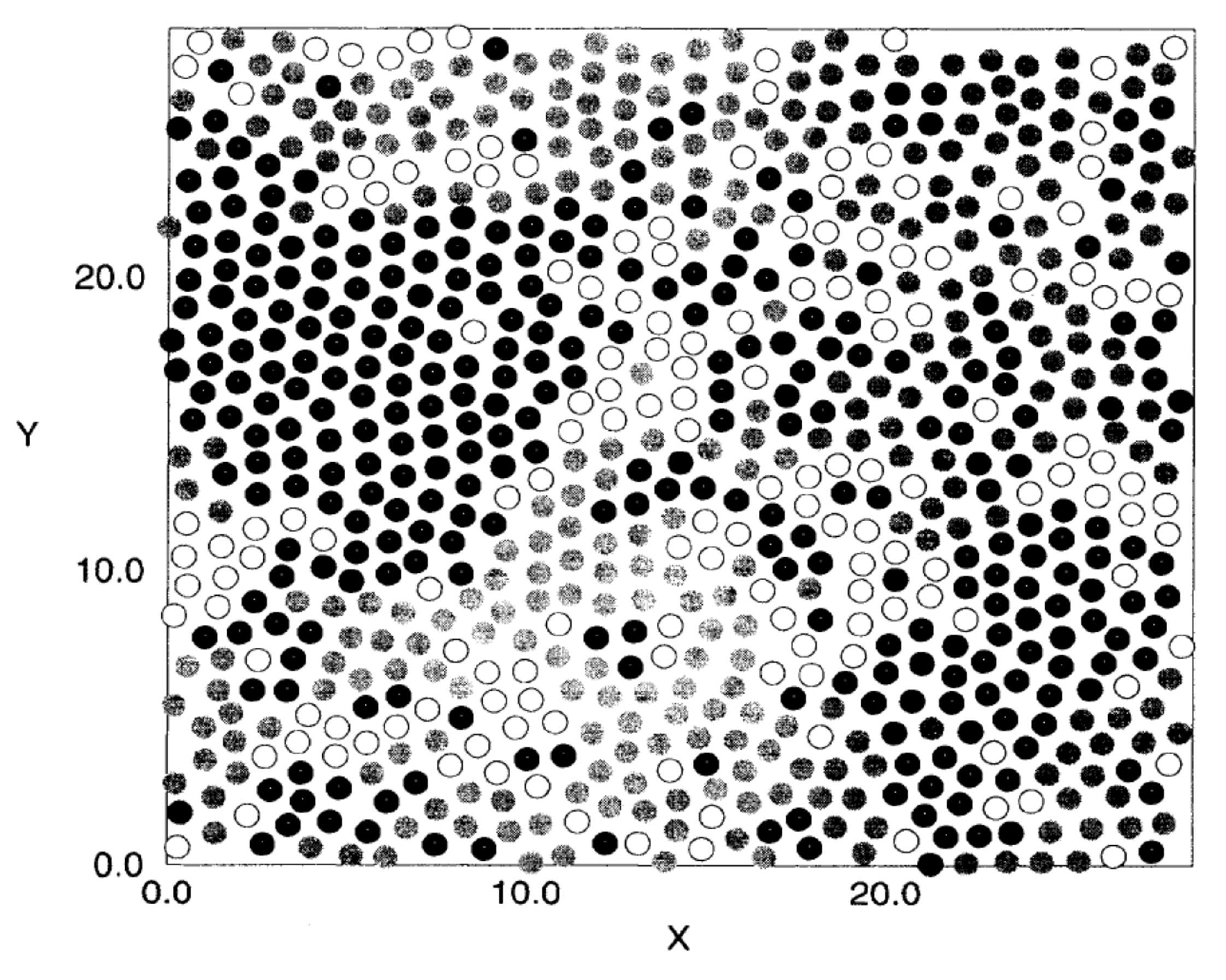} }
   \subfigure[]{\includegraphics[scale=0.3] {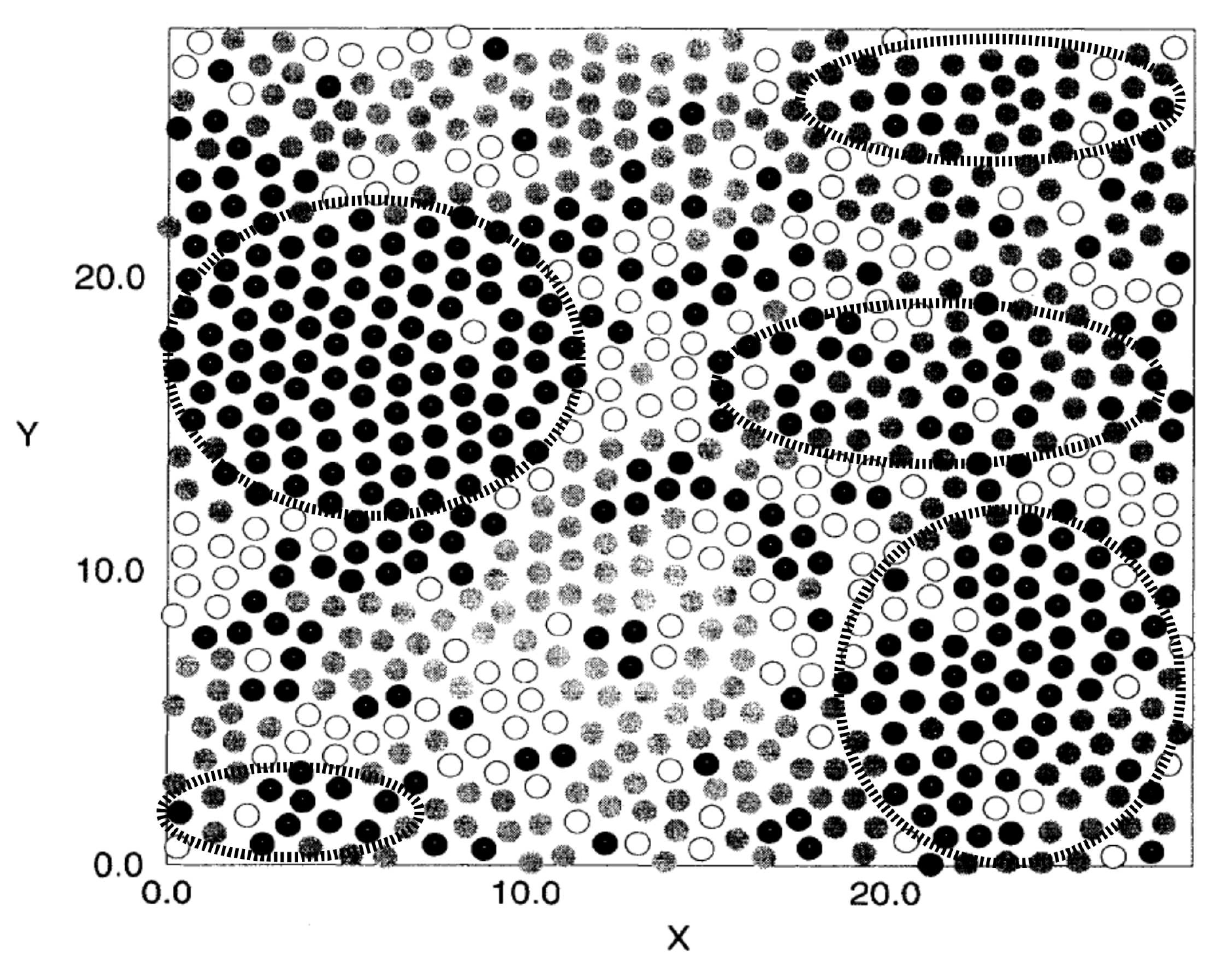} } \vskip -3mm
 }
\caption[2]{ (colour online) (a) Map of the relaxation times of one MD 
simulation run of a 2D system of 780 soft disks, interacting via a repulsive
$1/r^{12}$ potential cut off at some distance (from Ref.~\cite{Hur1995}).
Circles represent particle positions, the slowest 40\% coloured black, the 
fastest 40\% shaded gray and the intermediate 20\% are unfilled circles. Note
the distinct clustering of slow particles (dynamical heterogeneities, DHs). 
(b) The slower regions have been schematically highlighted to show their 
incipient cellular structure. }
\label{Fig1}
\end{figure}

\begin{figure}[h]
\centering
   \subfigure[]{\includegraphics[scale=0.50] {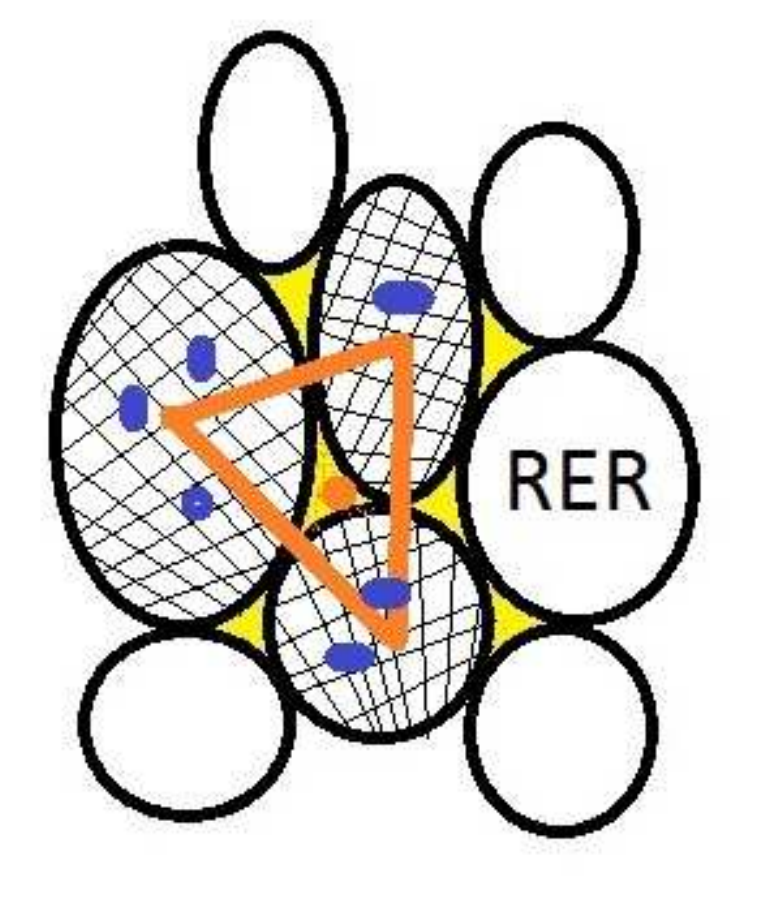} }
   \subfigure[]{\includegraphics[scale=0.25] {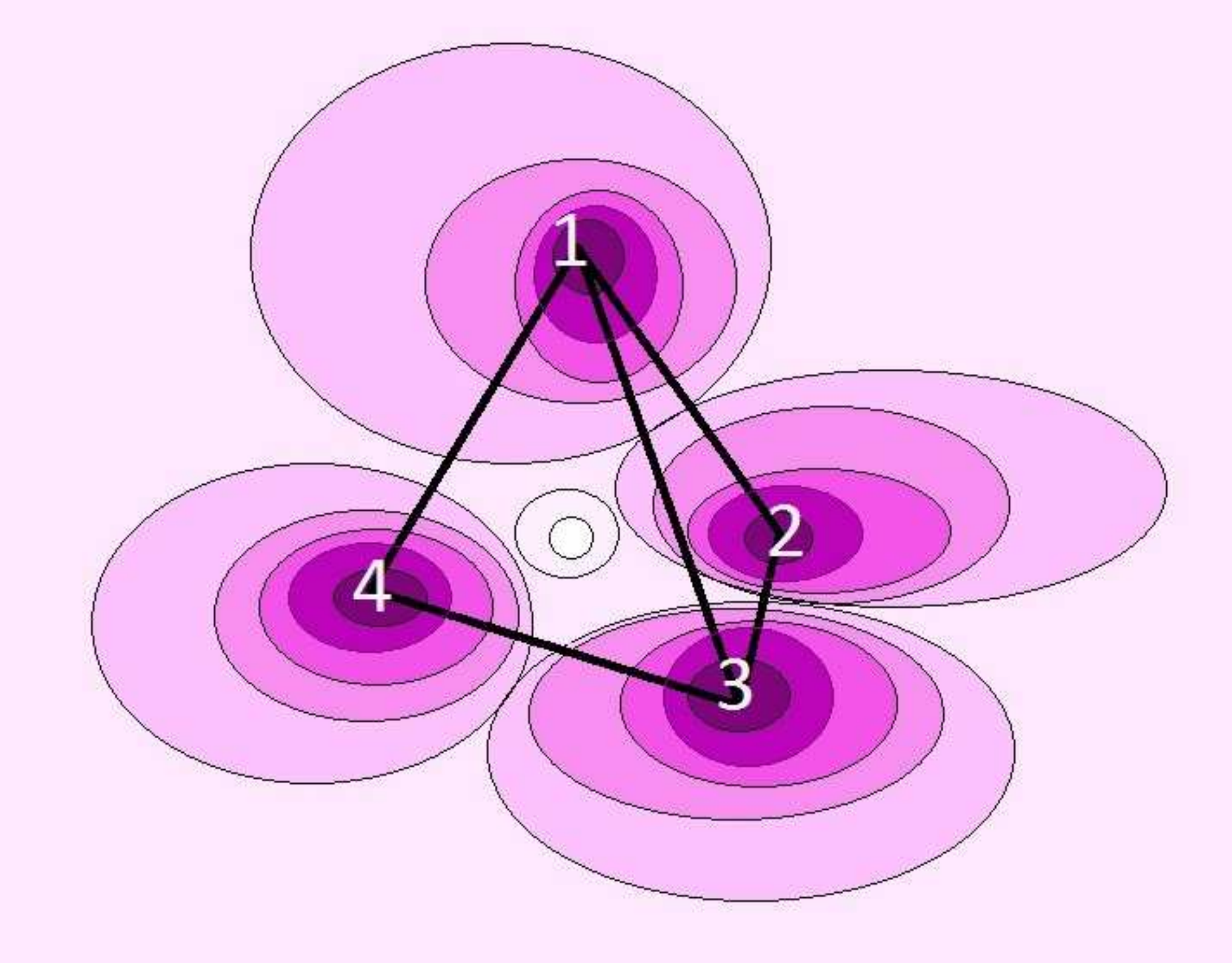} } 
\vskip -3mm
\caption[2]{ (colour online) (a) 2D cartoon of the cellular structure of an
amorphous solid just below $T_g$. The RERs (black-circled blobs, an 
oversimplified schematics for fractal-like, but compact objects) have grown 
to completely fill the space and enclose atomistic tunneling states of the 
2LS type (blue blobs). At the same time, in the RER interstitials (yellow 
regions, connecting to each other) the trapped, charged and faster particles 
of the DHs existing above $T_g$ and now probably charged dangling bonds, give 
rise to coherently tunneling large groups of ions to be represented by a 
single fictitious quasi-particle (orange dot) subjected to an effective 
tunneling potential having four natural wells in distorted tetrahedral 
configuration (b) for close-packed RERs. (b) The tetrahedral four-welled
potential (TFWP) in a 3D representation with colour-coded potential intensity
(dark=deepest, light=highest). }
\label{cellstructure}
\end{figure}

The structure proposed is a cellular-type arrangement of better-ordered regions
(regions of enhanced ordering, RERs) that can have complicated, maybe fractal,
but compact shapes with a narrow size distribution and interstitial regions 
between them populated by still fast-moving particles (normally charged, 
probably dangling bonded ions). We remark that a cellular-type structure was 
already embodied by the ``crystallite'' idea of Lebedev, but now no 
micro-crystals are here claimed to exists (except perhaps in the 
ceramic-glasses case, to an extent). The RERs are more like Wright's 
``cybotactic regions'' \cite{Wri2014} or Treacy's ``paracrystals'' 
\cite{Tre2012} and issue from the dynamical heterogeneities (DHs) picture of 
the glass-forming supercooled liquid phase above $T_g$ (see e.g. 
Fig.~\ref{Fig1}) \cite{Ber2011,Hur1995,Sil1999,Edi2000}. The DHs picture 
recognises the presence of regions of ``slower'' and ``faster'' particles,
and inspection of the slower-particle regions in the supercooled liquid 
reveals that these are also better ordered (solid-like) whilst the 
faster-particle regions are much more liquid-like. DHs are ubiquitous in 
most supercooled liquids \cite{Edi2000} and the claim (yet still somewhat 
speculative) here is that the slower-particle regions will grow on 
approaching $T_g$, but only up to a finite size and will be giving rise to 
the RERs in the frozen, glassy phase. Simulations in 
the frozen glassy state of the slower- and faster-particle regions confirm
that a DHs picture applies also below $T_g$ \cite{Vol2005} and with the 
slower-regions increasing in size as $T\to 0$. Earlier simulations (always for
model systems) \cite{Don1999} pointed out the difficulty of simulating the DHs 
picture below $T_g$ and came up with a picture of the slower-regions growing, 
possibly diverging in size, on approaching $T_g$. In this paper, however, the 
stance will be taken that the slower-regions' average size growths, in real
systems, but does not diverge at $T_g$ or at any other characteristic 
temperature. A full numerical proof of this fact is, however, still lacking.
We remark that a cellular structure for the glasses had been proposed in the 
past by de Gennes \cite{deG2002} and, in the context of the low-temperature 
anomalies, by Baltes \cite{Bal1973} who was able to explain the linear in $T$ 
anomaly in the heat capacity $C_p$ (but not those in the acoustic properties 
which require the introduction of 2LSs). A very similar picture is that of the
polyclusters of Bakai \cite{Bak1994}, as already mentioned, but here the 
thermal history of the cells, or RERs, is ascribed directly to the DHs 
situation already present above $T_g$.  In this approach the RERs, like 
grains in the frozen structure below $T_g$, are the thermal history 
continuation of the slower particle regions of the DHs above $T_g$ and contain 
the bulk of the 2LSs. These 2LSs are atomic tunneling states arising from the 
grains' own disorder causing a distortion in the natural bond angle between 
neighbouring cations with the anions becoming susceptible of taking up two 
nearly equivalent positions differing by a tiny energy (see Section 8). Some 
2LSs may be located at the meeting point between two grains, or cells (two 
RERs). In the interstitials between grains the faster particles of the DHs at 
$T>T_g$ give rise for $T<T_g$ to regions where a large number $N$ (on average) 
of charged atomistic tunneling particles are costrained to move in a coherent 
fashion due to the high Coulomb repulsion between them. 
Fig. \ref{cellstructure} shows in a
schematic way how the atomic/ionic matter can get organised below $T_g$ in a
real glass. 
Since the charged ions (dangling bonds, most likely) should act as a 
coherently tunneling ensemble, it seems natural to simplify the description
of the physics at the lower temperatures using phonons, propagating in the 
collection of cells now jammed against each other, and remnant localized 
degrees of freedom acting as TSs. These TSs will be the 2LSs within the cells
and at their points of contact (owing to inherent disorder in the cells' atomic 
arrangement) and effective quasi-particles sitting in the close-packed cells'
interstices and representing the collective motion of the coherently-
tunneling ions trapped in each interstice (Fig. \ref{cellstructure}). 
\begin{figure}[h]
\centering
   \subfigure[]{\includegraphics[scale=0.35] {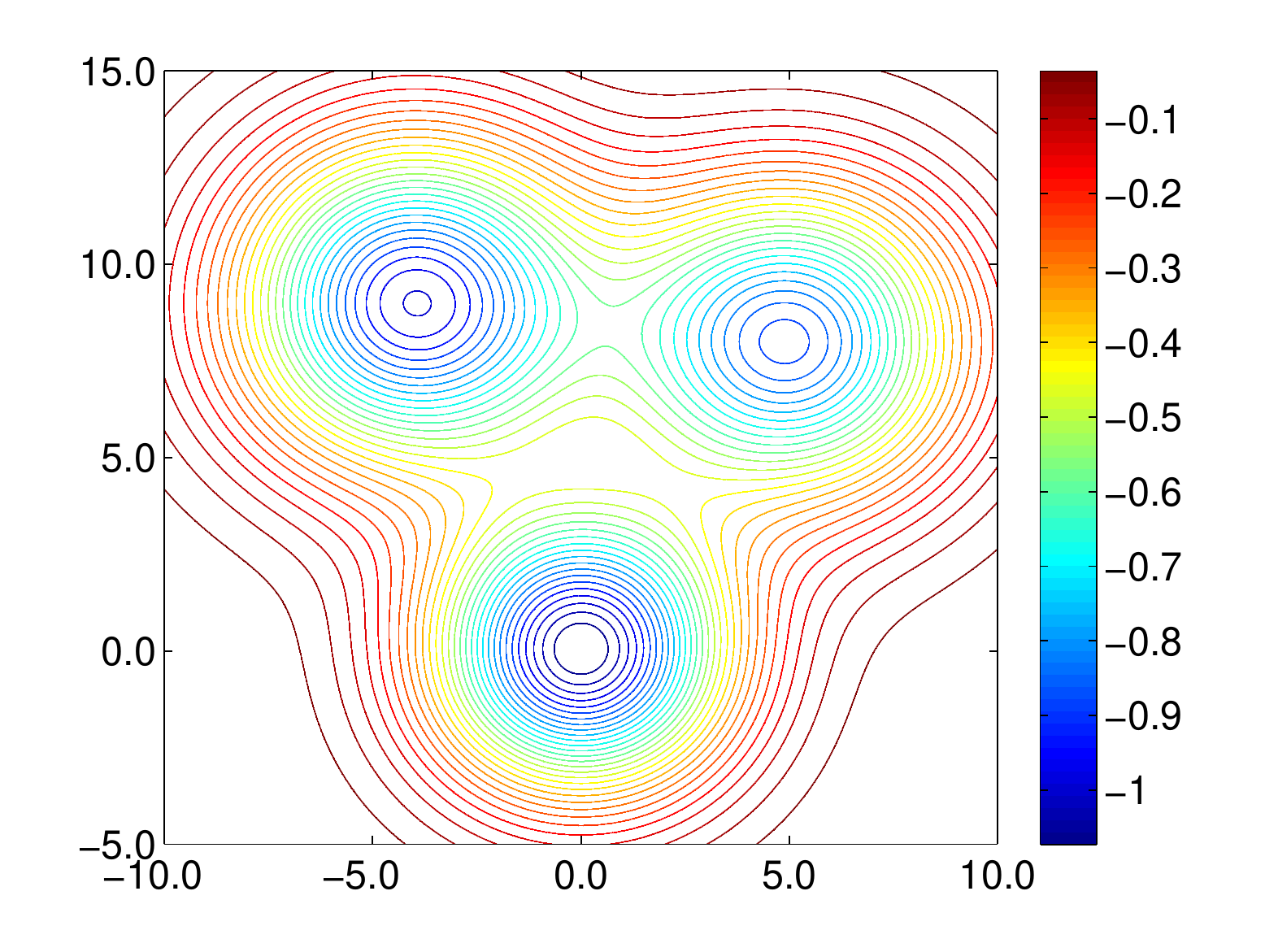} }
   \subfigure[]{\includegraphics[scale=0.35] {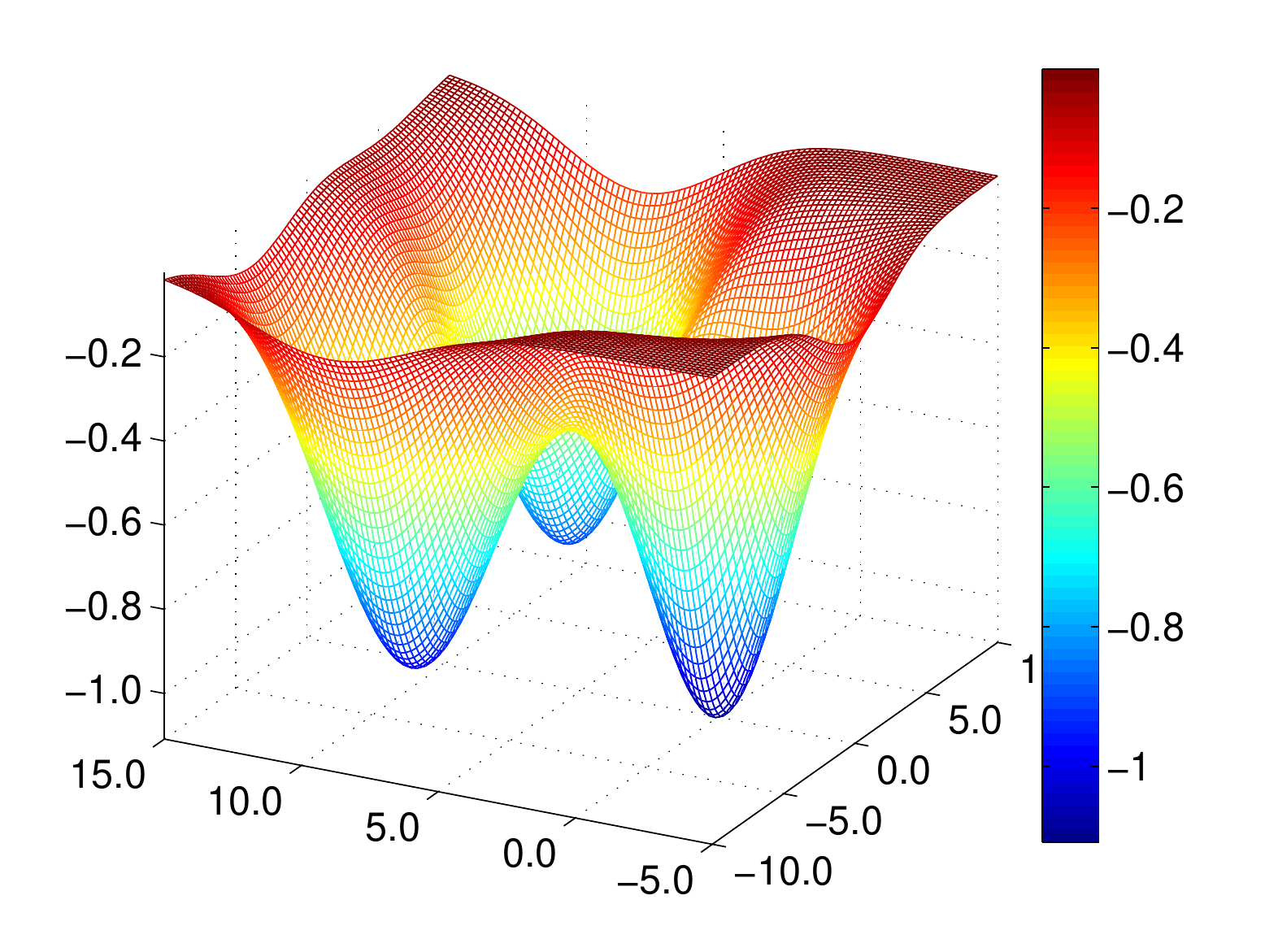} }
\vskip -3mm
\caption[2]{ (colour online) (a) Contour plot of a possible realization of the
2D effective three-welled potential (TWP) most likely felt by the 
quasi-particle of those charged real particles dangling from a group of three 
RERs on each one face of the tetrahedral configuration of an interstice formed 
by close-packed RERs well below $T_g$. (b) 3D visualization of the same example 
of a TWP potential. }
\label{threewells}
\end{figure}
The quasi-particle will be subjected to an effective potential of distorted-
tetrahedral shape characterized by four wells for each interstice, with a 
high barrier in the interstice's centre. {\it De facto} this 3D interstitial
TFWP potential can be replaced with four local 2D potentials for the four 
quasi-particles describing the coherent tunneling of the dangling-bond 
particles sitting near each face of the tetrahedron, close to a group of three
(on average) quasi-ordered cells (Fig. \ref{threewells}). Because of the better
ordering implicit in this model of the glassy intermediate-range atomic
structure and in each cell, the three wells of each effective local 2D 
potential for the tunneling quasi-particles (four per interstice, on average)
will be near-degenerate in terms of their ground-state energy asymmetries:
$E_1 \simeq E_2 \simeq E_3 \simeq 0$.
With this qualitative picture in mind, we now turn to the mathematical 
description of the physics of the remnant, still ergodic degrees of freedom
(phonons in the cells' network, 2LSs and ATSs (anomalous tunneling systems,
four in each interstice). For our model of a real glass, by construction the 
2LSs will be more numerous than the ATSs.

\section{The Quantum Mechanics of the Three Welled Potential}
In this approach \cite{Jug2004} the relevant degrees of freedom, beside the 
phonons, are dilute collections of independent 2LSs, described by the STM, and 
of fictitious quasi-particles tunneling in TWPs.
  
The formulation of the STM (the 2LS model) for the low temperature properties 
of glasses is well known. One assumes a collection of DWPs distributed in the 
substance and represented each by a 2$\times$2 Hamiltonian of the form, in the 
potential-well (or real-space) representation:
\begin{equation}
{\cal H}_{2LS}=-\frac{1}{2}\left( \begin{array}{cc}
\Delta & \Delta_0 \cr
\Delta_0 & -\Delta \end{array} \right).
\label{2lstunneling}
\end{equation}
Here the parameters $\Delta$ (the energy asymmetry) and $\Delta_0$ (twice the
tunneling parameter) are typically characterized by a probability
distribution that views $\Delta$ and $\ln(\Delta_0)$ (the latter linked to
the DWP energy barrier) broadly (in fact uniformly) distributed throughout
the disordered solid~\cite{Phi1987}:
\begin{equation}
{\cal P}_{2LS}(\Delta,\Delta_0)=\frac{\bar{P}}{\Delta_0}
\label{2lsdistribution}
\end{equation}
where some cutoffs are introduced when needed and where $\bar{P}$ is a
material-dependent parameter, like the cutoffs. In fact Eq. 
(\ref{2lsdistribution}) embodies the Zachariasen-Warren hypothesis for the 
intermediate atomic structure of a glass, assuming broadly distributed the
energy asymmetry $\Delta=E_2^{(0)}-E_1^{(0)}$, hence the single-well ground 
state energies $E_1^{(0)}$ and $E_2^{(0)}$ themselves, as well as the 
potential barrier height $V_0$ appearing in relations such as:
\begin{eqnarray}
&&\Delta_0\simeq\hbar\Omega e^{-\frac{d}{\hbar}\sqrt{2mV_0}} \\
&& \Delta_0=\frac{\hbar\Omega}{2} \left[3-\sqrt{\frac{8V_0}{\pi\hbar\Omega}}
\right]e^{-2\frac{V_0}{\hbar\Omega}} \label{overlap}
\end{eqnarray}   
where the first relation is the generic WKB result for an arbitrarily shaped 
DWP (where $m$ is the particle's mass, $\Omega$ its single-well harmonic
frequency (tunneling attempt frequency) and $d$ the tunneling distance) and 
the second formula refers to a symmetric ($\Delta=0$) DWP made up by two 
superimposed parabolic wells. In fact, the distribution (\ref{2lsdistribution})
in the end refers to the combination of parameters $\Delta_0/\hbar\Omega$. The
energies of the two levels $|0>$ and $|1>$ are then 
${\cal E}_{0,1}=\pm\frac{1}{2}\sqrt{\Delta^2+\Delta_0^2}$ and so on 
\cite{Phi1981,Phi1987}.

The tunneling Hamiltonian of a particle in a TWP is easily written down, in 
the same low-$T$ spirit as for a 2LS, as a generalization of the above matrix 
formulation to three levels~\cite{Jug2004}:
\begin{equation}
{\cal H}_{3LS}=\left( \begin{array}{ccc}
E_1 & D_0 & D_0 \cr
D_0 & E_2 & D_0 \cr
D_0 & D_0 & E_3 \end{array} \right)
\label{3lstunneling}
\end{equation}
where $E_1, E_2, E_3$ are the energy asymmetries between the wells and $D_0$ is
the most relevant tunneling amplitude (through saddles of the PEL, in fact).
This 3LS Hamiltonian has the advantage of readily allowing for the inclusion
of a magnetic field $B>0$, when coupling orbitally with a tunneling
``particle'' having charge $q$ ($q$ being some multiple of the electron's
charge $-e$)~\cite{Jug2004}:
\begin{equation}
{\cal H}_{3LS}(B)=\left( \begin{array}{ccc}
E_1 & D_0e^{i\varphi/3} & D_0e^{-i\varphi/3} \cr
D_0e^{-i\varphi/3} & E_2 & D_0e^{i\varphi/3} \cr
D_0e^{i\varphi/3} & D_0e^{-i\varphi/3} & E_3 \end{array} \right)
\label{3lsmagtunneling}
\end{equation}
where $\varphi/3$ is the Peierls phase for the tunneling particle through
a saddle in the field, and $\varphi$ is the Aharonov-Bohm (A-B) phase for a
tunneling loop and is given by the usual formula:
\begin{equation}
\varphi=2\pi\frac{\Phi}{\Phi_0}, \qquad \Phi_0=\frac{h}{\vert q\vert}
\label{ABphase}
\end{equation}
$\Phi_0$ being the appropriate flux quantum ($h$ is Planck's constant) and
$\Phi={\bf B}\cdot{\bf S}_{\triangle}$ the magnetic flux threading the area 
$S_{\triangle}$ formed by the tunneling paths of the particle in this simple 
(poor man's, yet as we have seen realistic) model. The energy
asymmetries $E_1, E_2, E_3$ typically enter through their combination
$D\equiv\sqrt{E_1^2+E_2^2+E_3^2}$. 

For $n_w$=3 wells an exact solution for the $k$=0, 1, 2 eigenvalues of the
multi-welled tunneling Hamiltonian Eq. (\ref{3lsmagtunneling}) is still 
possible:
\begin{eqnarray}
&&{\cal E}_k = 2D_0\sqrt{ 1-\frac{\sum_{i\not=j}E_iE_j}{6D_0^2} } ~
\cos\bigg( \frac{1}{3}\theta+\theta_k \bigg)
\label{3ls} \\
&&\cos\theta = \left( \cos\varphi+\frac{E_1E_2E_3}{2D_0^3} \right)
\left( 1-\frac{\sum_{i\not=j}E_iE_j}{6D_0^2} \right)^{-3/2}
\nonumber
\label{solution}
\end{eqnarray}
$\theta_k=0,+\frac{2}{3}\pi,-\frac{2}{3}\pi$ distinguishing the three
eigenstates. In the physically relevant limit, which we now discuss,
in which $\varphi\to 0$ (weak fields) and
$D=\sqrt{E_1^2+E_2^2+E_3^2}\to 0$ (near-degenerate distribution),
and at low temperatures, we can approximate (in a now simplified calculation) 
the $n_w=3$ - eigenstate system with an {\em effective 2LS} having
a gap $\Delta{\cal E}={\cal E}_1-{\cal E}_0$ widening with increasing 
$\varphi$ if $D_0>0$ (see below):
\begin{equation}
\lim\Delta{\cal E}\simeq
\frac{2}{\sqrt{3}}\sqrt{ D_0^2\varphi^2+\frac{1}{2}(E_1^2+E_2^2+E_3^2) }
\to \sqrt{D_0^2\varphi^2+D^2}
\label{3gap}
\end{equation}
(a trivial rescaling of $D_0$ and of the $E_i$ was applied).
One can easily convince oneself that if such a TWP is used with the standard 
parameter distribution, Eq.~(\ref{2lsdistribution}) with $D, D_0$ replacing 
$\Delta, \Delta_0$, for the description of the TS, one would then obtain 
essentially the same physics as for the STM's 2LS-description. In other words, 
there is no need to complicate
the minimal 2LS-description in order to study glasses at low temperatures,
unless structural inhomogeneities of the RER-type and a magnetic field are
present. Without the RERs, hence no distribution of the type
(\ref{atsdistribution}) below, the interference from separate tunneling paths is
only likely to give rise to an exceedingly weak A-B effect. Hence, it will be 
those TSs nesting between the RERs that will give rise to an enhanced 
A-B interference and these TSs can be minimally described -- most 
appropriately -- through Hamiltonian (\ref{3lsmagtunneling}) and with 
distribution of asymmetries
thus modified to favour near-degeneracy ($P^*$ being a dimensionless material
parameter)~\cite{Jug2004}:
\begin{equation}
{\cal P}_{3LS}^*(E_1,E_2,E_3;D_0)=\frac{P^{\ast}}{D_0(E_1^2+E_2^2+E_3^2)}.
\label{atsdistribution}
\end{equation}
We remark that the incipient ``crystallinity'' of the RERs calls for
near-degeneracy in $E_1, E_2, E_3$ simultaneously and not in a single one
of them, whence the correlated form of (\ref{atsdistribution}). We now have
three-level systems (3LSs) with energy levels
${\cal E}_0<{\cal E}_1\ll {\cal E}_2$, periodic in $\varphi$. The typical
spectrum, with $D_0>0$ (see below), is shown in Fig. \ref{spectrum} as a
function of $\varphi$ and one can see that the third, highest level 
${\cal E}_2$ can be safely neglected for most applications. Other
descriptions, with TFWPs or modified three-dimensional DWPs are possible for 
the TSs nested in the RERs and lead to the same physics as from 
Eqs.~(\ref{3lsmagtunneling}) and (\ref{atsdistribution}) above~\cite{Bon2015} 
(which describe what we call the anomalous tunneling
systems, or ATSs, nesting in the interstitials between the RERs).
\begin{figure}[h]
\centering
{
   {\includegraphics[scale=0.50] {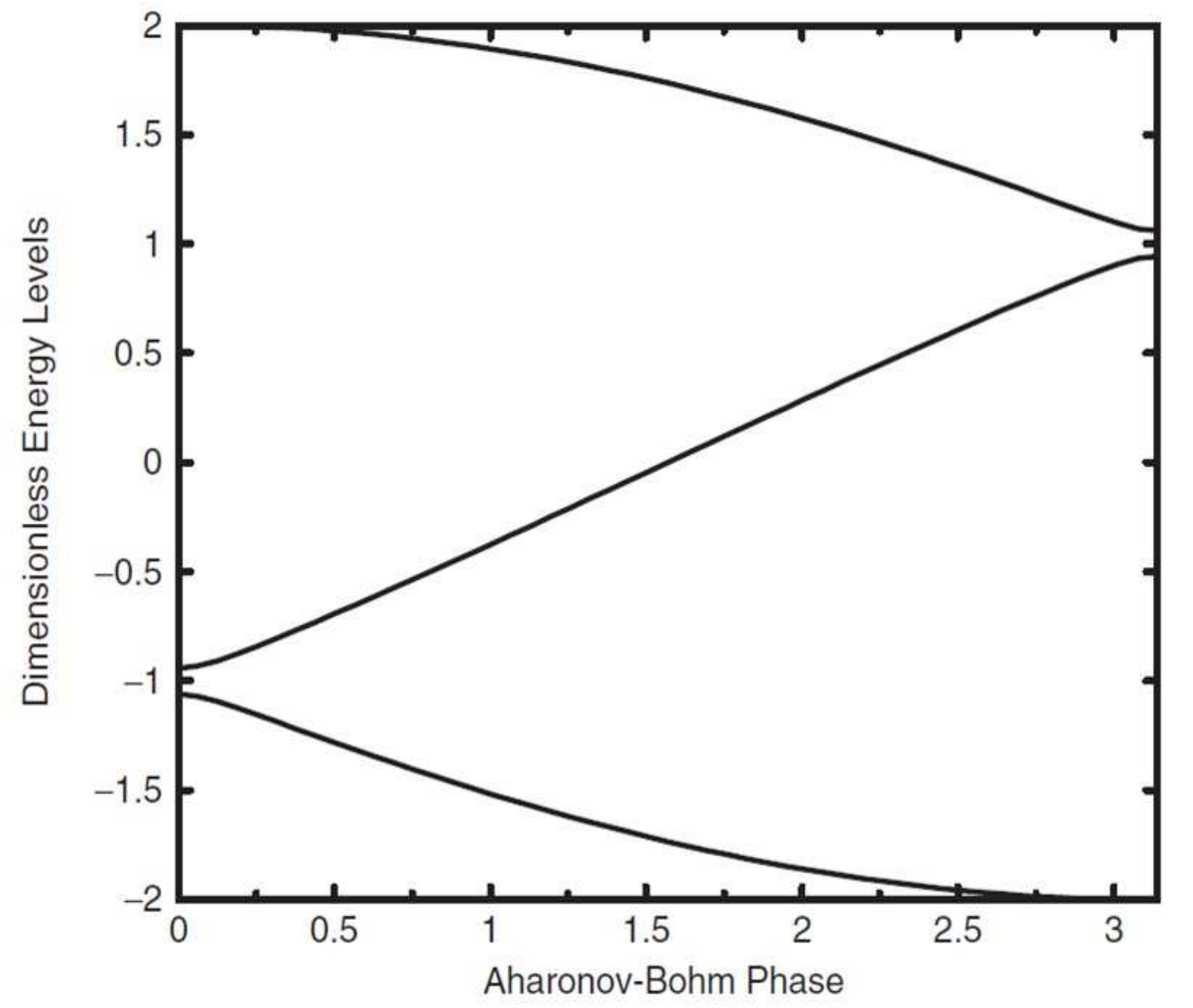} }
\vskip -3mm
}
\caption{ The energy spectrum ($D_0=1$ units) of the ATS (3LS, TWP) model, in 
the appropriate limits of weak field and near-degeneracy due to the embedding 
within RER's interstitials. On the horizontal axis the A-B phase 
$\varphi\propto B$. }
\label{spectrum}
\end{figure}
The final and most important consideration for the construction of a suitable 
mathematical model is that the TSs appear to be rather
diluted defects in the glass (indeed their concentration is of the order of
magnitude of that for trace paramagnetic impurities, as we shall see), hence
the tunneling ``particles'' are embedded in a medium otherwise characterized
only by simple acoustic-phonon degrees of freedom. This embedding, however,
means that the rest of the material takes a part in the making of the
tunneling potential for the TS's ``particle'', which itself is not moving
quantum-mechanically in a vacuum. Sussmann~\cite{Sus1962}
has shown that this leads to local trapping potentials that (for the case of
triangular and tetrahedral perfect symmetry) must be characterized by a
degenerate ground state. This means that, as a consequence of this TS
embedding, our poor man's model, Eq. (\ref{3lsmagtunneling}), for the ATSs 
must be chosen with a positive tunneling parameter~\cite{Jug2004}:
\begin{equation}
D_0>0
\label{degeneracy}
\end{equation}
where of course perfect degeneracy is always removed by weak disorder in the
asymmetries. The intrinsic near-degeneracy of (\ref{atsdistribution}) implies
that this model should be used in its $D/D_0\ll 1$ limit, which in turn reduces
the ATSs to effective magnetic-field dependent 2LSs and greatly simplifies the
analysis together with the limit $\varphi\to 0$ which we always take for
relatively weak magnetic fields. The ETM, first proposed in \cite{Jug2004}, 
consists then in a collection of independent, non-interacting 2LSs described 
by the STM and also 3LS TWPs, described by Eqs.~(\ref{3lsmagtunneling}) and 
(\ref{atsdistribution}) above, in the said $D/D_0\ll 1$ and $\varphi\to 0$ 
limits, the 3LSs nested in the interstitials between the close-packed RERs and 
the magnetic-field insensitive 2LSs distributed in the remaining 
homogeneously-disordered matrix of RERs and at their touch points or interfaces
\cite{Jug2013}.
\begin{figure}[!Htbp]
  \centering
  \includegraphics[scale=0.25] {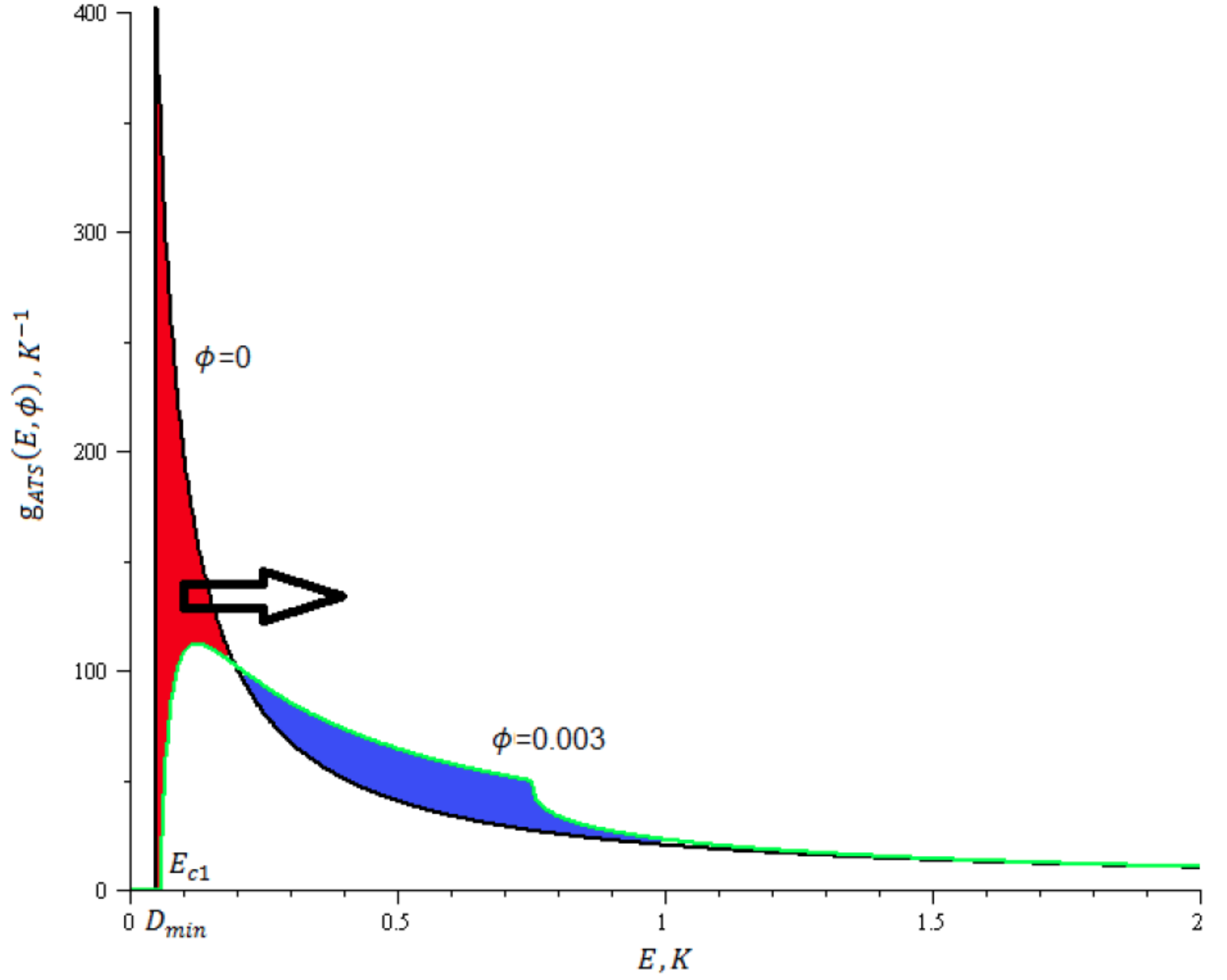}
\caption{(colour online) The magnetic-sensitive part of the density of states
(DOS) as a
function of the energy gap $E$ and different A-B phases $\varphi$ (proportional
to the magnetic field $B$) ($n_{ATS}P^{\ast}$ has been set to 1). The rapid
shift of quantum states to higher energy when a very weak $B$ is switched on
is the physical explanation for the origin of the magnetic effects. }  
  \label{dosfigphi}
\end{figure}
In Fig.~\ref{dosfigphi} we illustrate the behaviour of the density of states
(DOS) for this model as a 
function of the gap energy $E$ for different $\varphi$ values. This figure 
shows the physical origin of the magnetic effects: the quantum states being
conserved in number, they are very rapidly shifted towards high values of the
energy when a magnetic field, even very weak, is switched on.
\begin{figure}[!Htbp]
  \centering
  \includegraphics[scale=0.50] {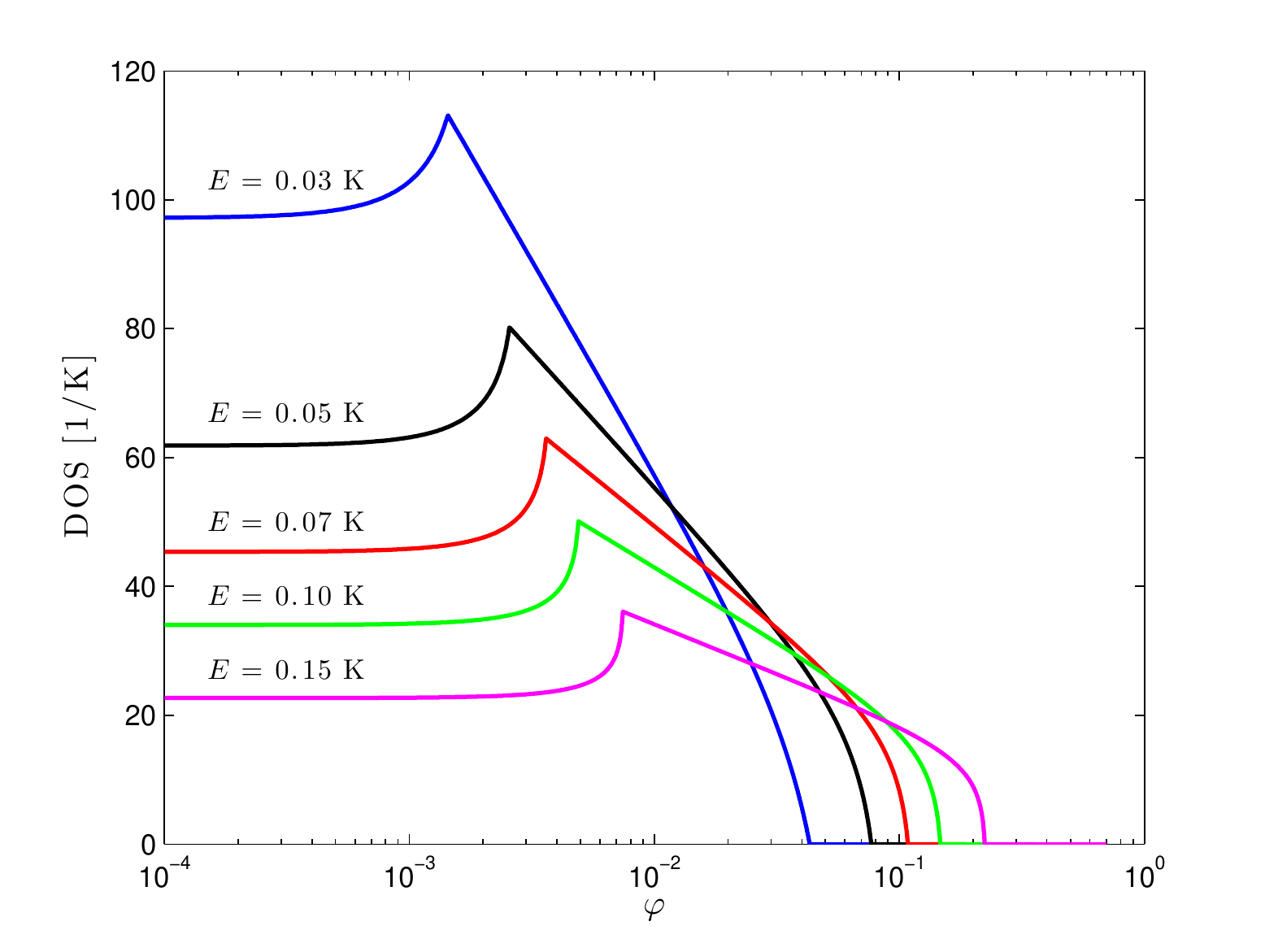}
\caption{(colour online) The magnetic-sensitive part of the density of states
(DOS) as a
function of the A-B phase $\varphi$ (proportional to the magnetic field $B$)
and different energies ($n_{ATS}P^{\ast}$ has been set to 1). The shape of
this part of the DOS (coming from the TWPs with a parameter distribution
(\ref{atsdistribution}) favouring near-degeneracy) is the ultimate source of
all the magnetic effects.
The cusp is an artifact of the effective 2LS approximation~\cite{Jug2004}, but
also of the existence of upper and lower bounds for $D_0$ owing to the nature
of the RER glassy atomic structure.}
  \label{dosfig}
\end{figure}
Our ETM has been able to explain the magnetic effects in the heat
capacity~\cite{Jug2004}, in the real~\cite{Jug2009} and
imaginary~\cite{Jug2014} parts of the dielectric constant and in the
polarization echo amplitude~\cite{Jug2014} measurements reported to date for
various glasses at low temperatures, as well as the composition-dependent
anomalies~\cite{Jug2010,Jug2013}.
The new physics is provided by the magnetic-field dependent TS DOS, which 
acquires a term due to the near-degenerate
TWPs~\cite{Jug2004} that gets added up to the constant DOS from the
STM 2LSs (having density $n_{2LS})$:
\begin{eqnarray}
g_{tot}(E,B)&=&n_{2LS}\bar{P}
+n_{ATS}\frac{P^{\ast}}{E}f_{ATS}(E,B)\theta(E-E_{c1}) \nonumber \\ 
&=&g_{2LS}(E)+g_{ATS}(E,B) 
\label{dos}
\end{eqnarray}
where $n_{ATS}$ is the ATSs' concentration,
$f_{ATS}$ is a magnetic-field dependent dimensionless function, already
described in previous papers~\cite{Jug2004}, and $E_{c1}$ is a material and
$B$-dependent cutoff:
\begin{eqnarray}
g_{ATS}(E,\varphi)=\int \Pi_i dE_i \delta(\Sigma_j E_j) \int d D_0
{\cal P}^*_{3LS}(\{E_k\},D_0) \delta(E-\Delta{\cal E}) \cr
=\begin{cases}
\frac{2\pi P^*}{E}\ln\left( \frac{D_{0max}}{D_{0min}}
\sqrt{ \frac{E^2-D_{0min}^2\varphi^2}{E^2-D_{0max}^2\varphi^2} }
\right) \qquad &{\rm if} \quad E>E_{c2}  \cr
\frac{2\pi P^*}{E}\ln \frac{ \sqrt{ (E^2-D_{0min}^2\varphi^2)
(E^2-D_{min}^2) } }{ D_{0min}D_{min}\varphi }
&{\rm if} \quad E_{c1}\leq E \leq E_{c2}   \cr
0 &{\rm if} \quad E<E_{c1}. \cr
\end{cases}
\label{3dos}
\end{eqnarray}
Here, after a suitable renormalization of parameters: 
$E_{c1}=\sqrt{D_{min}^2+D_{0min}^2\varphi^2}$, 
$E_{c2}=\sqrt{D_{min}^2+D_{0max}^2\varphi^2}$, $D_{min}$, $D_{0min}$ and
$D_{0max}$ being suitable cutoffs (material parameters).
The $1/E$ dependence of the ATS DOS is a consequence of the chosen tunneling 
parameter distribution, Eq.~(\ref{atsdistribution}), and gives rise to a peak 
in $g_{tot}$ near $E_{c2}$ that is rapidly eroded away as soon as a weak 
magnetic field is switched on. The form and evolution of the magnetic part 
of the DOS is shown in Fig.~\ref{dosfig} for some typical parameters, as a 
function of $\varphi\propto B$ for different values of $E$.
This behaviour of the DOS with $B$ is, essentially, the underlying mechanism
for all of the experimentally observed magnetic field effects in the cold
glasses within this model: the measured physical properties are convolutions 
of this DOS (with
appropriate $B$-independent functions) and in turn reproduce its shape as
functions of $B$. 

\section{The Magnetic Field Dependent Heat Capacity}
\subsection{Theory}
As a first example, the total TS heat capacity is given by
\begin{equation}
C_{pTS}(T,B)=\int_0^{\infty}dE~g_{tot}(E,B)C_{p0}(E,T)
\label{convolution}
\end{equation}
where
\begin{equation}
C_{p0}(E,T)=k_B\left( \frac{E}{2k_{B}T} \right)^2\cosh^{-2}\left(
\frac{E}{2k_{B}T} \right)
\end{equation}
is the heat capacity contribution from a single TS having energy gap $E$ and
where $g_{tot}(E,B)$ is given by Eq. (\ref{dos}).

In this Section we re-analyze some of the available data~\cite{Sie2001} for 
the magnetic effect in the heat capacity of two multi-component glasses,
commercial borosilicate Duran and barium-allumo-silicate (AlBaSiO, or BAS in 
short) glass, in order to show that the ATS model works well for the 
magnetic-field dependent $C_p$. A systematic experimental
study of $C_p(T,B)$ around and below 1 K in these multi-silicate glasses was
carried out by Siebert~\cite{Sie2001} and those data have been used, upon
permission, by one of us~\cite{Jug2004} as the very first test of the above
(Sections 2, 3) ETM~\cite{Jug2004}. That earlier analysis best-fitted the $C_p$
data by Siebert with the sum of Einstein's $\gamma_{ph}T^3$ phonon term plus
the 2LS $\gamma_{2LS}T$ non-magnetic contributions, as well as with Langevin's
paramagnetic and the ATS contributions (see below). The analysis came up with
concentrations $\bar{n}_J\simeq$48 ppm and, respectively, $\bar{n}_J\simeq$ 20
ppm instead of the quoted~\cite{Sie2001} 126 ppm (or 180 ppm in a different
study~\cite{Her2000}) and 102 ppm for Duran and for BAS glass, respectively.

In order to better understand this large discrepancy we re-analysed Siebert's 
data for $C_p(T,B)$, after subtraction of the data taken at the same 
temperatures for the same glass, but in the presence of the strongest applied 
magnetic field (8~T) \cite{Bon2015b}. In this way only the magnetic-field
dependent contributions should remain in the data for
$\bar{C}_p(T,B)\equiv C_p(T,B)-C_p(T,\infty)$. The parameters involved when 
fitting data are the cutoff $D_{min}$ and combinations of
cutoffs, charge and area $D_{0min}qS$ and $D_{0max}qS$~\cite{Jug2004}, as
well as: $n_{Fe^{2+}}$ (Fe$^{2+}$ impurity concentration), $n_{Fe^{3+}}$ 
(Fe$^{3+}$ impurity concentration) and $n_{ATS}$ (ATS concentration (always 
multiplied by $P^{\ast}$)).
The data from \cite{Sie2001} (we have
restricted the best fit to the three temperatures having the most data points
around the peak of $C_p(B)$) have been best-fitted by using the following
magnetic-dependent contributions:
\begin{enumerate}
\item the known Langevin contribution of the paramagnetic Fe impurities
(Fe$^{2+}$ and Fe$^{3+}$) having concentration $n_J$:
\begin{equation}
C_J(T,B)=n_J \frac{k_B z^2}{4} \bigg(\bigg(\frac{1}{\sinh \frac{z}{2}}\bigg)^2-\bigg(\frac{2J+1}{\sinh \frac{(2J+1)z}{2}}\bigg)^2\bigg)
\label{param_impu_formula_cp}
\end{equation}
where $z=\frac{g\mu_B JB}{k_B T}$ and where $g$ is Land\`e's factor for the
paramagnetic ion in that medium, $\mu_B$ is Bohr's magneton and $J$ the ion's
total angular momentum (in units $\hbar$=1); $k_B$ is Boltzmann's constant.
We have assumed the same values the parameters $g$ and $J$ take for Fe$^{2+}$
and Fe$^{3+}$ in crystalline SiO$_2$: $J$=2 with $g$=2 and, respectively,
$J=5/2$ with $g$=2 (we have adopted, in other words, complete quenching of
the orbital angular momentum~\cite{Abr1961}, consistent with other Authors'
analyses~\cite{Sie2001,Lud2003}).
\item the averaged contribution of the ATSs~\cite{Jug2004}, written in
terms of a sum of individual contributions from each ATS of lowest energy
gap $E$ (making use of Eqs. (\ref{dos}), (\ref{3dos}) and (\ref{convolution})): 
\begin{equation}
\begin{split}
&C_{ATS}(T,\varphi)=\frac{\pi}{4} \frac{P^{\ast}~n_{ATS}}{k_{B}T^2}\\
&\times\Big\lbrace
\int_{E_{c1}}^{E_{c2}}~dE~\frac{E}{\cosh^2(\frac{E}{2k_{B}T})}~
\ln \Big[ \frac{(E^2-D_{0min}^2\varphi^2)(E^2-D_{min}^2)}
{D_{min}^2D_{0min}^2\varphi^2} \Big] \\
&+\int_{E_{c2}}^{\infty}~dE~\frac{E}{\cosh^2(\frac{E}{2k_{B}T})}~
\ln \Big[ \bigg( \frac{D_{0max}}{D_{0min}} \bigg)^2
\frac{E^2-D_{0min}^2\varphi^2}{E^2-D_{0max}^2\varphi^2} \Big]
\Big\rbrace\\
\end{split}
\end{equation}
or, re-written in a dimensionless form as
\begin{equation}
\begin{split}
&C_{ATS}(T,\varphi)=\widetilde{C}_0(T,\varphi)+2\pi P^{\ast} n_{ATS} k_B \Big\lbrace\big[I(x_{c1})-I(x_{c2})\big]\ln(x_{min}x_{0min}\varphi)\\
&+\frac{1}{2}\big[\mathcal{I}(x_{c1},x_{min})-\mathcal{I}(x_{c2},x_{min})+\mathcal{I}(x_{c1},x_{0min}\varphi)-\mathcal{I}(x_{c2},x_{0max}\varphi)\Big\rbrace\\
\end{split}
\label{heat_capa_tot_formula}
\end{equation}
where:
\begin{itemize}
\item $E_{c1}=\sqrt{D_{min}^2+D_{0min}^2 \varphi^2}$ and $E_{c2}=\sqrt{D_{min}^2+D_{0max}^2 \varphi^2}$;
\item $x_{c1,2}=\frac{E_{c1,2}}{2k_{B}T}$, $x_{min}=\frac{D_{min}}{2k_{B}T}$, etc.;
\item $I(x) \equiv x \tanh x-\ln \cosh x$;
\item $\mathcal{I}(x,a) \equiv \int_x^\infty \mathrm{d}y \frac{y}{\cosh^2 y}\ln(y^2-a^2)$.
\end{itemize}
and the following expression:
\begin{equation}
\widetilde{C}_0(T,\varphi)=2\pi P^{\ast} n_{ATS} k_B \ln\bigg(\frac{D_{0max}}{D_{0min}}\bigg)\big\lbrace \ln(2)-I(x_{c2}). \big\rbrace\\
\end{equation}
The angular average over the ATS orientations is performed by replacing
$\varphi\rightarrow\frac{\varphi}{\sqrt{3}}$ (in other words averaging
$\cos^2\theta$,
$\theta$ being the orientation of ${\bf S}$ with respect to ${\bf B}$).
\end{enumerate}
The above formula for $C_{ATS}$ is actually correct only for weak magnetic
fields (up to about 1 T). For higher magnetic fields one must make
use of an improved form~\cite{Pal2011} for the ATS lower energy gap. In
practice, this consists in making the following replacement for the A-B phase
$\varphi$
\begin{equation}
\varphi^2\to\varphi^2\Big( 1-\frac{1}{27}\varphi^2 \Big)\to\frac{1}{3}
\varphi^2\Big\{1- \frac{1}{45}\bigg( \frac{B}{B^{\ast}} \bigg)^2 \Big\}
\label{correction}
\end{equation}
where the second expression holds after orientational averaging and where
$B^{\ast}$ is the upturn value of the magnetic field for, e.g., the
$B$-dependence of the dielectric constant $\epsilon'(T,B)$~\cite{Pal2011}.
\subsection{Comparison with available data} 
\begin{figure}[!Htbp]
\centering
  \subfigure[]{\includegraphics[scale=0.50] {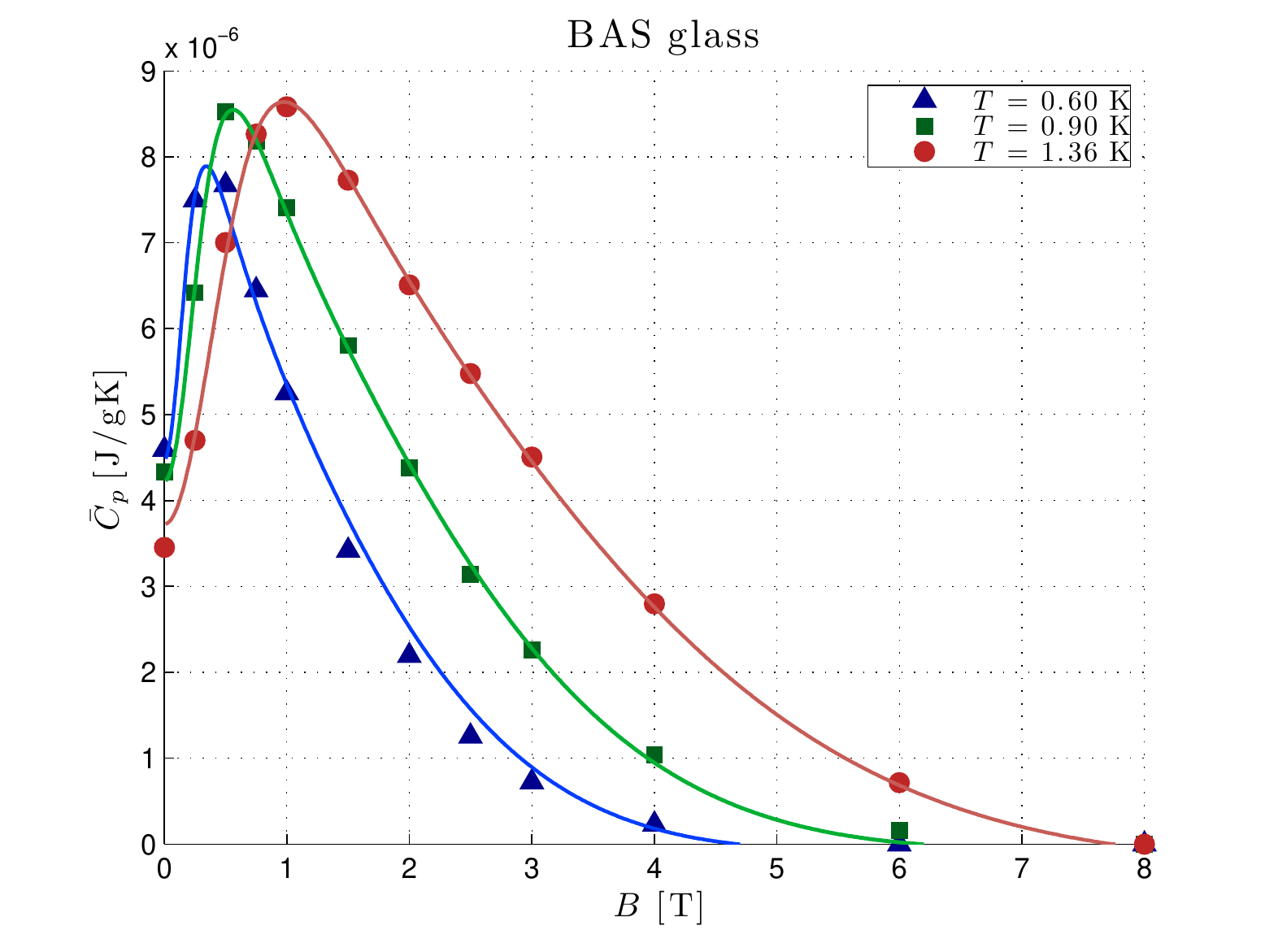} \label{c_tot_albasi}}
  \subfigure[]{\includegraphics[scale=0.50] {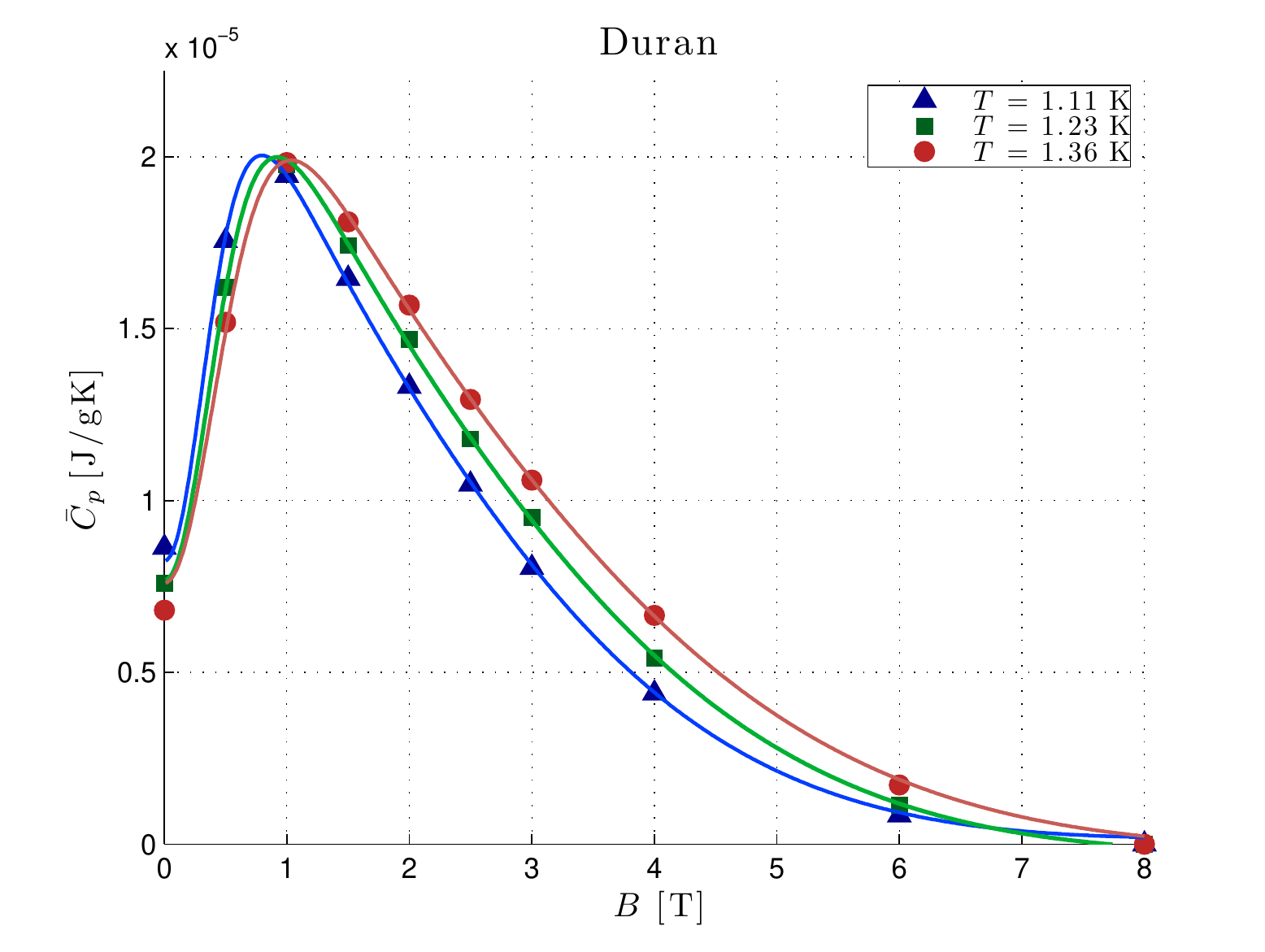} \label{c_tot_duran} }
\caption{(colour online) The heat capacity best fit for the a)~BAS (or AlBaSiO)
and b)~Duran glasses. The continuous lines are the predictions from the ETM.}
\label{heat_capa_fit}
\end{figure}

The concentrations of the ATSs and Fe-impurities extracted from the best fit of
the heat capacity as a function of $B$, for the BAS glass, are reported in
Table~\ref{tab_imp_extr}; having fixed the concentrations, it was possible to
extract the other parameters for the BAS glass (Table~\ref{tab_d0_ALBASI_extr}).
The best fit of the chosen data is reported in Fig.~\ref{heat_capa_fit}(a).
\begin{table}[!Htbp]
\begin{center}
\begin{tabular}{|c|c|c|}
\hline
BAS glass & Concentration $\mathrm{[g^{-1}]}$ & Concentration [ppm] \\
\hline
\hline
\textbf{$n_{Fe^{2+}}$} & 1.06$\times10^{17}$ & 14.23 \\
\textbf{$n_{Fe^{3+}}$} & 5.00$\times10^{16}$ &  6.69\\
\textbf{$P^{\ast}n_{ATS}$} & 5.19$\times10^{16}$ & - \\
\hline
\end{tabular}
\caption{Extracted parameters (from the heat capacity data) for the
concentrations of ATSs and Fe-impurities for the BAS glass.}
\label{tab_imp_extr}
\end{center}
\end{table}
\begin{table}[!Htbp]
\begin{center}
\begin{tabular}{|c|c|c|c|}
\hline
Temperature [K] & $D_{min}$ [K] & $D_{0min}\vert\frac{q}{e}\vert S$ [K$\AA^2$] & $D_{0max}\vert\frac{q}{e}\vert S$ [K$\AA^2$]\\
\hline
\hline
0.60 & 0.49 & 4.77$\times 10^{4}$ & 3.09$\times 10^{5}$ \\
0.90 & 0.53 & 5.07$\times 10^{4}$ & 2.90$\times 10^{5}$ \\
1.36 & 0.55 & 5.95$\times 10^{4}$ & 2.61$\times 10^{5}$ \\
\hline
\end{tabular}
\caption{Extracted tunneling parameters (from the $C_p$ data) for the BAS
glass.}
\label{tab_d0_ALBASI_extr}
\end{center}
\end{table}
The concentrations of the ATSs and Fe-impurities extracted from the best fit of
the heat capacity as a function of $B$, for Duran, are reported in
Table~\ref{tab_imp_extr_dur}; having fixed the concentrations, it was possible
to extract the other parameters for Duran (Table~\ref{tab_d0_duran_extr}). The
fit of the chosen data is reported in Fig.~\ref{heat_capa_fit}(b).
The first comment that we make is that these good fits, with a smaller set of
fitting parameters, repropose concentrations and tunneling parameters very
much in agreement with those previously obtained by one of us~\cite{Jug2004}.
The problem with
the concentrations of the Fe-impurities reported in the literature is that they
do not allow for a good fit of the $\bar{C}_p(B)=C_p(B)-C_p(\infty)$ data in
the small field region when the Langevin contribution alone is employed
(Eq.~(\ref{param_impu_formula_cp}) without Eq.~(\ref{heat_capa_tot_formula})).
See for instance Section 5, Fig.~\ref{spec_heat_contri_duran}.
The Langevin contribution drops to zero below the peak, whilst Siebert's
data definitely point to a non-zero value of $\bar{C}_p(0)=C_p(0)-C_p(\infty)$ 
at $B=0$ for any $T>0$. This non-zero difference is well accounted for by our 
ETM and it comes from the ATS contribution to the DOS in Eq.~(\ref{dos}).
\begin{table}[!Htbp]
\begin{center}
\begin{tabular}{|c|c|c|}
\hline
Duran & Concentration $\mathrm{[g^{-1}]}$ & Concentration [ppm] \\
\hline
\hline
\textbf{$n_{Fe^{2+}}$} & 3.21$\times10^{17}$ &  33.01 \\
\textbf{$n_{Fe^{3+}}$} & 2.11$\times10^{17}$ &  21.63 \\
\textbf{$P^{\ast}n_{ATS}$} & 8.88$\times10^{16}$ & - \\
\hline
\end{tabular}
\caption{Extracted parameters (from the heat capacity data) for the
concentration of ATSs and Fe-impurities for Duran.}
\label{tab_imp_extr_dur}
\end{center}
\end{table}
\begin{table}[!Htbp]
\begin{center}
\begin{tabular}{|c|c|c|c|}
\hline
Temperature [K] & $D_{min}$ [K] & $D_{0min}\vert\frac{q}{e}\vert S$ [K$\AA^2$] & $D_{0max}\vert\frac{q}{e}\vert S$ [K$\AA^2$]\\
\hline
\hline
1.11 & 0.34 & 4.99$\times 10^{4}$ & 2.68$\times 10^{5}$ \\
1.23 & 0.32 & 5.30$\times 10^{4}$ & 2.50$\times 10^{5}$ \\
1.36 & 0.32 & 5.54$\times 10^{4}$ & 2.46$\times 10^{5}$ \\
\hline
\end{tabular}
\caption{Extracted tunneling parameters (from the $C_p$~data) for Duran.}
\label{tab_d0_duran_extr}
\end{center}
\end{table}
The parameter $B^{\ast}$ was left indetermined in these fits, owing to the fact
that the measured $C_p$ becomes very small at higher fields.

The results of our $C_p$ analysis definitely indicate that the concentration of
paramagnetic impurities in the multi-silicate glasses is much lower than
previously thought and extracted from SQUID-magnetometry measurements of the
magnetization $M(T,B)$ at moderate to strong field values and as a function of
$T$. Therefore we will turn our attention to a re-analysis of the
SQUID-magnetometry data in Section 7.

\section{The Magnetic Field Dependent Dielectric Constant}
\subsection{Dielectric properties of cold glasses: general theory in zero
magnetic field}
To measure the frequency-dependent dielectric properties of the cold glasses
one applies an ac electric field to the sample, typically at radio
frequencies (RF). The linear response regime only will be considered in this 
work. The TSs then couple to this field via the
electric charge or dipole moment of the tunneling ``particle''. The applied
electric field both modulates the energy splitting of the tunneling states
and excites them from thermodynamic equilibrium. The electric field only
affects the asymmetry energy $\Delta $ \cite{Phi1981}. The influence of the 
electric field on the tunnel splitting $\Delta_{\mathrm{0}}$ is usually 
neglected \cite{Jac1972}. The coupling of a 2LS to the external field 
therefore causes resonant processes like resonant absorption and stimulated 
emission. 
In the presence of the external electric field F the Hamiltonian matrix
takes the form (coordinate representation):
\begin{equation}
H=H_{0}+{\bf p}_{0}\cdot{\bf F}\sigma_{z}=-\frac{1}{2}\left(
{\begin{array}{*{20}c}
\Delta -2{\bf p}_{0}\cdot{\bf F} & \Delta_{0}\\
\Delta_{0} & -\Delta +2{\bf p}_{0}\cdot{\bf F}\\
\end{array} } \right)
\label{eq213}
\end{equation}
Here ${\bf p}_{0}$ denotes the electric dipole moment of the fictitious 
particle, ${\bf F}={\bf F}_{\omega }\mathrm{\cos}\omega t$ is the 
time-dependent electric field. Diagonalizing the Hamiltonian 
(Eq.~(\ref{eq213})) one can get (energy representation):
\begin{equation}
H=-\left( {\begin{array}{cc}
\frac{1}{2}E & 0\\
0 & -\frac{1}{2}E\\
\end{array} } \right)+\left( {\begin{array}{cc}
\frac{\Delta }{E} & \frac{\Delta_{0}}{E}\\
\frac{\Delta_{0}}{E} & -\frac{\Delta }{E}\\
\end{array} } \right){\bf p}_{0}\cdot{\bf F}\mathrm{\cos}\omega t
\label{eq214}
\end{equation}
The dynamics of the 2LSs is given by the change in the expectation values 
through the Bloch equations, which were first derived by Bloch in the context 
of magnetic resonance \cite{Blo1946}:
\begin{equation}
\begin{split}
&\frac{dS_x}{dt}=\ -\frac{1}{T_2}S_x+\ \widetilde{\gamma }\left(S_yB_z-S_zB_y\right)\\
& \frac{dS_y}{dt}=-\frac{1}{T_2}S_y+\widetilde{\gamma }\left(S_zB_x-S_xB_z\right)\\
&\frac{dS_z}{dt}=-\frac{1}{T_1}\left(S_z-\left\langle S_z\right\rangle \right)+\widetilde{\gamma }\left(S_xB_y-S_yB_x\right)\\
\end{split}
\label{bloch}
\end{equation}
Here we have introduced the pseudo-spin $\frac{1}{2}$ operator 
\textbf{\textit{S}}$=\boldsymbol{\sigma}/2$, where $\boldsymbol{\sigma}$ are 
Pauli's matrices (in the 2LS energy representation), $T_1$ is a characteristic 
time for the equilibration of the level populations of the 2LS, and $T_2$ is 
the transverse dephasing time due to spin-spin (i.e. 2LS-2LS) interactions. 
Also, $\left\langle S_z\right\rangle$ is the thermal equilibrium value of
$S_z$ given by 
$\left\langle S_z\right\rangle =\frac{1}{2}\tanh\left(\frac{\widetilde{\gamma }\hbar B_z\left(t\right)}{2k_{B}T}\right)$, 
$\widetilde{\gamma }B=\omega_0$, $\widetilde{\gamma }$ is the appropriate 
(fictitious) gyromagnetic ratio and 
\textbf{\textit{B}}= \textbf{\textit{B}}$_{dc}$+ \textbf{\textit{B}}$_{ac}$
is a fictitious effective field made up of a static (dc) part and of an 
oscillating (ac) one, both proportional to the electric field with frequency 
$\omega $. The dimensionless fictitious spin \textbf{\textit{S}} processes 
around this fictitious effective field \textbf{\textit{B}}, given by 
($\mu=\hbar\widetilde{\gamma }$:
\begin{equation}
\mu {\bf B}=\left(\frac{2{\Delta }_0}{E}{{\bf p}}_0\cdot{\bf F},\ 0,\ E+\frac{2\Delta }{E}{{\bf p}}_0\cdot{\bf F}\right)
\label{fictive}
\end{equation}
Since the ac field is a small perturbation, one can expand (in linear response 
theory) $\left\langle S_z\right\rangle$ in a Taylor series by keeping terms up to 
the first order in \textbf{\textit{B}}$_{ac}$. The solution to the Bloch equations 
takes the form ${\bf S}(t)={\bf S}^0(t)+{\bf S}^1(t)$, where ${\bf S}^0(t)$ is of 
zeroth order and ${\bf S}^1(t)$ is of first order in \textbf{\textit{B}}$_{ac}$. 
Thus the linearised Bloch equations become, for the zero-order and first-order 
contributions, respectively, to the ${\bf S}$-components:
\begin{equation}
\begin{split}
&\frac{dS^0_z}{dt}+\frac{1}{T_1}\left[S^0_z\left(t\right)-S^0_z\left(\infty\right)\right]=0\\
&\frac{dS^1_x}{dt}-{\omega }_0S^1_y+\frac{1}{T_2}S^1_x=0\\
&\frac{dS^1_y}{dt}+{\omega }_0S^0_x+\frac{1}{T_2}S^1_y-\delta \alpha S^0_z\left(t\right)\cos\omega t=0\\
&\frac{dS^1_z}{dt}+\frac{1}{T_1}S^1_z-\delta \alpha S^0_z\left(t\right)\cos\omega t=0
\end{split}
\label{linbloch}
\end{equation}
where we introduced the resonance frequency 
${\omega }_0=\widetilde{\gamma }B_{z,dc}=-E/\hbar$, which depends on the level 
splitting $E$, 
$\delta =-\left(\frac{2\Delta }{\hbar E}\right)p_0F_{\omega }{\cos \theta}$ which is 
a first-order term in the ac field $F_\omega$, $\theta$ being the angle between the 
ac field and the dipole moment. $S_z^0(0)$ is the initial value of $S_z^0(t)$ 
shortly after the field is applied, 
$S^0_z\left(\infty\right)=-{\tanh \left(E/2k_{B}T\right)}/2$ is the equilibrium value 
of the aligned spin, $\alpha=\Delta_0/\Delta$, and we define 
$\lambda =\hbar \left[1-4{\left(S^0_z\left(\infty\right)\right)}^2\right]/{4k}_{B}T$.
If one introduces raising and lowering operators $S^{\pm}=S_x^1\pm iS_y^1$ , then 
the equations for $S^+$ and $S^-$ separate. The equations for $S^+$ becomes:
\begin{equation}
\label{eq1}
\frac{dS^+\left(t\right)}{dt}+i\left({\omega }_0-\frac{i}{{T }_2}\right)S^+\left(t\right)-i\alpha \delta S^0_z\left(t\right)\cos\omega t=0
\end{equation}
and the equations for $S^-$ is the complex conjugate of the above.
The solutions of these equations are given by the following expressions 
\cite{Car1994}:
\begin{equation}
\begin{split}
S^0_z\left(t\right)&=S^0_z\left(\infty\right)+\left[S^0_z\left(0\right)-S^0_z\left(\infty\right)\right]e^{-t/T_1}\\
S^1_z\left(t\right)&=\frac{\delta \lambda }{1+T_1^2{\omega }^2}\left[\cos\omega t
+T_1\omega \sin\omega t\right]\\
S^+\left(t\right)&=\frac{\delta \alpha \left[\left({\omega }_0
-i/T_2\right)\cos\omega t-i\omega \sin\omega t\right]S^0_z\left(\infty\right)}
{{\left({\omega }_0-i/T_2\right)}^2-{\omega }^2}\\
&+\frac{\delta \alpha \left[\left({\omega }_0+i/T_1-i/T_2\right)\cos\omega t
-i\omega \sin\omega t\right]\left[S^0_z\left(0\right)
-S^0_z\left(\infty\right)\right]e^{-t/T_1}}{{\left({\omega }_0+i/T_1
-i/T_2\right)}^2-{\omega }^2}\\
\end{split}
\label{eq218}
\end{equation}
the result for $S^-(t)$ being the complex conjugate of the above equation for 
$S^+(t)$. The Bloch spins should now be related to the 2LS polarization in the 
electric field direction. The component $p_\parallel$ of the dipole moment 
along the direction of the electric field, in the diagonal basis as in 
Eq. (\ref{eq214}), is now given by 
$p_\parallel=-\langle\frac{\Delta}{E}\sigma_z
+ \frac{\Delta_0}{E}\sigma_x\rangle p_0 \cos \theta$ 
and using the average values of 
$\langle \sigma_z \rangle$ and $\langle \sigma_x \rangle$ from the solutions 
of the Bloch equations one can obtain the dipole moment in the energy 
representation 
\cite{Car1994}:
\begin{equation}
p_\parallel = -p_0 \cos \theta \bigg(\frac{2\Delta S_z^1(t)}{E}+\frac{\Delta_0(S^+ +S^-)}{E} \bigg)
\label{eq219}
\end{equation}
Then, one must insert the deduced pseudo-spin values $S_z(t)$, $S^+(t)$ and 
$S^-(t)$, Eqs.~(\ref{eq218}) to Eq.~(\ref{eq219}). Eq.~(\ref{eq219}) depends 
on the electric field $F_\omega$ linearly and can be easily differentiated 
with respect to the electric field, and this gives a formula for the complex 
dielectric constant 
$\epsilon=\frac{\mathrm{d}p_\parallel}{\mathrm{d}F_\omega}\bigg\vert_{F_\omega=0}$. 
For convenience, one may separate the resulting formulae writing 
$\epsilon=(\epsilon_{RES}'+\epsilon_{REL}')+i(\epsilon_{RES}''+\epsilon_{REL}'')$. 
These are the real ($\epsilon'$) and imaginary ($\epsilon''$) parts of the 
dielectric constant. The imaginary part is interpreted as a dielectric loss 
(loss tangent, $\tan\delta=\epsilon''/ \epsilon'$) - a parameter of the 
dielectric material that quantifies its inherent dissipation of electromagnetic 
energy (much like in a RLC circuit). One gets:
\begin{equation}
\begin{split}
&{\epsilon'}_{RES}= \frac{{p_0}^2{{\cos}^2 \theta }}{\hbar }{\left(\frac{{\Delta }_0}{E}\right)}^2\bigg[\left(\frac{\left({\omega }_0+\omega \right){T }^2_2}{1+{\left({\omega }_0+\omega \right)}^2{T }^2_2}+\frac{{(\omega }_0-\omega ){T }^2_2}{1+{\left({\omega }_0-\omega \right)}^2{T }^2_2}\right){\tanh \left(\frac{E}{2k_{B}T}\right)}\\
&-\left(\frac{\left({\omega }_0+\omega \right){T }^2_{12}}{1+{{T }^2_{12}\left({\omega }_0+\omega \right)}^2}+\frac{\left({\omega }_0-\omega \right){T }^2_{12}}{1+{{T }^2_{12}\left({\omega }_0-\omega \right)}^2}\right)\left(2{S^0_z\left(0\right)+\tanh \left(\frac{E}{2k_{B}T}\right)\ }\right)e^{-t/{T }_1}\bigg]\\
&{\epsilon{''}}_{RES}=\ \frac{{p_0}^2{{\cos}^2 \theta \ }}{\hbar }{\left(\frac{{\Delta }_0}{E}\right)}^2\bigg[\left(\frac{{T }_2}{1+{\left({\omega }_0-\omega \right)}^2{T }^2_2}-\frac{{T }_2}{1+{\left({\omega }_0+\omega \right)}^2{T }^2_2}\right){\tanh \left(\frac{E}{2k_{B}T}\right) }\\
&+\left(\frac{{T }_{12}}{1+{{T }^2_{12}\left({\omega }_0+\omega \right)}^2}-\frac{{T }_{12}}{1+{{T }^2_{12}\left({\omega }_0-\omega \right)}^2}\right)\left(2{S^0_z\left(0\right)+\tanh \left(\frac{E}{2k_{B}T}\right)\ }\right)e^{-t/{T }_1}\bigg]
\end{split}
\label{eq220}
\end{equation}
\begin{equation}
\begin{split}
&{\epsilon'}_{REL}=\frac{{p_0}^2{{\cos}^2 \theta \ }}{k_{B}T}{\left(\frac{\Delta }{E}\right)}^2{{\cosh}^{-2} \left(\frac{E}{2k_{B}T}\right)\frac{1}{1+{T }^2_1{\omega }^2}}\\
&{\epsilon'}_{REL}=\frac{{p_0}^2{{\cos}^2 \theta \ }}{k_{B}T}{\left(\frac{\Delta }{E}\right)}^2{{\cosh}^{-2} \left(\frac{E}{2k_{B}T}\right)\frac{1}{1+{T }^2_1{\omega }^2}}\\
&{\epsilon''}_{REL}=\frac{{p_0}^2{{\cos}^2 \theta \ }}{k_{B}T}{\left(\frac{\Delta }{E}\right)}^2{{\cosh}^{-2} \left(\frac{E}{2k_{B}T}\right)\frac{{T }_1\omega }{1+{T }^2_1{\omega }^2}}\\
\end{split}
\end{equation}
where $T_{12}^{-1}=T_2^{-1}-T_1^{-1}$. In the adiabatic limit the initial 
value of the pseudo-spin is $S_z^0(0)=-\frac{1}{2}\tanh(E/2k_{B}T)$, when the 
2LS energy eigenvalues are $\pm \frac{1}{2}E$. That makes the time dependent 
terms of Eq.~(\ref{eq220}) equal to 0 shortly after applying the field.
For an ensemble of 2LS, from the manipulation of the Bloch equations for the 
motion of the spatial components of a pseudo-spin 1/2 under periodic electric 
(and/or elastic) perturbations and taking into account the phonon relaxation 
mechanism, one then finds the explicit form of the expression for the 
dielectric constant \cite{Hun1976}: 
\begin{equation}
\epsilon=\epsilon_{RES}+\epsilon_{REL}\frac{1}{1+i\omega \tau }=\left(\epsilon_{RES}+\epsilon_{REL}\frac{1}{1+{\omega }^2{\tau }^2}\right)-{i}\epsilon_{REL}\frac{\omega \tau }{1+{\omega }^2{\tau }^2}
\label{relaxapprox}
\end{equation}
where we have put $\tau=T_1$. The above equation is the fundamental result of 
the low-frequency, relaxation-time approach to the complex dielectric constant 
of glasses. The typical energy splittings of the 2LSs in low temperature 
experiments correspond to frequencies in the range of 
$\frac{\omega_0}{2\pi}\approx10^8$~Hz (GHz range, see Section 6), when the 
electric field frequency $\omega$ is about 10$^{3}$~Hz. This justifies a 
low-frequency approximation $\omega\ll\omega_0$. To obtain the resonant part 
we can also set $T_2^{-1}=0$, which simplifies expressions (Eq.~\ref{eq220}), 
remembering that $E=\hbar \omega_0$. From the averaging over the dipole 
orientation angle $\theta$ we get a prefactor 1/3: 
$\overline{p_0^2\cos^2\theta}=\frac{1}{3}\overline{p_0^2}$, where $\overline{p_0^2}$ 
is the configurationally averaged square 2LS electric-dipole moment.
The real part of the relative dielectric constant for 2LS shows the 
temperature-dependent contributions 
($\epsilon(T)=\epsilon'(0)+\Delta\epsilon'(T)$, with
$\vert\Delta\epsilon'\vert\ll\epsilon'$):
\begin{equation}
{\left.\frac{\Delta\epsilon'}{\epsilon'}\right|}_{2LS\ RES}=\frac{2}{3}\overline{p^2_0}\frac{\Delta^2_0}{E^3}{\tanh \left(\frac{E}{2k_{B}T}\right)}
\label{eq221}
\end{equation}
\begin{equation}
{\left.\frac{\Delta\epsilon'}{\epsilon'}\right|}_{2LS\ REL}=\frac{1}{3k_{B}T}\overline{p^2_0}\frac{\Delta^2}{E^2}{\cosh^{-2} \left(\frac{E}{2k_{B}T}\right)\frac{1}{1+{\omega }^2{\tau }^2}}
\label{eq222}
\end{equation}
Eq.~(\ref{eq221}) corresponds to the resonant tunneling contribution to the 
dielectric constant, and Eq.~(\ref{eq222}) is the relaxational contribution. We
neglect for now, for low $\omega$, the frequency dependence in the RES part so 
long as $\omega\ll\omega_0$. The dielectric loss is described by the following 
formula (the resonant contribution being vanishingly small):
\begin{equation}
\Delta \tan\delta \bigg\vert_{2LS\ REL}={\left.\frac{\Delta\epsilon''}{\epsilon'}\right|}_{2LS\ REL}=\frac{1}{3k_{B}T}\overline{p^2_0}\frac{\Delta^2}{E^2}{\cosh^{-2} \left(\frac{E}{2k_{B}T}\right)\frac{1}{1+{\omega }^2{\tau }^2} }
\label{eq223}
\end{equation}
Integrating Eqs.~(\ref{eq221})-(\ref{eq223}) over the parameter distribution 
of the 2LS and over the dipole orientation angle $\theta$, using the 
expressions for the 2LS $(E,\tau)$-parameter distribution (see e.g. 
\cite{Jac1972}), one can find the temperature-dependent contributions to the 
dielectric constant and dielectric loss:
\begin{equation}
\begin{split}
&{\left.\frac{\Delta \epsilon'}{\epsilon'}\right|}_{2RES}=\frac{2\overline{P}\overline{p^2_0}}{3\epsilon_0\epsilon_r}\int^{E_{max}}_{{\triangle }_{0,min}}{\frac{dE}{E}\tanh\left(\frac{E}{2k_{B}T}\right)\sqrt{1-{\left(\frac{{\triangle }_{0,min}}{E}\right)}^2}}\\
&{\left.\frac{\Delta \epsilon'}{\epsilon'}\right|}_{2REL}=\frac{\overline{P}\overline{p^2_0}}{3\epsilon_0\epsilon_rk_{B}T}\int^{E_{max}}_{{\triangle }_{0,min}}{dE\int^{{\tau }_{max}\left(E\right)}_{{\tau }_{min}\left(E\right)}{\frac{d\tau }{\tau }\sqrt{1-\frac{{\tau }_{min}(E)}{\tau }}{\cosh}^{-2}\left(\frac{E}{2k_{B}T}\right)\frac{1}{1+{\omega }^2{\tau }^2}}}\\
&\left. \Delta \tan\delta \right|_{2REL}=\frac{\overline{P}\overline{p^2_0}}{3\epsilon_0\epsilon_rk_{B}T}\int^{E_{max}}_{{\Delta }_{0,min}}{dE\int^{{\tau }_{max}\left(E\right)}_{{\tau }_{min}\left(E\right)}{\frac{d\tau }{\tau }\sqrt{1-\frac{{\tau }_{min}(E)}{\tau }}{\cosh}^{-2}\left(\frac{E}{2k_{B}T}\right)\frac{\omega \tau }{1+{\omega }^2{\tau }^2}}}
\end{split}
\end{equation}
If one extends, when appropriate and as a further approximation, the 
integration limits ($E_{max}\rightarrow\infty$ and 
$\Delta_{0,min}\rightarrow 0$), then calculating (estimating, in fact) the 
$E$-integral one gets a characteristic logarithmic variation of the real part 
of the dielectric constant as a function of temperature:
\begin{equation}
{\left.\frac{\Delta \epsilon'}{\epsilon'}\right|}_{2RES}\approx \left\{ \begin{array}{c}
-\frac{2\overline{P}\overline{p^2_0}}{3\epsilon_0\epsilon_r}\ln\left(\frac{T}{T_0}\right),T<\frac{{\Delta }_{0max}}{2k_B} \\
0,~~~~~~~~~~~~~~~~~~T>\frac{{\Delta }_{0max}}{2k_B} \end{array}
\right.
\label{eq227}
\end{equation}
For $\omega\tau_{min}\gg1$ and at low temperatures, the contribution from 
relaxation to the real part as compared to the resonant contribution is 
negligible. Under the condition $\omega\tau_{min}\ll1$, however, the term 
$1/\tau$ dominates in the $\tau$-integral and we obtain again a logarithmic 
variation with temperature:
\begin{equation}
{\left.\frac{\Delta \epsilon'}{\epsilon '}\right|}_{2REL}\approx \left\{ \begin{array}{c}
0,~~~~~~~~~~~~~~~~~\omega {\tau }_{min}\gg 1 \\
\frac{1\overline{P}\overline{p^2_0}}{3\epsilon_0\epsilon_r}\ln\left(\frac{T}{T_0}\right),~\omega {\tau }_{min}\ll 1 \end{array}
\right.
\label{eq228}
\end{equation}
A crossover between the resonant (low temperature) and relaxation (high $T$)
regimes occurs at a characteristic temperature \cite{Hun1976}
\begin{equation}
T_0\left(\omega \right)\simeq\frac{1}{k_B}\sqrt[3]{\frac{\omega \pi \rho {\hbar }^4}{{\gamma }^2_l/v^5_l\ +{2\gamma }^2_t/v^5_t}}
\end{equation}
which for a thermal 2LS with $E=\Delta=k_B T$ satisfies the condition
$\omega\tau(T_0)=1$. The above equation arises from the celebrated expression 
(see e.g. \cite{Jac1972,Hun1976}) for the one-phonon relaxation time:
\begin{equation}
\tau^{-1}=\left( \frac{\gamma_l^2}{v_l^5}+2\frac{\gamma_t^2}{v_t^5} \right)
\frac{\Delta_0^2E}{2\pi\rho\hbar^4}\coth\left( \frac{E}{2k_{B}T} \right), 
\label{tau}
\end{equation}
arising from the longitudinal ($l$) and transverse ($t$) modes with 
deformation potentials $\gamma_{l,t}$ and sound speeds $v_{l,t}$, 
respectively. In terms of $(E,\tau)$ the 2LS distribution can be re-written as:
\begin{equation}
{\cal P}(E,\tau)=\frac{\bar{P}}{2\tau\sqrt{1-\tau_{min}(E)/\tau}},
\quad \tau_{min}(E)\le\tau\le\tau_{max}(E)
\label{taudistrib}
\end{equation}
where $\tau_{min}(E)=\frac{\gamma}{E^3}\tanh\left( \frac{E}{2k_{B}T} \right)$ 
and 
$\tau_{max}(E)=\frac{\gamma}{\Delta_{0min}^2E}\left( \frac{E}{2k_{B}T} \right)$ 
are the smallest ($\Delta=0$) and largest ($\Delta_0=\Delta_{0min}$) 
relaxation times, respectively. 
$\gamma=2\pi\rho\hbar^4\left( \frac{\gamma_l^2}{v_l^5}+2\frac{\gamma_t^2}{v_t^5} \right)^{-1}$ 
is an elastic parameter of the solid and $\rho$ is its density. 

Eqs.~(\ref{eq227}) and (\ref{eq228}) show that with increasing temperature the 
$T$ dependence changes from a decrease in the resonant regime to an increase 
in the relaxation one. At the temperature $T_0(\omega)$ there is a minimum. 
Thus, the sum of the two contributions has a characteristic V-shaped form, in 
a semi-logarithmic plot, with the minimum occurring at a $T_{0}$ roughly given 
by the condition $\omega\tau_{min}(k_{B}T_0)\cong1$, or 
$k_{B}T_0(\omega)\cong\bigg(\frac{1}{2}\gamma \omega\bigg)^{1/3}$. 
$\epsilon_0\epsilon_r$ is here the bulk of the solid's dielectric constant 
and we see that a -2:1 characteristic behavior is predicted by the STM with 
the slope for $T>T_{0}$ given by Eq. (\ref{eq228}). This behavior is indeed 
observed, but only in very pure (impurity-free) $a$-SiO$_{2}$ \cite{Wie1987}. 
However, in most multi-component glasses (chemically made up of GGFs as well 
as of GCFs, for example AlBaSiO (or BAS) glass:
a-Al$_{\mathrm{2}}$O$_{\mathrm{3}}$-BaO-SiO$_{\mathrm{2}}$), or for a 
contaminated mono-component glass, it is rather a V-shaped curve with a 
(roughly) -1:1 slope ratio that is often observed. This has been explained by 
our theory \cite{Jug2010}.

\subsection{Results for the dielectric constant $\epsilon'$ in a magnetic
field}
We now derive the contribution to the dielectric constant $\epsilon(\omega)$ 
from the TWPs or ATSs sitting in the interstices between the RERs. One can 
treat the ATS again as an {\it effective} 2LS having lowest energy gap 
$\Delta{\cal E}={\cal E}_1-{\cal E}_0=E=\sqrt{D^2+D_0^2\varphi^2}$ for ``weak'' 
fields. Within this picture, the linear-response quasi-static resonant and 
relaxational contributions to the polarizability tensor $\alpha_{\mu\nu}$ are 
extracted according to the general 2LS approach described in the previous 
Section 5.1, to get \cite{Jug2009,JugXXXX} 
\begin{equation}
\alpha^{RES}_{\mu\nu}=\int_0^{\infty}\frac{dE}{2E}
{\cal G}_{\mu\nu}\left ( \left\{ \frac{E_i}{E} \right\};{{\bf p}_i}\right )
\tanh\big( \frac{E}{2k_{B}T} \big)\delta(E-\Delta{\cal E})
\label{respolariz}
\end{equation}
and
\begin{equation}
\alpha^{REL}_{\mu\nu}=\frac{1}{4k_{B}T}\int_0^{\infty} dE \left( \sum_{i,j=1}^{3}
\frac{E_iE_j}{E^2}p_{i\mu}p_{j\nu} \right )
\cosh^{-2}\left( \frac{E}{2k_{B}T} \right) \delta(E-\Delta{\cal E})
\label{relpolariz}
\end{equation}
where
\begin{equation}
{\cal G}_{\mu\nu}\left ( \left\{ \frac{E_i}{E} \right\} ;{{\bf p}_i}\right )=
\sum_{i=1}^{3}p_{i\mu}p_{i\nu}-\sum_{i,j}\frac{E_iE_j}{E^2}p_{i\mu}p_{j\nu}
\label{geomcorr}
\end{equation}
contains the single-well dipoles ${\bf p}_i=q{\bf a}_i$. This
expression assumes vanishing electric fields and no TS-TS interactions, a
situation which does not wholly apply to the experiments. To keep the theory
simple one can still use Eq. (\ref{respolariz}) and the analogous one for
the relaxational contribution to the polarizability. Eq. (\ref{respolariz})
must be averaged over the random energies' distribution (\ref{atsdistribution})
($[\dots]_{av}$, responsible for the high sensitivity to weak fields) and
over the dipoles' orientations and strengths ($\overline{(\dots)}$). For a
collection of ATS with $n_w>2$ wells this averaging presents serious 
difficulties and one must resort to the decoupling:
\begin{equation}
\overline{{\cal G}_{\mu\nu}\delta(E-\Delta{\cal E})}\simeq
\overline{{\cal G}_{\mu\nu}}\cdot\overline{\delta(E-\Delta{\cal E})},
\label{decoupl}
\end{equation}
where $\overline{[\delta(E-\Delta{\cal E})]_{av}}=g_{ATS}(E,B)$ is the
fully-averaged density of states. To calculate $\overline{{\cal G}_{\mu\nu}}$,
one can envisage a fully isotropic distribution of planar $n_w$-polygons to
obtain \cite{JugXXXX}:
\begin{equation}
\overline{{\cal G}_{\mu\nu}}=\frac{1}{3}\left ( \frac{n_w}{n_w-1} \right )
\overline{p_i^2}\frac{(n_w-2)E^2+D_0^2\varphi^2}{E^2}\delta_{\mu\nu}.
\label{isoglass}
\end{equation}
The second term in the numerator of Eq. (\ref{isoglass}) gives rise to a
peak in $\delta\epsilon/\epsilon$ at very low $B$, while the first term
(present only if $n_w>2$) gives rise to a {\it negative} contribution to
$\delta\epsilon/\epsilon$ at larger $B$ which can win over the enhancement
term for all values of $B$ if $D_{0max}\gg D_{0min}$ ($D_{0min}$, $D_{0max}$
corresponding to cutoffs in the distribution of ATS energy barriers).
The observations in Duran and BK7 indeed show a significant depression of
$\epsilon'(B)$ for weak fields \cite{Woh2001}, thus giving direct evidence for
the existence of ATSs with $n_w>2$ in the multisilicate glasses. Carrying out
the averaging $[\dots]_{av}$ one gets analytical expressions for the
polarizability; the uniform average over orientation angles $\theta$ must be
performed numerically (although a very good approximation is the replacement 
of $\varphi^2$ with $\frac{1}{3}\varphi^2$ in the averaged expression, 
corresponding to the replacement $\overline{\cos^2\theta}\to\frac{1}{3}$).

The expression for the resonant part of the polarizability Eq. 
(\ref{respolariz}) should be averaged over the probability distribution of 
parameters Eq. (\ref{atsdistribution}). Averaging over the probability 
distribution \\
$\int^{D_{0max}}_{D_{0min}}{dD_0\int{{dE}_1dE_2dE_3{\cal P}_{ATS}^{\ast}\left({\{E}_i\};D_0\right)\delta \left(\sum_i~E_i\right)}}$ 
can be done as in the previous Sections, to get to the expressions 
\begin{equation}
\begin{split}
{\alpha }_{RES}(T,B)=&\frac{\pi }{2}P^*\overline{p^2_1}\int^{\infty }_0{dE\frac{1}{E^3}{\tanh  \left(\frac{E}{2k_{B}T}\right)\ }\int^{\infty }_{D_{min}}{\frac{dD}{D}\int^{D_{0max}}_{D_{0min}}{\frac{dD_0}{D_0}}}\left[E^2-D^2_0\varphi^2\right]}\\
& \times \delta \left(E-\Delta {\mathcal E}\right)={\alpha }_1\left(T,B\right)+{\alpha }_0\left(T,B\right)\\
{\alpha }_0\left(T,B\right)=&\varphi^2\frac{\pi }{2}P^*\overline{p^2_1}\int^{\infty }_0{\frac{dE}{E^3}{\tanh  \left(\frac{E}{2k_{B}T}\right)\ }\int^{\infty }_{D_{min}}{\frac{dD}{D}\int^{D_{0max}}_{D_{0min}}{dD_0}}D_0}\delta \left(E-\Delta {\mathcal E}\right)\\
{\alpha }_1\left(T,B\right)=&\frac{\pi }{2}P^*\overline{p^2_1}\int^{\infty }_0{\frac{dE}{E}{\tanh  \left(\frac{E}{2k_{B}T}\right)\ }\int^{\infty }_{D_{min}}{\frac{dD}{D}\int^{D_{0max}}_{D_{0min}}{\frac{dD_0}{D_0}}}}\delta \left(E-\Delta {\mathcal E}\right)\\
\end{split}
\label{eq72}
\end{equation}
with the energy gap $\Delta{\cal E} =\sqrt{D^2+D^2_0\varphi^2}$.

\begin{figure}[!Htp]
  \centering
 \includegraphics[scale=0.25] {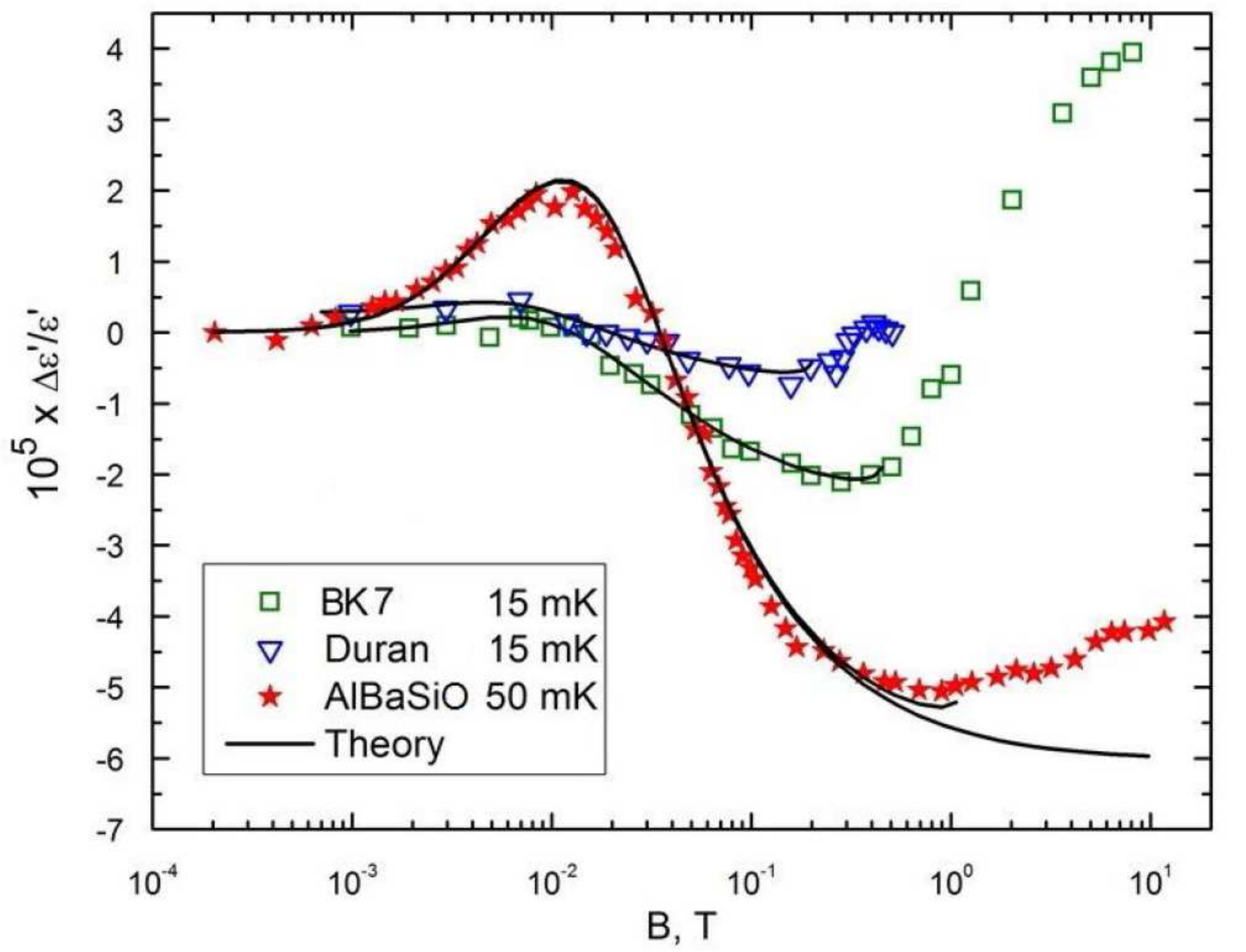}
 \caption[]{(colour online) The relative dielectric constant variation as a function of the magnetic field for AlBaSiO (BAS) glass \cite{Woh2001}b, BK7 \cite{Woh2001} and Duran \cite{Woh2001}b glasses. With best-fit parameters as in Table~\ref{tab71}, the curves are the results of our theory in the ``weak field'' approximation with (and, for AlBaSiO, without) higher order correction. From \cite{Pal2011}.}
\label{fig71}
\end{figure}

\begin{figure}[!Htp]
  \centering
 \includegraphics[scale=0.25] {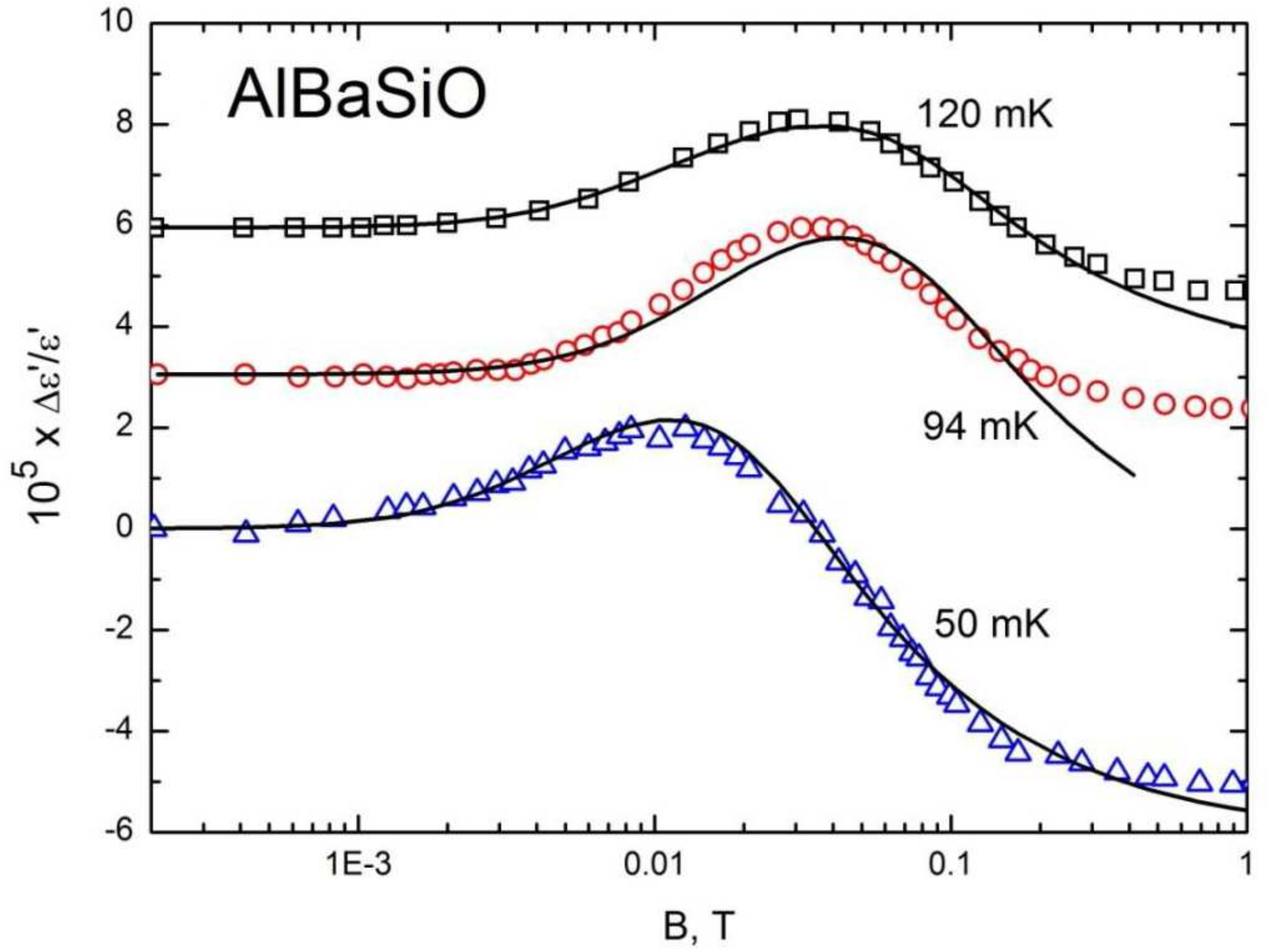}
 \caption[]{(colour online) Relative dielectric constant variation as a function of the magnetic field and temperature for AlBaSiO (BAS) glass \cite{Woh2001}b. With fitting parameters as in Table~\ref{tab71}, the curves are the result of our theory in the ``weak field'' approximation. From \cite{Pal2011}.}
\label{fig72}
\end{figure}

\begin{figure}[!Htp]
  \centering
 \includegraphics[scale=0.25] {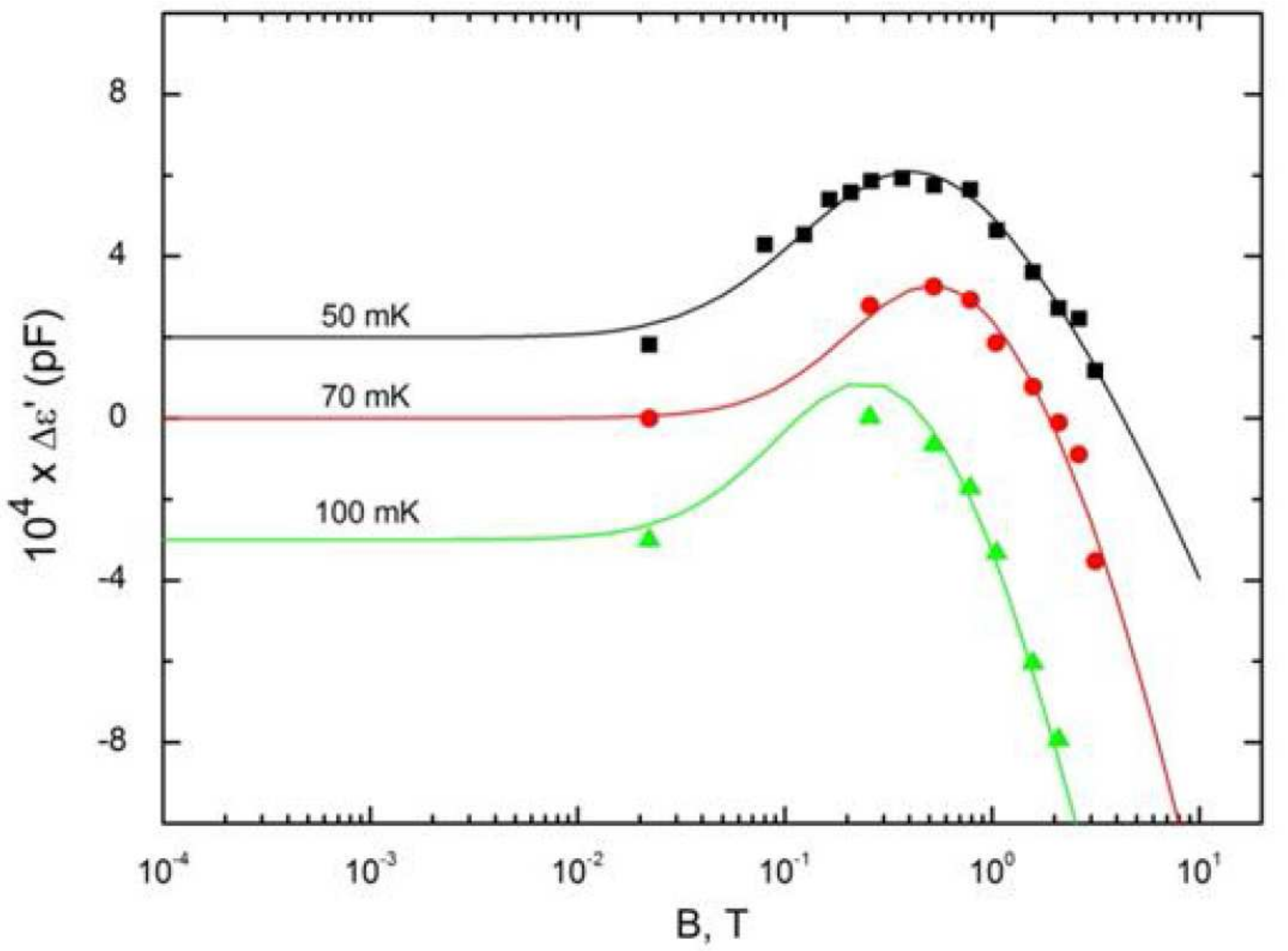}
 \caption[]{(colour online) Dielectric constant variation as a function of the magnetic field and temperature for the SiO${}_{2+}$\textit{${}_{x}$}C\textit{${}_{y}$}H\textit{${}_{z}$} glass \cite{LeC2002}. Fitting parameters as in Table~\ref{tab72}. 
From \cite{Pal2011}.}
\label{fig73}
\end{figure}

Similarly to the calculation for the density of states Eq. (\ref{3dos}), the 
integrals in Eq.~(\ref{eq72}) can be reduced as follows \cite{JugXXXX}:
\begin{equation}
\begin{split}
{\alpha }_0\left(T,B\right)=&\varphi^2\frac{\pi }{2}P^*\overline{p^2_1}\int^{E_{c2}}_{E_{c1}}\frac{dE}{E^2}{\tanh  \left(\frac{E}{2k_{B}T}\right)}\int^{\frac{1}{\varphi}\sqrt{E^2-D^2_{min}}}_{D_{0min}}{\frac{D_0 dD_0}{E^2-D^2_0     \varphi^2}} \\
&+\frac{\pi }{2}P^*\overline{p^2_1}\int^{\infty }_{E_{c2}}{\frac{dE}{E^2}{\tanh  \left(\frac{E}{2k_{B}T}\right)}\int^{D_{0max}}_{D_{0min}}{\frac{D_0 ~dD_0}{E^2-D^2_0\varphi^2}}}\\
=&\frac{\pi }{2}P^*\overline{p^2_1}\bigg[\int^{E_{c2}}_{E_{c1}}\frac{dE}{E^2}{\tanh  \left(\frac{E}{2k_{B}T}\right)}\frac{1}{2}{\ln  \left(\frac{E^2-D^2_{0min}\varphi^2}{D^2_{min}}\right)} \\
& + \int^{\infty}_{E_{c2}}{\frac{dE}{E^2}{\tanh  \left(\frac{E}{2k_{B}T}\right)}\frac{1}{2}{\ln  \left(\frac{E^2-D^2_{0min}\varphi^2}{E^2-D^2_{0max}\varphi^2}\right)}}\bigg] \\
\end{split}
\label{eq74}
\end{equation}
and likewise for ${\alpha }_1\left(T,B\right)$:
\begin{equation}
\begin{split}
{\alpha }_1\left(T,B\right)=&\frac{\pi }{2}P^*\overline{p^2_1}\int^{\infty }_0{\frac{dE}{E}{\tanh  \left(\frac{E}{2k_{B}T}\right)\ }\int^{D_{0max}}_{D_{0min}}{\frac{dD_0}{D_0}\frac{E}{\sqrt{E^2-D^2_0\varphi^2}}}\int^{\infty }_{D_{min}}{\frac{dD}{D}}}\times \delta \left(D-D_a\right)\\
{\alpha }_1\left(T,B\right)=&\frac{\pi }{2}P^*\overline{p^2_1}\int^{E_{c2}}_{E_{c1}}{dE~{\tanh  \left(\frac{E}{2k_{B}T}\right)\ }\int^{\frac{1}{\varphi}\sqrt{E^2-D^2_{min}}}_{D_{0min}}{\frac{dD_0}{D_0}\frac{1}{E^2-D^2_0\varphi^2}+\ }}\\
&+\ \frac{\pi }{2}P^*\overline{p^2_1}\int^{\infty }_{E_{c2}}{dE~{\tanh  \left(\frac{E}{2k_{B}T}\right)\ }\int^{D_{0max}}_{D_{0min}}{\frac{dD_0}{D_0}\frac{1}{E^2-D^2_0\varphi^2}}}\\
=&\frac{\pi }{2}P^*\overline{p^2_1}\bigg[\int^{E_{c2}}_{E_{c1}}\frac{dE}{E^2}{\tanh  \left(\frac{E}{2k_{B}T}\right)}\frac{1}{2}{\ln  \left(\frac{\left(E^2-D^2_{0min}\varphi^2\right)\left(E^2-D^2_{min}\right)}{{D^2_{0min}\varphi^2D}^2_{min}}\right)}\\
&+\int^{\infty }_{E_{c2}}{\frac{dE}{E^2}{\tanh  \left(\frac{E}{2k_{B}T}\right)}\frac{1}{2}{\ln  \left(\frac{D^2_{0max}\left(E^2-D^2_{0min}\varphi^2\right)}{D^2_{0min}\left(E^2-D^2_{0max}\varphi^2\right)}\right)\ }}\bigg] \\
\end{split}
\label{eq75}
\end{equation}
with $E_{c1}, E_{c2}$ as in Section 4.1. For the whole mass of the glass:
\begin{equation}
\frac{1}{V}\sum^{N_{ATS}}_1{\left({\alpha }_1\left(T,B\right)+{\alpha }_0\left(T,B\right)-\alpha \left(T,0\right)\right)=\frac{{{\mathcal N}}_{ATS}}{V}\Delta \alpha =x_{ATS}\Delta \alpha}
\end{equation}
where $x_{ATS}$ is the volume concentration of ATS and 
$\alpha \left(T,0\right)$ is:
\begin{equation}
\alpha \left(T,0\right)=\frac{\pi }{2}P^*\overline{p^2_1}{\ln  \left(\frac{D_{0max}}{D_{0min}}\right)\ }\int^{\infty }_{E_{c2}}{\frac{dE}{E^2}{\tanh  \left(\frac{E}{2k_{B}T}\right)\ }}
\label{eq76}
\end{equation}
The relative change of the dielectric constant is expressed by 
Eq.~(\ref{eq74}), Eq.~(\ref{eq75}) and Eq.~(\ref{eq76}):
\begin{equation}
\frac{\Delta \epsilon'\left(T,B\right)}{\epsilon'}=x_{ATS}\frac{\Delta \alpha \left(T,B\right)}{\epsilon_0\epsilon_r}=\frac{x_{ATS}}{\epsilon_0\epsilon_r}\left({\alpha }_1\left(T,B\right)+{\alpha }_0\left(T,B\right)-\alpha \left(T,0\right)\right)
\label{eq77}
\end{equation}
Eq.~(\ref{eq77}) with Eqs.~(\ref{eq74})-(\ref{eq75}) describes well the 
experimental data for different glasses, as is shown in 
Figs.~\ref{fig71}-\ref{fig73} and with the fitting parameters presented 
in Tables~\ref{tab71},~\ref{tab72}.

For the sake of clarity, the data and curves in Figs. \ref{fig72} and 
\ref{fig73} have been shifted apart vertically.
\begin{table}[!Htbp]
\begin{center}
\begin{tabular}{ |c|c|c|c|c|}
\hline
Material and Temperature & ${\pi x_{ATS}P}^*\overline{p^2_1}/\epsilon_r\epsilon_0$ & $D_{min}$,K & $D_{0min}\left|\frac{q}{e}\right|S_{\Delta }$,KÅ$^{2}$ & $D_{0max}\left|\frac{q}{e}\right|S_{\Delta }$,KÅ$^{2}$  \\
\hline
BK7 15 mK & $0.089\cdot {10}^{-5}$ & $0.03$ & $1.668\cdot {10}^5$ & $4.576\cdot {10}^5$  \\
\hline
Duran 15 mK & $0.052\cdot {10}^{-5}$ & 0.021 & $2.457\cdot {10}^5$ & $4.151\cdot {10}^5$ \\
\hline
AlBaSiO 50 mK & $0.89\cdot {10}^{-5}$ & $0.015$ & $2.440\cdot {10}^5$ & $3.080\cdot {10}^5$ \\
\hline
AlBaSiO 94 mK & $3.75\cdot {10}^{-5}$ & 0.025 & $1.225\cdot {10}^5$ & $1.589\cdot {10}^5$ \\
\hline
AlBaSiO 120 mK & $3.09\cdot {10}^{-5}$ & 0.0227 & $1.767\cdot {10}^5$ & $2.248\cdot {10}^5$ \\
 \hline
\end{tabular}
\caption{Fitting parameters for the dielectric constant in a magnetic field for three different types of glasses.}
 \label{tab71}
\end{center}
\end{table}

\begin{table}[!Htbp]
\begin{center}
\begin{tabular}{ |c|c|c|c|c|}
\hline
Temperature & ${\pi x_{ATS}P}^*\overline{p^2_1}/\epsilon_r\epsilon_0$ & $D_{min}$\textit{${}_{,}$, }K & $D_{0min}\left|\frac{q}{e}\right|S_{\Delta }$\textit{, }KÅ${}^{2}$ & $D_{0max}\left|\frac{q}{e}\right|S_{\Delta }$\textit{, }KÅ${}^{2}$ \\
\hline
50 mK & $4.38\cdot {10}^{-5}$ & $0.015$ & $0.076\cdot {10}^3$ & $3.047\cdot {10}^4$ \\
\hline
70 mK & $12.22\cdot {10}^{-5}$ & 0.0486 & $0.600\cdot {10}^3$ & $2.662\cdot {10}^4$ \\
\hline
100 mK & $13.63\cdot {10}^{-5}$ & 0.0486 & $3.035\cdot {10}^3$ & $7.616\cdot {10}^4$ \\
\hline
\end{tabular}
\caption{Fitting parameters for the SiO${}_{2+}$\textit{${}_{x}$}C\textit{${}_{y}$}H\textit{${}_{z}$} glass for different temperatures.}
 \label{tab72}
\end{center}
\end{table}


\subsection{Results for the dielectric loss in a magnetic field} 
The dielectric loss (or loss angle $\delta $) for a dielectric substance is a 
measure of the power lost in dissipation and is obtained as the following 
expression:
\begin{equation}
{\tan  \delta \ }\equiv \frac{\epsilon''}{\epsilon'}\cong \frac{\epsilon''}{\epsilon_0\epsilon_r}
\label{eq81}
\end{equation}
where $\epsilon''=\epsilon_{REL}\frac{\omega \tau }{1+{\omega }^2{\tau }^2}$ 
is the imaginary part of the dielectric constant, typically evaluated in the 
relaxation time approximation (Section 5.1). It should be pointed out that the 
2LS STM does not describe well the temperature- and frequency-dependence of 
$\epsilon''$ in glasses. The reason is that the theory works well only in the 
low-frequency regime. Since $\epsilon''\left(\omega \right)$ should be linked 
to the real part $\epsilon'$ of the dielectric constant through the 
Kramers-Kroning relation, one can see that the knowledge of the low-frequency 
behaviour of $\epsilon'\left(\omega \right)$ is not enough to reproduce the 
correct form of $\epsilon''\left(\omega ,T\right)$. However, inclusion of the 
ATS contribution does seem to improve the agreement between theory and 
experiment \cite{Pal2011}, as least in the case of BK7 (optical glass) which 
is probably the best approximation to a fully networked glass.

The relaxation part of the dielectric constant depends on the volume 
concentration $x_{ATS}=n_{ATS}\rho$ of ATS's and is expressed by:
\begin{equation}
\epsilon_{REL}-1\cong x_{ATS}~{\alpha }_{REL}
\label{eq83}
\end{equation}
where ${\alpha }_{REL}$ should be averaged over all parameters by means of the 
distribution function Eq.~(\ref{atsdistribution}); following similar 
calculations as in the previous Sections we get: 
\begin{equation}
\begin{split}
{\alpha }_{REL}=&\frac{\pi P^*\overline{p^2_1}}{4k_{B}T}\int^{\infty }_0{\frac{dE}{E^2}}~{{\cosh }^{-2} \left(\frac{E}{2k_{B}T}\right)}\int^{D_{0max} }_{D_{0min}}{\frac{dD_0}{D_0}\int^{\infty }_{D_{min}}{\frac{dD}{D}D^2}} \\
&\times \frac{\omega {\tau }_{ATS}}{1+{\omega }^2{\tau }^2_{ATS}}\delta \left(E-\sqrt{D^2+D^2_0\varphi^2}\right)\\
\end{split}
\label{eq84}
\end{equation}
The relaxation time for ATSs at low temperature and in a magnetic field is now 
found (after a long calculation that will be reported elsewhere) to be given 
by the following expression \cite{JugXXXX}:
\begin{equation}
{\tau }^{-1}_{ATS}={\tau }^{-1}\left(E,\varphi\right)=\frac{E^3\left(D^2_0\varphi^2+\frac{5}{6}D^2\right)}{{\Gamma {\rm \ tanh} \left(\frac{E}{2k_{B}T}\right)\ }}=\frac{E^3\left(E^2-\frac{1}{6}D^2\right)}{{\Gamma {\rm \ tanh} \left(\frac{E}{2k_{B}T}\right)\ }}={\tau }^{-1}\left(E,D\right)
\label{eq85}
\end{equation}
where as usual the A-B phase $\varphi$ is directly proportional to the magnetic 
field $B$. It appears, therefore, that the total dielectric relaxation time, 
obtained through its inverse:
\begin{equation}
\frac{1}{{\tau }_{tot}}=\frac{1}{{\tau }_{2LS}}+\frac{1}{{\tau }_{ATS}(\varphi)}
\label{eq86}
\end{equation}
must diminish in a non-trivial manner as the magnetic field is switched on. 
This very interesting prediction of the present theory appears to be confirmed 
explicitly, albeit only qualitatively, in the laboratory and for some 
multi-silicate glasses so far only via the work of a Russian group at 
liquid-He temperatures \cite{Smo1979}. A systematic study of the 
magnetic-field dependence of ${\tau }_{tot}$ in the multi-component glasses is 
still lacking.

The probability distribution function for the ATS dielectric relaxation times 
turns out to be rather different from that of the standard 2LS case, one finds 
indeed \cite{JugXXXX}:
\begin{equation}
P\left(E,\tau \right)=\frac{\pi P^*{\tau }_{min}}{5E\left(\tau -{\tau }_{min}\right)({\tau }_{max}-\tau )}
\label{eq87}
\end{equation}
(with suitable $\varphi$-dependent boundaries) where (with $\Gamma$ a new
ATS-related elastic constant). 
${\tau }_{max}(E)=E^5/\left[ \Gamma\tanh\left( \frac{E}{2k_{B}T} \right)\right]$ 
is the maximum ATS relaxation time and the minimum allowed is 
$\tau_{min}(E)=\left({5}/{6}\right)\tau_{max}(E)$.
The ATS relaxation-time distribution is therefore very narrow-ranged in 
$\tau $ and also very singular (albeit in the $B=0$ case only). This has 
important experimental consequences that will be discussed elsewhere.

\begin{figure}[!Htbp]
  \centering
  \subfigure[]{\includegraphics[scale=0.25] {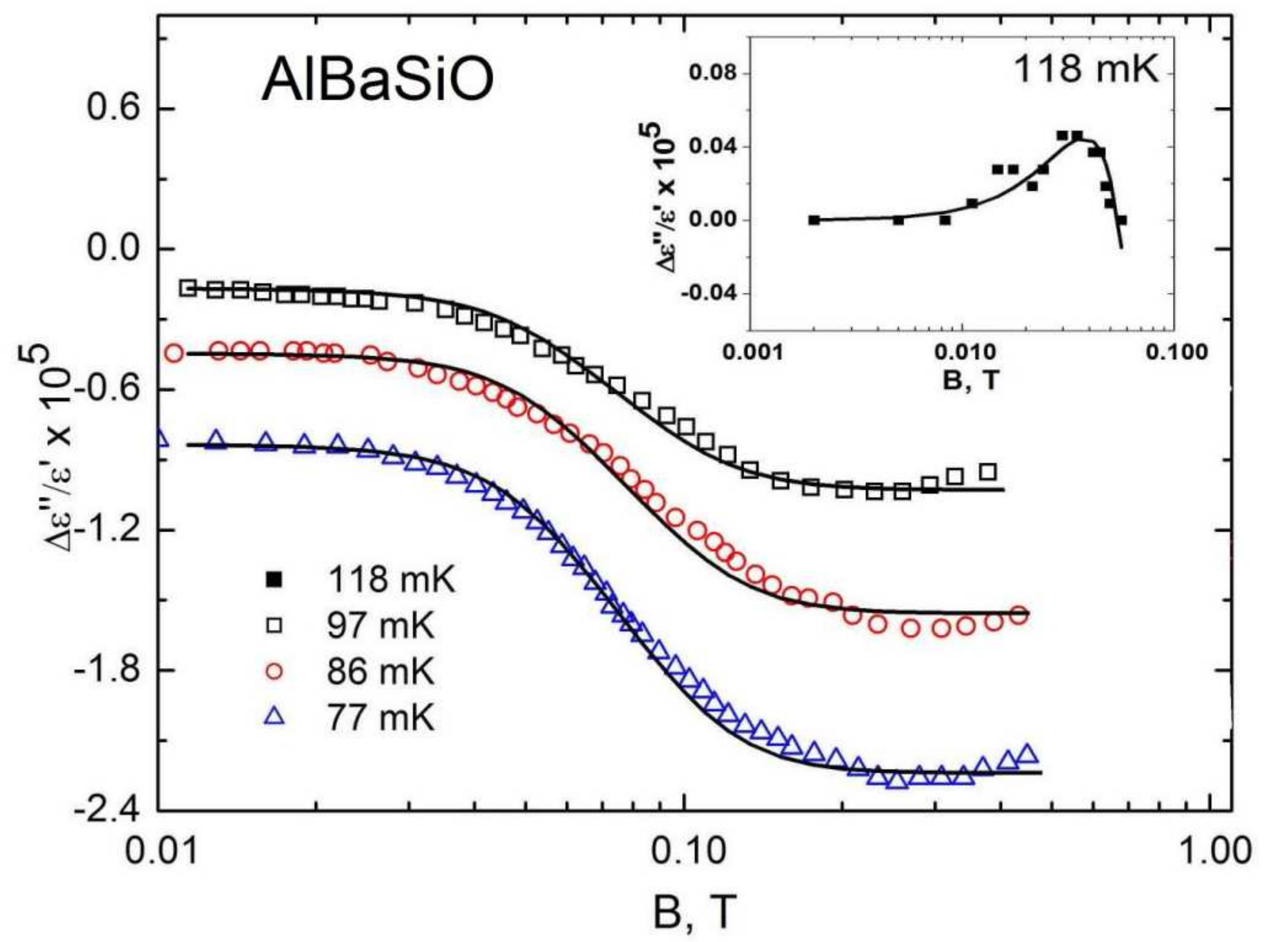} 
\label{image3l}}
  \subfigure[]{\includegraphics[scale=0.30] {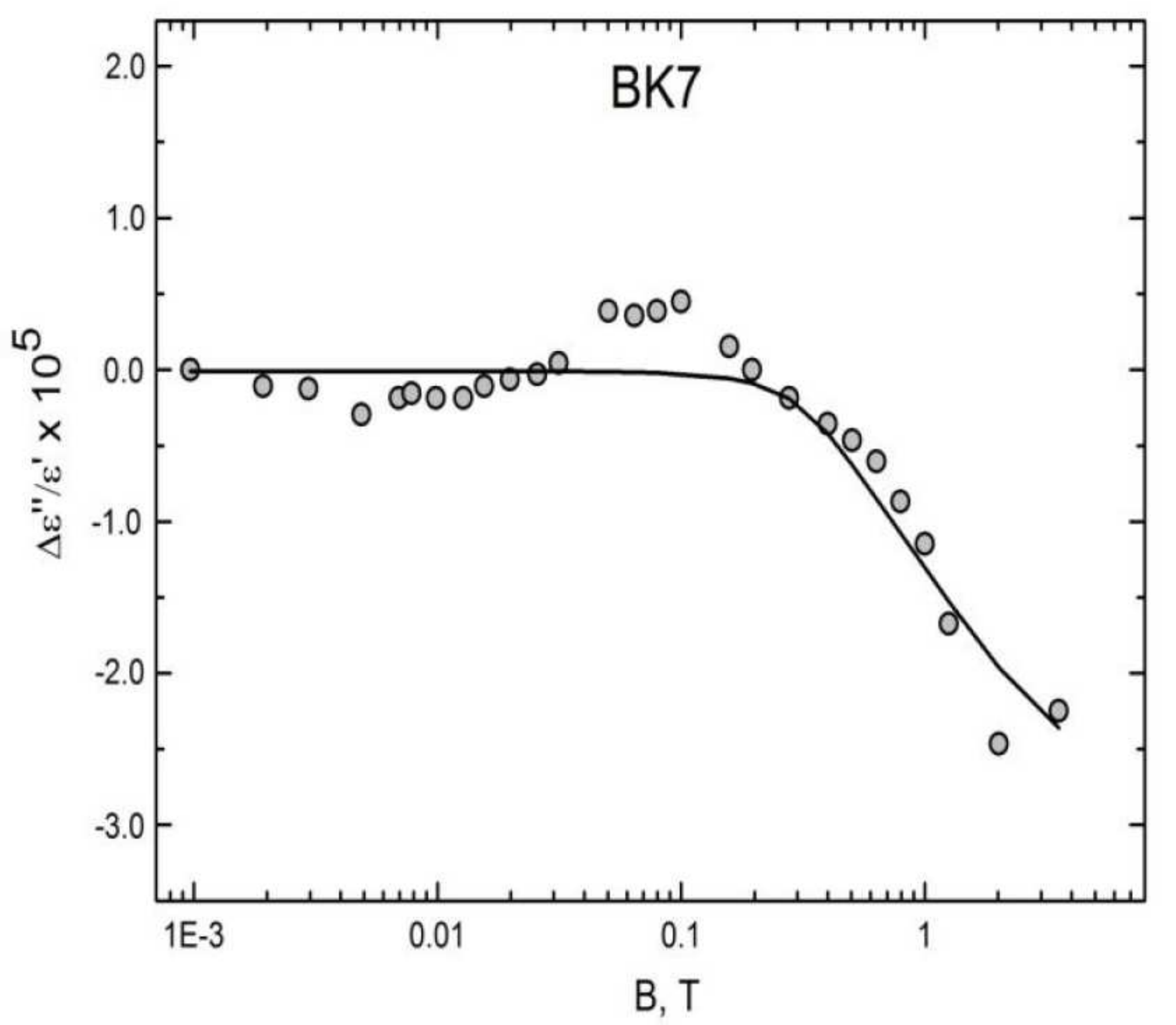} 
\label{image4l}}
  \caption[]{(colour online) The relative dielectric loss as a function of the magnetic field and temperature in the AlBaSiO (BAS) (a) and BK7 (b) glasses (data from \cite{Woh2001}b). The continuous curves are from the present theory. In the inset of (a) we show that a faint peak seen experimentally at very weak fields can also be explained by the theory. From \cite{Pal2011}.}
\label{im3-4}
\end{figure}

It is however not convenient to switch to $\tau $ as an integration variable. 
Then one writes, integrating over $D_0$ first via:
\begin{equation}
\delta \left(E-\sqrt{D^2+D^2_0\varphi^2}\right)=\frac{E}{\varphi\sqrt{E^2-D^2}}\delta \left(D_0-\frac{1}{\varphi}\sqrt{E^2-D^2}\right)
\label{eq88}
\end{equation}
Thus the integral in Eq.~(\ref{eq84}) becomes:
\begin{equation}
\begin{split}
{\alpha }_{REL}=&\frac{\pi P^*\overline{p^2_1}}{4k_{B}T}\int^{\infty }_0{\frac{dE}{E}}{{\cosh }^{{\rm -2}} \left(\frac{E}{2k_{B}T}\right)\ } \int^{\infty }_0{dD\frac{D}{E^2-D^2}}\frac{\omega {\tau }_{ATS}}{1+{\omega }^2{\tau }^2_{ATS}}\\
& \times \theta \left(D-D_{min}\right)\theta \left(D-D_1\left(E\right)\right)\theta \left(D_2\left(E\right)-D\right)\\
\end{split}
\label{eq89}
\end{equation}
The integral in Eq.~(\ref{eq89}) has two special points 
$D_1\left(E\right)=\sqrt{E^2-D^2_{0max}\varphi^2}$ and 
$D_2\left(E\right)=\sqrt{E^2-D^2_{0min}\varphi^2}$, hence 
$D_1\left(E\right)<D_2\left(E\right)$. The $\theta$-conditions divide the 
integral in Eq.~(\ref{eq89}) into two terms with different intervals in the 
energy value: $E_{c1}\le E\le E_{c2}$ and $E\ge E_{c2}$, and the integal for 
$E\le E_{c1}$ vanishes:
\begin{equation}
\begin{split}
{\alpha }_{REL}=&\frac{\pi P^*\overline{p^2_1}}{4k_{B}T}\bigg[\int^{E_{c2}}_{E_{c1}}{\frac{dE}{E}}{{\cosh }^{{\rm -2}} \left(\frac{E}{2k_{B}T}\right)}  \int^{D_2(E)}_{D_{min}}{dD\frac{D}{E^2-D^2}}\frac{\omega \tau \left(E,D\right)}{1+{\omega }^2{\tau }^2\left(E,D\right)}\\
&+\int^{\infty }_{E_{c2}}{\frac{dE}{E}}{\ {{\cosh }^{{\rm -2}} \left(\frac{E}{2k_{B}T}\right)}\int^{D_2(E)}_{D_1(E)}{dD\frac{D}{E^2-D^2}}\frac{\omega \tau \left(E,D\right)}{1+{\omega }^2{\tau }^2\left(E,D\right)}}\bigg]\\
\end{split}
\label{eq810}
\end{equation}
Substituting Eq.~(\ref{eq85}), Eq.~(\ref{eq810}) into Eq.~(\ref{eq83}) and 
then into Eq.~(\ref{eq81}) we obtain the final formula for the dielectric loss 
variation in a magnetic field:
\begin{equation}
\begin{split}
{\Delta \tan \delta}=&\frac{{x_{ATS}\pi P}^*\overline{p^2_1}}{4\epsilon_0\epsilon_r}\frac{1}{k_{B}T}\bigg[\int^{E_{c2}}_{E_{c1}}{\frac{dE}{E}}{{\cosh }^{{\rm -2}} \left(\frac{E}{2k_{B}T}\right)}\int^{D_2(E)}_{D_{min}}{dD\frac{D}{E^2-D^2}\ }\frac{\omega \tau \left(E,D\right)}{1+{\omega }^2{\tau }^2\left(E,D\right)}\\
&+\int^{\infty }_{E_{c2}}{\frac{dE}{E}}{{{\cosh }^{{\rm -2}} \left(\frac{E}{2k_{B}T}\right)}\int^{D_2(E)}_{D_1(E)}{dD\frac{D}{E^2-D^2}}\frac{\omega \tau \left(E,D\right)}{1+{\omega }^2{\tau }^2\left(E,D\right)}}\bigg]\\
\end{split}
\end{equation}
withe $E_{c1}$ and $E_{c2}$ as in previous Sections.
The fitting of relative dielectric loss variation in a magnetic field is shown 
in Figure~\ref{im3-4}, using the best-fit parameters from Table~\ref{table81}.
One can see that, once again, the experimental data are very well reproduced 
by the present theory and with fitting parameters very similar to those 
extracted from the study of the (real part of the) dielectric constant 
(Section 5.2).

\begin{table}[!Htbp]
\begin{center}
\begin{tabular}{ |c|c|c|c|c|c| }
\hline
Temperature & ${\pi x_{ATS}P}^*\overline{p^2_1}/\epsilon_r\epsilon_0$ & $D_{min}$\textit{${}_{ }$, }K & $D_{0min}\left|\frac{q}{e}\right|S_{\Delta }$\textit{, }KÅ${}^{2}$ & $D_{0max}\left|\frac{q}{e}\right|S_{\Delta }$\textit{, }KÅ${}^{2}$ & ${\Gamma {\rm =}\gamma }'k^5_B$, (sK${}^{5}$)${}^{-1}$ \\
  \hline
  \multicolumn{6}{|c|}{{\textbf{AlBaSiO}}} \\
  \hline
77 mK & $1.54·{10}^{-5}$ & $0.0209$ & $1.98·{10}^4$ & $4.96·{10}^5$ & $5.0·{10}^9$ \\
\hline
88 mK & $1.35·{10}^{-5}$ & 0.$0206$ & $1.98·{10}^4$ & $4.96·{10}^5$ & $4.0·{10}^9$ \\
\hline
96 mK & $1.10·{10}^{-5}$ & 0.$0213$ & $1.98·{10}^4$ & $4.96·{10}^5$ & $4.4·{10}^9$ \\
\hline
 \multicolumn{6}{|c|}{{\textbf{BK7}}} \\
\hline
15 mK & $2.02·{10}^{-5}$ & 0.$0287$ & $0.69·{10}^3$ & $0.66·{10}^4$ & $3.34·{10}^9$ \\
   \hline
\end{tabular}
\caption{ Fitting parameters for the dielectric loss in a magnetic field in the AlBaSiO (BAS) and BK7 glasses.}
 \label{table81}
\end{center}
\end{table}


\section{The Magnetic Field Dependent Polarization Echo Amplitude}
\subsection{The Polarization Echo Experiment}
The experimental detection of electric and phonon echoes in glasses is one 
strong convincing argument for the 2LSs' existence. Echoes in glasses are 
similar to other echo phenomena such as spin echo, photon echo and so on. 
But only at very low temperatures the relaxation of the TSs becomes so slow 
that coherent phenomena like polarization echoes become observable in the 
insulating glasses.

\begin{figure}[h]
\centering
   \subfigure[]{\includegraphics[scale=0.90] {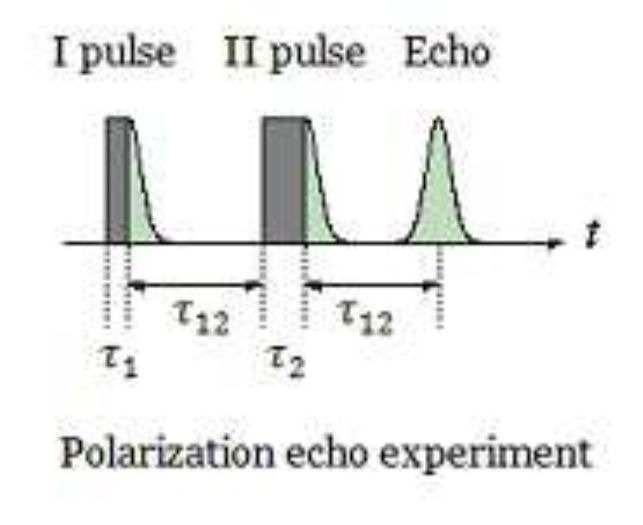}}   
    \subfigure[]{\includegraphics[scale=0.70] {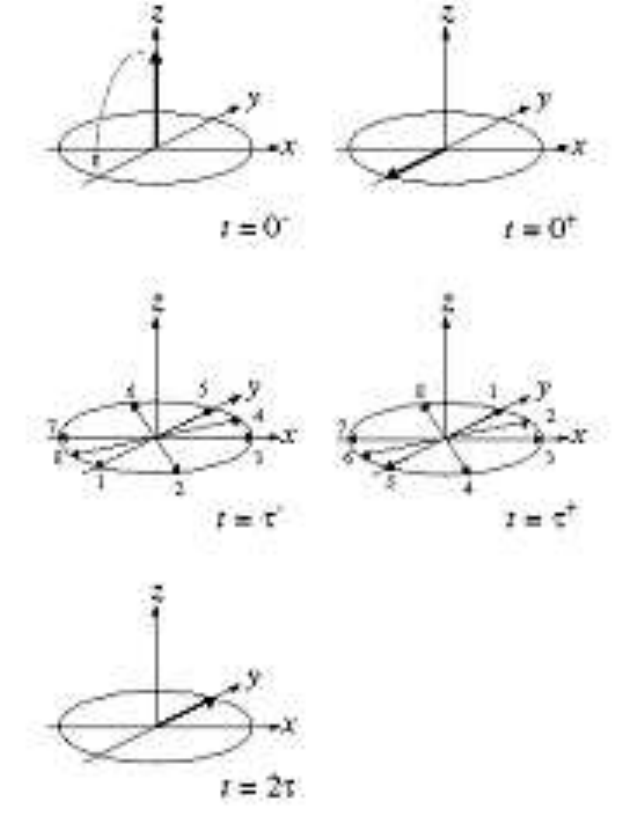}}
\caption{(colour online) The two-pulse polarization echo experiment. 
Hahn's vector interpretation on the right hand side is for NMR's spin-echo 
experiment.}
\label{image1-2e}
\end{figure}

The essence of the effect is the following (see Fig. \ref{image1-2e}). A glass 
sample placed in a reentrant resonating cavity (``Topfkreisresonator'') is 
subjected to two short ac electromagnetic pulses at the frequency of about 
1 GHz separated by a time interval $\tau_{12}$. The duration $\tau_1$ and 
$\tau_2$ of these pulses should be much shorter that all relaxation processes 
in the observed system. The macroscopic polarization produced by the first 
pulse vanishes rapidly due to the distribution of parameters of the TSs 
in glasses. This phenomenon is similar to the well-known free-induction 
decay observed in nuclear magnetic resonance (NMR) experiments. The ``phase'' 
(energy-level populations) of each TS develops freely between the two 
exciting pulses. The second pulse causes an effective time reversal for the 
development of the phase of the TSs. The initial macroscopic polarization of 
the glass is recovered roughly at a time $\tau_{12}$ after the second pulse. 
Since the thermal relaxation processes and (see later) spectral diffusion are 
strongly temperature dependent, polarization echoes in glasses can be observed 
in practice only at very low temperatures, typically below 100 mK. The echo 
amplitude is proportional to the number of TSs that are in or near resonance 
with the exciting microwave pulse and that do not loose their phase coherence 
during the time $2\tau_{\rm 12}$ \cite{Phi1987}.
It should be pointed out that, due to the wide distribution for the parameters 
of the two-level systems in glasses, the description of polarization echoes in 
glasses is much more complicated than in the case of nuclear spin systems. In 
analogy to the two-pulse echo in magnetic resonance experiments this phenomenon 
is referred to as the spontaneous echo.

The polarization echo phenomenon can help to understand more about the 
microscopic structure of TSs in glasses and gives different kinds of 
information. The analysis of these experiments follows that for the equivalent 
magnetic case, except that the TS problem is complicated by three factors. 
First, the elastic or electric dipoles are not aligned with respect to the 
driving field and a calculation of the echo signal involves an average over 
their orientations. 
Secondly, for a given pumping frequency $\omega$ there exists a distribution of 
induced moments (electric or elastic) and relaxation times, which should be 
included in the analysis. Finally, in electric echo experiments the local field 
seen by the TSs is not equal to the applied field, and a local-field correction 
factor must be used when evaluating absolute values of the dipole moment 
\cite{Phi1987}.

In the polarization echo experiments at radio frequencies and at very low 
temperature of about 10 to 100 mK it has been shown that the TSs 
in glasses couple directly to the magnetic field \cite{Lud2003,Lud2002}. 
Unexpectedly, the amplitude of two-pulse echoes in the BAS glass was found to 
be strongly dependent on the applied magnetic field showing a non-monotonic 
(even oscillatory) field variation. In subsequent papers 
\cite{Nag2004,Wur2002}, such behavior was attributed to the existence of 
nuclear electric quadrupole moments (NEQM) for some tunneling particles 
(having nuclear spin $I>\frac{1}{2}$) interacting with 
the magnetic field and with gradients of the internal microscopic electric 
field. The NEQM model is based on the consideration that the levels of 
tunneling particles with non-zero nuclear quadrupole moment exhibit a 
quadrupole splitting, 
which is different in the ground state and in the excited state of a tunneling 
2LS. The magnetic field causes an additional Zeeman splitting of these levels 
giving rise to interference effects. In turn, these effects cause the 
non-monotonic magnetic field variation of the echo amplitude.

\begin{figure}[h!]
\centering
   \subfigure[]{\includegraphics[scale=0.25] {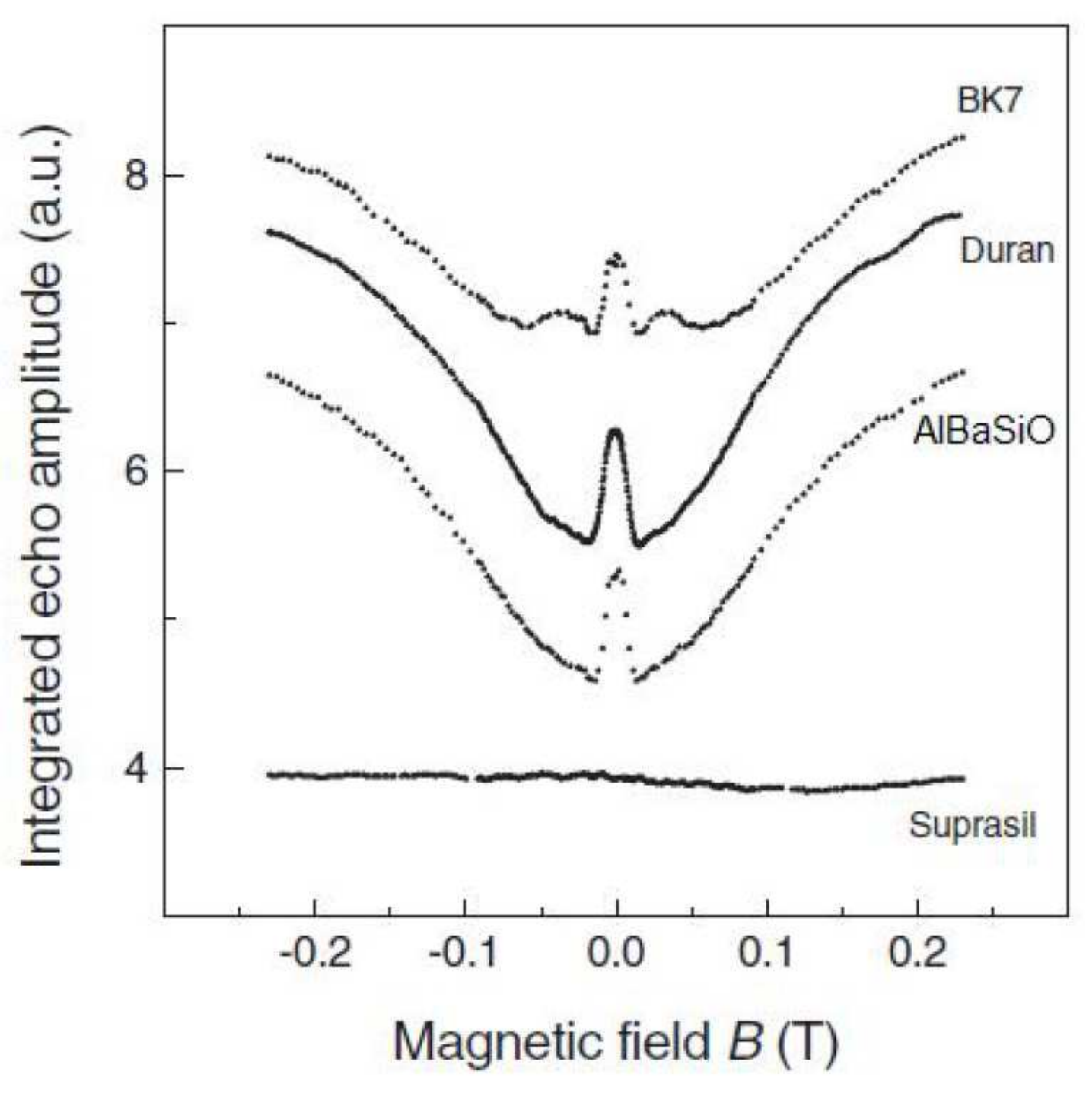}}
   \subfigure[]{\includegraphics[scale=0.255] {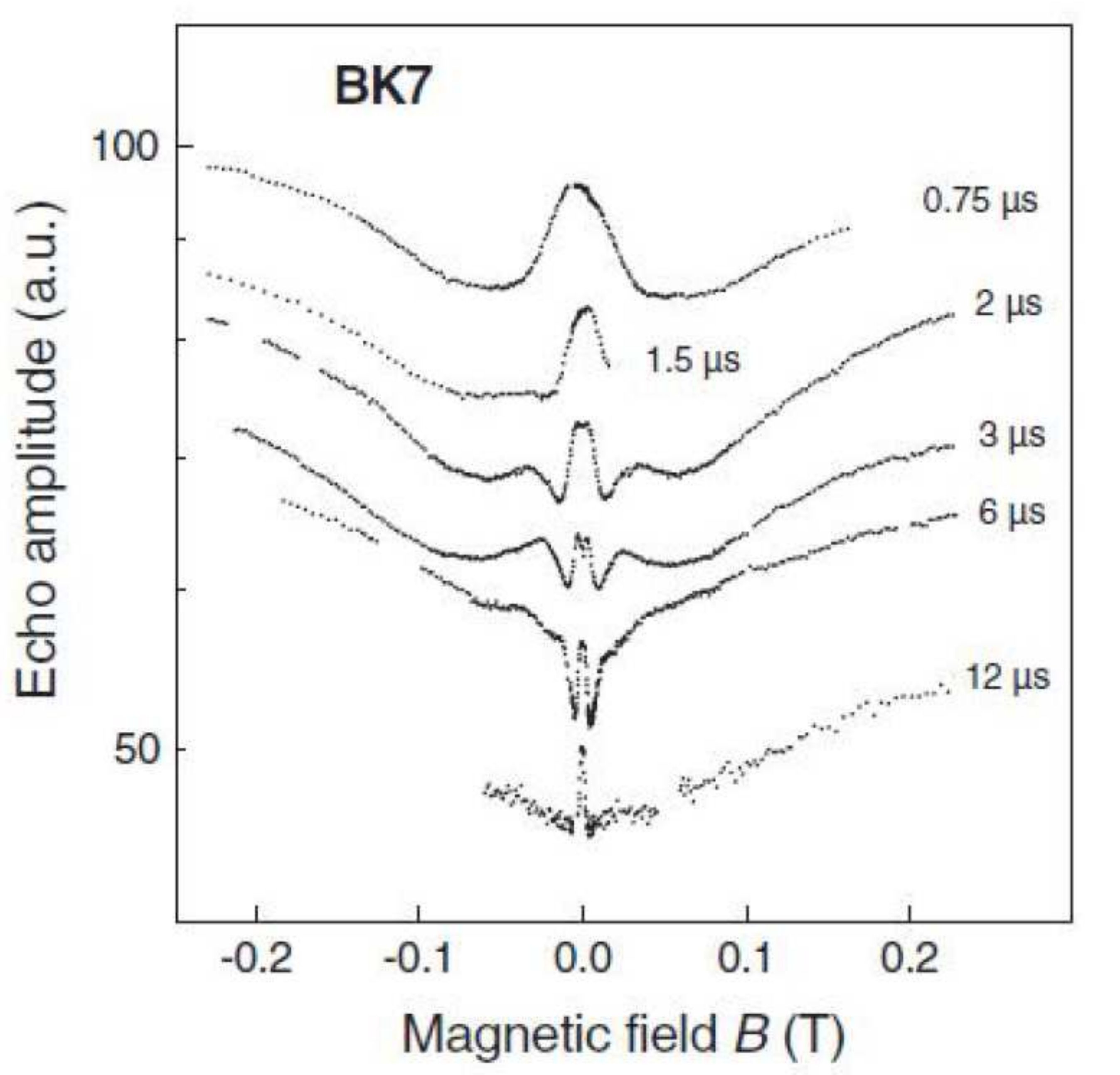}}
\caption{a) The integrated echo amplitude as a function of the magnetic field 
for different silicate glasses: BK7, Duran, AlBaSiO (BAS) and Suprasil I. All 
data were taken at $T$=12 mK, $\tau_{12}$=2 ms, and roughly 1 GHz, except for 
Duran, where the delay time was $\tau_{12}$=1.7 ms \cite{Lud2002}. b) The 
amplitude of two-pulse echoes in BK7 glass as a function of the magnetic field 
for different values of the waiting time $\tau_{12}$ between pulses. All data 
sets were taken at 4.6 GHz and 12 mK except that for $\tau_{12}$=2 ms which 
was taken at 0.9 GHz. }
\label{image3-4e}
\end{figure}

The amplitude (or integrated amplitude) of two-pulse polarization echoes of 
four types of silicate glasses is shown in Fig.~\ref{image3-4e}(a) as a 
function of magnetic field \cite{Lud2002}. In contrast to many other 
low-temperature properties of glasses the influence of the magnetic field on 
the amplitude of spontaneous echoes is obviously not universal. BK7 and Duran 
show similar effects, although the concentration of magnetic impurities 
differs by at least a factor of 20. Perhaps a most remarkable result of the 
measurements is the fact that Suprasil I (very pure $a$-SiO${}_2$) shows no 
measurable magnetic field effect. While Duran, BAS and BK7 contain nuclei 
with non-zero nuclear quadrupole moment, Suprasi I is virtually free of such 
nuclei. This fact is used to provide justification for the nuclear quadrupole 
model. The variation of the echo amplitude with the applied magnetic field is 
similar for Duran, BK7 and BAS, but not identical. 
All three samples exhibit a principal maximum at very weak fields, 
$B\sim$ 10 mT, but only BK7 has a relevant second maximum and a hint to an 
oscillation in $B$. At high fields the amplitude of the echo rises well above 
its value at zero magnetic field and seemingly saturates (yet, see 
Fig.~\ref{heat_capa_fit}, this is very similar to what happens to the inverted heat 
capacity, $-C_p$, as a function of $B$).
In Figure~\ref{image3-4e}(b) the amplitude of spontaneous echoes in the BK7 
glass is shown as a function of the applied magnetic field for different delay 
times $\tau_{12}$ between the exciting pulses. We can see obvious differences 
for different values of $\tau_{\rm 12}$ and that a second maximum (the 
``oscillation'') is not always present. These findings necessitate a good 
theory for spectral diffusion in real glasses and this theory remains to be 
accomplished. 
\begin{figure}[h!]
\centering
   \subfigure[]{\includegraphics[scale=0.25] {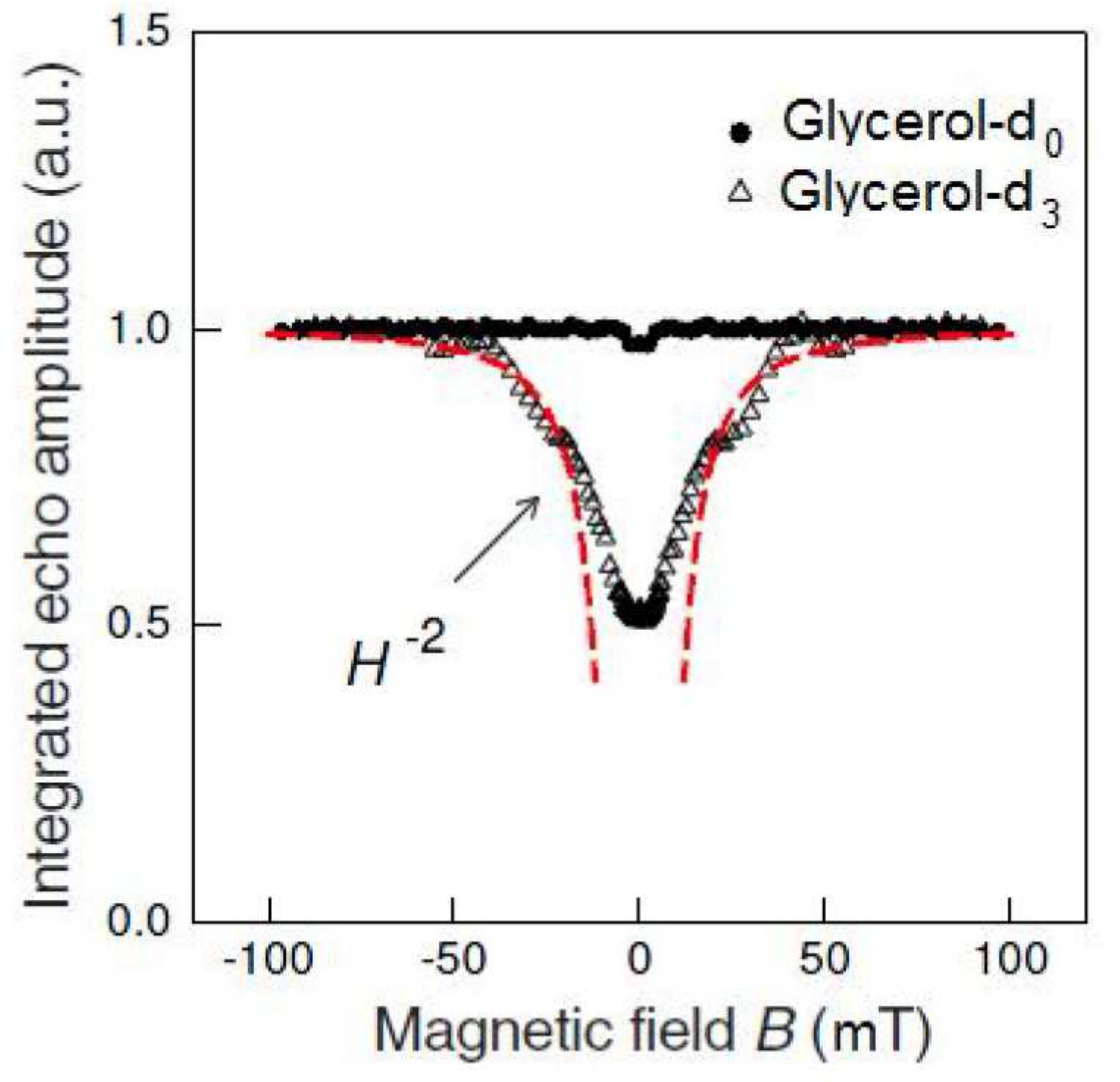}}
    \subfigure[]{\includegraphics[scale=0.25] {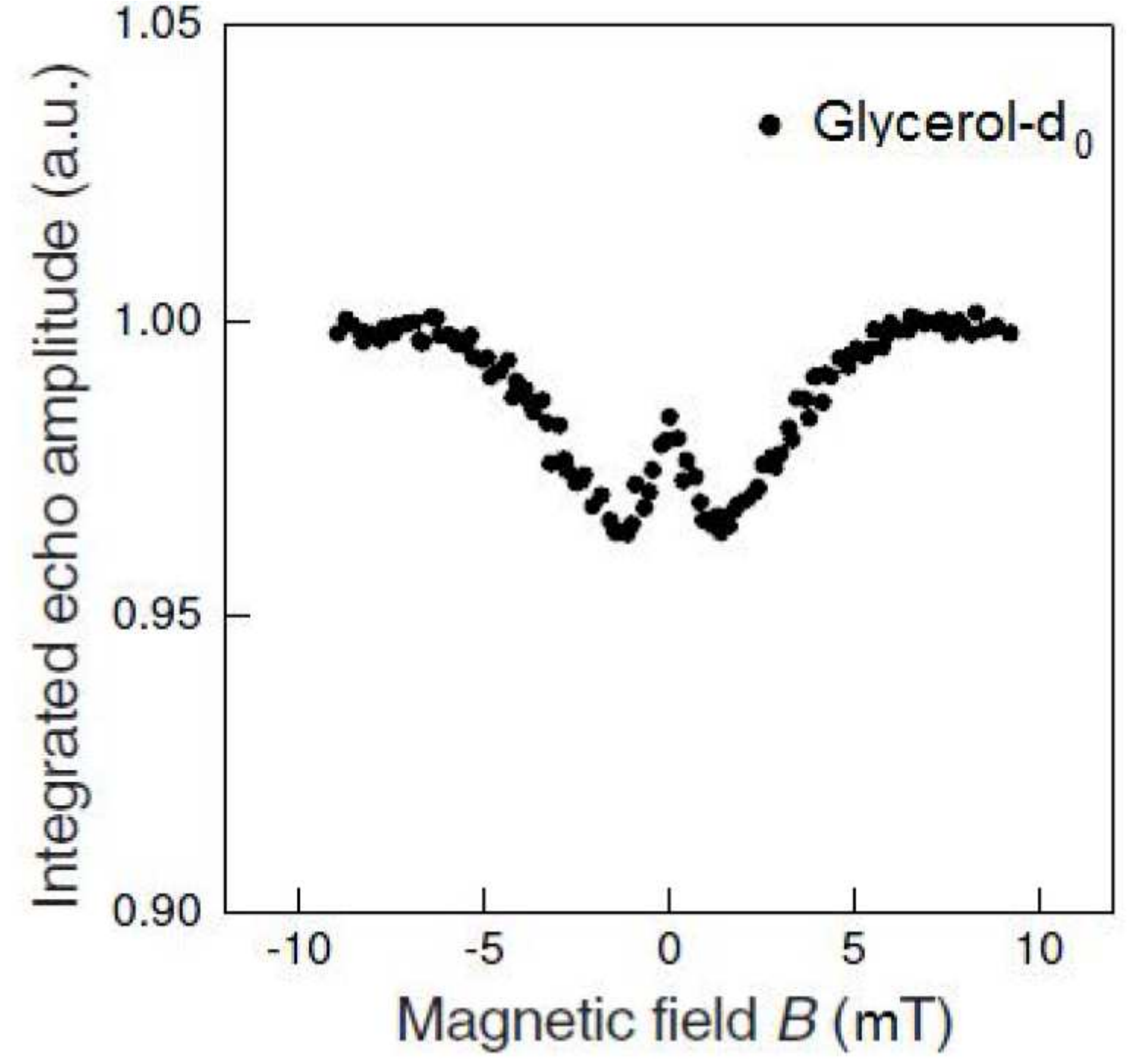}}
\caption{(colour online) The integrated echo amplitude as a function of the 
magnetic field at $T$=13 mK, generated in partially deuterated glycerol 
(glycerol-d$_3$) and in ordinary glycerol (glycerol-d$_0$). On the left-hand 
side the figure shows that the echo amplitude for deuterated glycerol-d$_3$ is 
much more sensitive to the magnetic field in comparison with non-deuterated 
glycerol-d$_0$, shown at the right-hand side (from \cite{Bra2004}). }
\label{image5e}
\end{figure}
The most remarkable fact about these experiments on echoes from glasses in a 
weak magnetic field is that the strong magnetic effect is not confined to the 
inorganic, silicate glasses. Figure~\ref{image5e} shows the amplitude of 
spontaneous echoes in partially deuterated and in ordinary amorphous glycerol 
as a function of the weak magnetic field $B$. In the case of ordinary glycerol 
(d${}_0$) there is very small change of the echo amplitude with $B$. However, 
for partially deuterated glycerol (d${}_3$) a change is much more noticeable, 
of a different shape and duration. This experiment seemingly provides proof 
that the magnetic effect is of nuclear origin, for the two amorphous glycerol 
samples differ in the content of nuclei carrying a NEQM. Glycerol-d$_0$ has 
none, other than the natural abundance of deuterium, some 125 ppm, and of 
$^{17}$O, about 500 ppm, concentrations which are however a factor 10 too weak 
to account for the observed magnetic effect. Glycerol-d$_3$ contains instead 
37.5\% D ($I$=1) and 62.5\% H ($I$=1/2). However, glycerol-d$_8$ (nominally 
100\% D) displays a magnetic effect that is only 10\% larger than in 
glycerol-d$_3$ while glycerol-d$_5$ (62.5\% D) displays a smaller magnetic 
effect than glycerol-d$_3$ \cite{Nag2004,Bra2004b}. All this hints to the fact 
that the effect does not scale with NEQM concentration.

\begin{figure}[h!]
  \centering
  \includegraphics[scale=0.25] {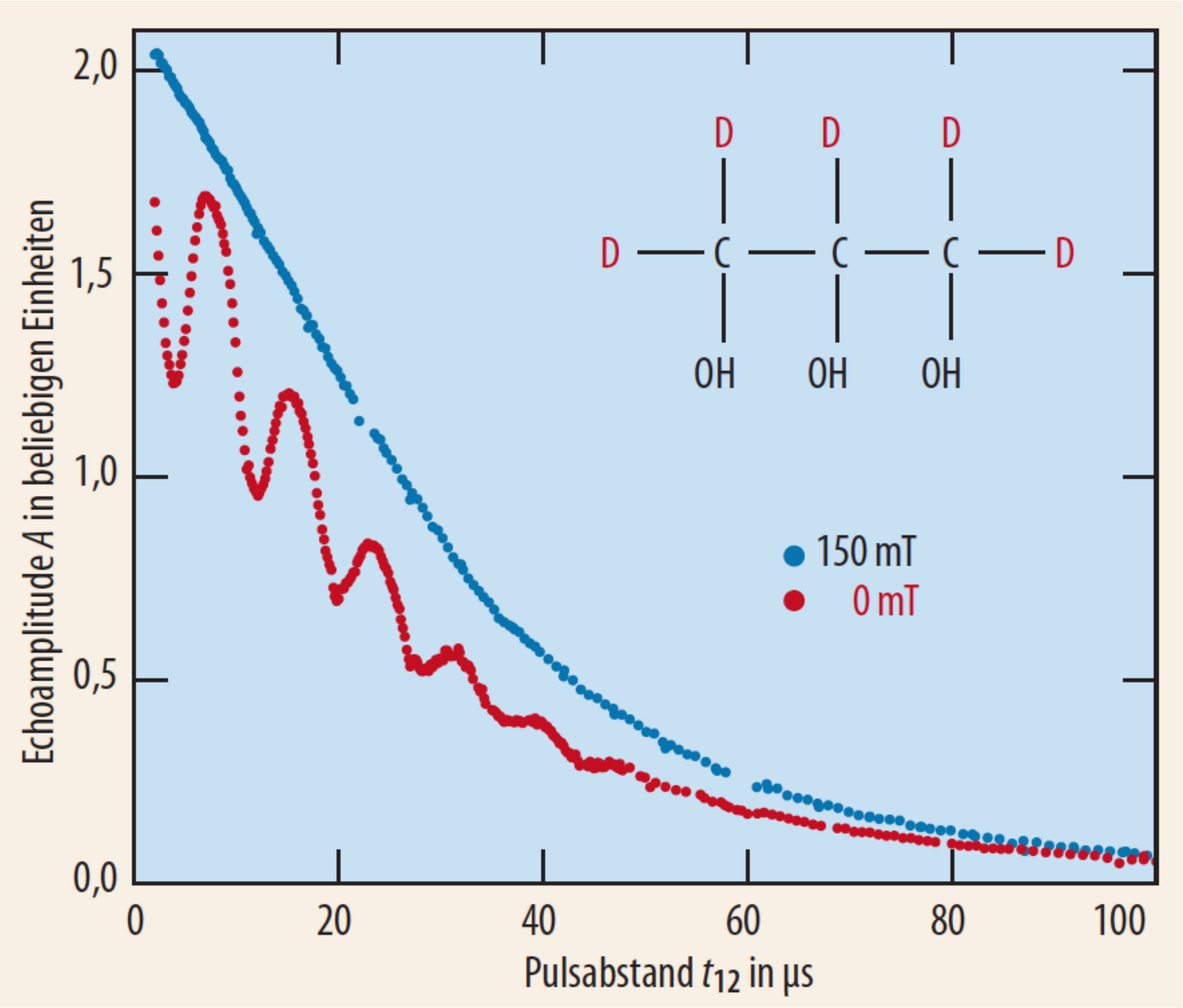}
\caption{(colour online) The integrated echo amplitude as a function of the 
waiting time $\tau_{12}$ at zero magnetic field (red curve) and in weak 
magnetic field of 150 mT as generated in deuterated glycerol-d$_5$ (from 
\cite{Fle2007,Bar2013}).}
  \label{image6e}
\end{figure}

As reproduced in Fig. \ref{image6e}, the integrated echo amplitude as a 
function of the waiting time $\tau_{12}$ in amorphous partially deuterated 
glycerol-d$_5$ (that is C$_3$O$_3$H$_3$D$_5$ instead of ordinary 
C$_3$O$_3$H$_8$) shows exponential decay with spectacular oscillations at zero 
applied magnetic field. On the other hand, all oscillations disappear for the 
relatively weak magnetic field of 150 mT \cite{Nag2004,Fle2007,Bar2013}. 
These findings are extraordinary, especially when combined with the 
observation that for all multi-silicate glasses the oscillations of the echo 
amplitude in $\tau_{12}$ are absent for all values of the magnetic field (a 
fact that the NEQM approach cannot explain). 
All these findings are collectively hard to explain on the basis of the 2LS 
STM or starting from more microscopic models and so far only the NEQM has been 
able to provide the beginning of a rationale for some of these startling 
experimental results.

\subsection{The polarization echo in a magnetic field: Schr\"odinger equation 
formalism}
Using the density matrix formalism of quantum mechanics, it is possible to 
obtain, for a collection of 2LSs only, the following expression for the 
average dipole moment near $t=2\tau_{12}$: \cite{Pal2011}
\begin{eqnarray}
P_{\parallel}&=&\int{dE}\int{d{\Delta }_0\int{d\Delta} \frac{\overline{P}{\Delta }^3_0F^3_0{{\rm p}}^{{\rm 4}}_0}{{\hbar }^3E}}\int^1_0{dy\ y^4}{\rm Im}\left[ \frac{e^{i\omega t}}{{E\Omega }^3_G}{\tanh  \left(\frac{E}{2k_{B}T}\right)\ }{{\sin }^{{\rm 2}} \frac{{\Omega }_G{\tau }_2}{2}\ }\right.\\
&\times& \left({\sin  {\Omega }_G{\tau }_{{\rm 1}}\ }-2i\frac{z}{{\Omega }_G}{{\sin }^{{\rm 2}} \frac{{\Omega }_G{\tau }_1}{2}\ }\right){\exp  \left(-\overline{\frac{\gamma }{2}}t\right)\ }{\exp  \left(iz\left({t-2\tau }_{12}\right)\right)\ }\\
&{\rm \times}& \left. { \exp \left(-i\int^t_0{\Delta \omega \left(t'\right)s\left(t'\right)dt'}\right)\ }\right] \delta (E-\sqrt{{\Delta }^{{\rm 2}}{\rm +}{\Delta }^{{\rm 2}}_0})
\label{echo_densitymat}
\end{eqnarray}
where: ${\bf F}={\bf F}_0(t)\cos(\omega t)$ is the pulsed external electric 
field (having intensity $F_0$ and MW frequency $\omega$), ${\bf p}_0$ is the
local bare electric 2LS dipole moment, $y=\cos\theta$, $\theta$ is the 
dipole's orientation angle wrt ${\bf F}_0$, 
$\Omega_R=\Delta_0{\bf p}_0\cdot{\bf F}_0/\hbar E$ is the Rabi frequency of
the resonant 2LS, $\Omega_G=\sqrt{\Omega_R^2+(\omega-E/\hbar)^2}$ is the
generalized (non-resonant) Rabi frequency, $z=E/\hbar-\omega$ and where 
$\bar{\gamma}=2\tau^{-1}$ is the 2LS intrinsic phonon damping ($\tau$ being
the 2LS phonon-induced relaxation time (Section 5.1)). We have also defined a 
fluctuating 2LS energy gap $E(t)=E+\hbar\Delta\omega(t)$ with 
$\hbar\Delta\omega(t)$ the contribution due to interactions with the 
surrounding thermal 2LSs, and a service function:
\begin{eqnarray}
s(t)=\begin{cases}
+1 &{\rm if} \quad 0<t<\tau_{12} \cr
-1 &{\rm if} \quad t>\tau_{12} \cr 
\end{cases}
\end{eqnarray}
This result reproduces (and improves) the derivation by Gurevich et al. 
\cite{Gur1990}. It should be stressed that the (normalized in $\omega _0=E/h$) 
spectral function 
$\sigma \left({\omega }_0\right)=\frac{1}{2}{{\Omega }^2_R}/{{\Omega }^3_G}$ 
is in practice, for ${\Omega }_R\ll \omega$, very narrowly centered at the 
pumping frequency $\omega$ and can therefore be replaced by a Dirac's 
$\delta \left(\omega _0-\omega \right)$, thus resulting in the 
strictly-resonant approximation which is normally employed. Things change 
considerably, however, when the density of states ${\rm g}(E)$ is non-uniform, 
as is the case for the ATS model.

The above result for the echo signal from a collection of 2LS can also be 
obtained - and from first principles - from a lengthy but straightforward 
Schr\"odinger equation treatment in which high-frequency modes are neglected 
and phonon-damping is treated in a phenomenological way. In the most rigorous 
way, one obtains for the echo signal:
\begin{equation}
{\wp }_\parallel \left(t\right)={\wp }_+\left(t\right){\cos \left(\omega \Delta \tau \right)}-{\wp }_-\left(t\right){\sin \left(\omega \Delta \tau \right)}
\label{echo_schr}
\end{equation}
where $\Delta \tau ={\tau }_2-\ {\tau }_1$ (whence $\omega \Delta \tau $ typically as large as ${10}^2$) and where:
\begin{equation}
\begin{split}
{\wp }_+\left(t\right)=&{{\rm p}}_{0\parallel}\frac{{\Delta }_0}{E}\tanh  \left(\frac{E}{2k_{B}T}\right)e^{-\frac{\overline{\gamma }t}{2}}{\left(\frac{{\Omega }_R}{{\Omega }_G}\right)}^3\\
&{\rm Im}\bigg\{{{\rm sin}}^2\left(\frac{{\Omega }_G{\tau }_2}{2}\right)\left[{\sin  \left({\Omega }_G{\tau }_1\right)}-2i\frac{{\omega }_0-\omega }{{\Omega }_R}{{\rm sin}}^2\left(\frac{{\Omega }_G{\tau }_1}{2}\right)\right] \\
&\exp\{i{\omega }_0\left(t-2{\overline{\tau }}_{12}\right)-i\int^t_0{\Delta \omega \left(t'\right)s\left(t'\right)dt'}\}\bigg\}\\
{\wp }_-\left(t\right)=&{{\rm p}}_{0\parallel}\frac{{\Delta }_0}{E}\tanh  \left(\frac{E}{2k_{B}T}\right)e^{-\frac{\overline{\gamma }t}{2}}{\left(\frac{{\Omega }_R}{{\Omega }_G}\right)}^3\\
&{\rm Re}\bigg\{{{\rm sin}}^2\left(\frac{{\Omega }_G{\tau }_2}{2}\right)\left[{\sin  \left({\Omega }_G{\tau }_1\right)}-2i\frac{{\omega }_0-\omega }{{\Omega }_R}{{\rm sin}}^2\left(\frac{{\Omega }_G{\tau }_1}{2}\right)\right]\\
&\exp\{i{\omega }_0\left(t-2{\overline{\tau }}_{12}\right)-i\int^t_0{\Delta \omega \left(t'\right)s\left(t'\right)dt'}\}\bigg\}\\
\end{split}
\end{equation}
in which 
$2{\overline{\tau }}_{12}={\tau }_1+{\tau }_{12}+{\tau }_2
+{\tau }_{12}\approx 2{\tau }_{12}$ 
is the total elapsed time at the echo signal's centre. The above 
Eq.~(\ref{echo_schr}) gives the expectation value of the polarization per TS 
in the direction of the applied electric field. It still needs to be averaged 
wrt all STM parameter distribution and over a uniform orientational 
distribution of 2LS dipoles ${\bf p}_0$. 
${\frac{\overline{\gamma }}{2}=\tau }^{-1}$ is again the phonon relaxation 
rate. If $\omega \Delta \tau $ is neglected, then Eq.~(\ref{echo_densitymat}) 
(after averaging) is recovered for $P_\parallel=\overline{{\wp }_\parallel}$.

We are now in the position to extend the polarization echo's calculation to 
the case of the ATS model describing glasses in a magnetic field; the point of 
view will be taken that a background of ordinary 2LS's - insensitive to the 
magnetic field - also exists in the glass. This is in line with our generic
cellular model for the real glasses (Section 2).

One starts with a collection of 3LS ($n_w=3$ is not only computationally 
convenient, but physically correct as explained in Section 2), but with the 
single ATS Hamiltonian written in the energy representation:
\begin{equation}H'=SHS^{-1}=\left( \begin{array}{ccc}
{\cal E }_0 & 0 & 0 \\
0 & {\cal E }_1 & 0 \\
0 & 0 & {\cal E }_2 \end{array}
\right)+S\left( \begin{array}{ccc}
-{{\mathbf p}}_1\cdot {\mathbf F} & 0 & 0 \\
0 & -{{\mathbf p}}_2\cdot {\mathbf F} & 0 \\
0 & 0 & -{{\mathbf p}}_3\cdot {\mathbf F} \end{array}
\right)S^{-1}
\label{3lselectric}
\end{equation}
where the diagonalizing matrix $S=S(\varphi)$ is magnetic-field dependent, the 
${\cal E }_i$\textit{ }are the (\textit{B}-dependent) ATS energy levels and the 
${{\mathbf p}}_i$ are the wells' electric dipoles. As in the treatment of 
Gurevich \textit{et al.} \cite{Gur1990} there is also a phonon bath, but this 
will be treated - as always - phenomenologically and resulting in a 
phonon-damping exponential. The second term in Eq.~(\ref{3lselectric}) causes 
irrelevant energy-level shifts and produces an extra matrix term 
$\Delta H'(t)=(A_{ij})$ of which the only relevant element (see below) is
\begin{equation}
A_{01}=A^*_{10}=\sum^3_{k=1}{-{{\mathbf p}}_k\cdot {{\mathbf F}}_0{\rm \ }{{\rm S}}_{{\rm 0k}}{\rm (}\varphi{\rm )}{{\rm S}}^{{\rm *}}_{{\rm 1k}}{\rm (}\varphi{\rm )}}\ {\cos  \omega t\ }
\end{equation}
The $A_{ij}$ cause transitions between the ATS levels 0, 1, 2 when the pulses 
are applied. In the weak magnetic field limit (most appropriate for the echo 
experiments) and in the approximation $D\ll D_0$ that we always use (and that 
is always confirmed by out best fits to the data), one quickly discovers that 
the second excited level remains unperturbed and one can make use of the 
``effective 2LS approximation'' (where, however, the ground-state 
wavefunctions of the three wells mix). One can then repeat the Schr\"odinger 
equation (or density-matrix, for that matter) calculation carried out for the 
2LS case, at the cost of introducing a complex Rabi frequency:
\begin{equation}
{\Omega }_0= \frac{A_{01}}{\hbar}
\end{equation}
The evolution of the generic ATS during and in the absence of pulses can then 
be followed exactly, in much the same way as before, except that in order to 
simplify the formalism it is convenient to introduce from the outset an 
orientationally-averaged Rabi frequency (now a real quantity):
\begin{equation}
{\Omega }_R=\sqrt{\overline{{|{\Omega }_0|}^2}}
\end{equation}
the bar denoting the average wrt 3LS base-triangle's orientations. Replacing 
${\Omega }_0$ with ${\Omega }_R$ before carrying out the averaging of the 
sample's polarization is our main approximation, allowing for a considerably 
simplified treatment and leading to the magnetic-field dependent expression 
\cite{JugXXXX}:
\begin{equation}
{\Omega }_R=\frac{{{\rm p}}_{{\rm 1}}{{\rm F}}_0}{\hbar }\sqrt{\frac{D^2_0\varphi^2+\frac{5}{6}D^2}{{6E}^2}}
\end{equation}
Here, ${\bf p}_1$ is a single-well (averaged) electric dipole and 
$E=\hbar\omega_0=\sqrt{D^2+D^2_0\varphi^2}$ is the usual magnetic-field 
dependent lower energy gap in the weak field approximation. The above 
approximation for $\Omega_0$ treats incorrectly the ATS's that have 
${\bf F}_0$ roughly orthogonal to the ATS base triangle; luckily these have 
$\Omega_0\approx 0$ and do not contribute to the echo signal.

Proceeding as for the derivation of Eq.~(\ref{echo_schr}) one finds that there 
is a magnetic contribution to the (partly averaged) polarization of the sample 
from the generic ATS given by \cite{JugXXXX}:
\begin{equation}
\begin{split}
\Delta {\wp }_\parallel(t)\cong& -\frac{\hbar }{{{\rm F}}_0}\tanh  \bigg( \frac{E}{2k_B T} \bigg) e^{-\frac{\overline{\gamma }}{2}t}\frac{\Omega^4_R}{\Omega^3_G} \\
&{\rm Im}~\left\{ {\sin}^2 (\frac{{\Omega }_G{\tau }_2}{2})\bigg[{\sin ({\Omega }_G{\tau }_1) }-2i\frac{{\omega }_0-\omega }{{\Omega }_R}{{\rm sin}}^2(\frac{{\Omega }_G{\tau }_1}{2})\bigg] \right\} e^{i\Phi(t)-i\int^t_0{\Delta \omega (t')s(t')dt'}} \\
\end{split}
\label{eq930}
\end{equation}
Now, ${\frac{\overline{\gamma }}{2}=\tau }^{-1}$ is the magnetic ATS phonon 
relaxation rate given by Eq. (\ref{eq84}), the generalized Rabi frequency is 
again given by 
${\Omega }_G{\rm =}\sqrt{{\Omega }^{{\rm 2}}_R{\rm +}{{\rm (}{\omega }_0{\rm -}\omega {\rm )}}^{{\rm 2}}}$ 
and:
\begin{equation}
\Phi\left(t\right)={\omega }_0\left(t-2{\overline{\tau }}_{12}\right)+\omega \Delta \tau 
\end{equation}
is the appropriate time argument. From this, it is obvious that the time at 
which all ATS (regardless of their energy gap $E=\hbar {\omega }_0$) will be 
refocused is $t=2{\overline{\tau }}_{12}$ and this determines the echo's peak 
position (if the echo signal has a reasonable shape, which is not always 
true \cite{Lud2003}). 
The measured echo amplitude's contribution from the magnetic ATS is therefore 
(allowing for an arbitrary amplification factor $A_0$):
\begin{equation}
\begin{split}
\Delta A\left(\varphi\right)=&A_0\frac{d}{{\varepsilon }_0{\varepsilon }_r}x_{ATS}2\pi P^*\int^{\infty }_0{dE}\int{\frac{dD}{D}}\int{\frac{dD_0}{D_0}\Theta (}D,D_0)\\
&\times \delta \left(E-\sqrt{D^2+D^2_0\varphi^2}\right)\Delta {\wp }_\parallel\left(2{\overline{\tau }}_{12}\right)
\end{split}
\label{eq932}
\end{equation}
where $d$ is the sample's thickness, $\Theta (D,D_0)$ is the usual 
theta-function restriction for the integration domain (previous Sections) and 
where a final orientational averaging wrt the angle 
$\beta =\widehat{{\mathbf B}{{\mathbf S}}_{\triangle }}$ (defining the 
A-B phase $\varphi$, see Eq. (\ref{ABphase})) is in order. At this point 
one deals with the delta-function's constraint and the energy parameters 
integrations in the usual way, to arrive at, after a lengthy calculation
\cite{JugXXXX}:
\begin{equation}
\begin{split}
\Delta A(\varphi)\cong & {-A}_0\frac{d}{{\varepsilon }_0{\varepsilon }_r}x_{ATS}\frac{4\pi {\hbar }^2P^*}{{{\rm F}}_0}{\cos (\omega \Delta \tau) }\\
& \times \int^{E_{c_2}}_{E_{c_1}}{\frac{dE}{E}}\int^{D_2(\varphi)}_{D_{min}}{\frac{dD}{D}{\tanh (\frac{E}{2k_{B}T})}\frac{E^2}{E^2-D^2}}\\
& \cdot e^{-w2{\overline{\tau }}_{12}}{\Omega }^{2}_R\sigma (E)\left[{\rm S}({\theta }_{{1}},{\theta }_{{2}}){\tan (\omega \Delta \tau)}{\rm +C}({\theta }_{{1}},{\theta }_{{2}})\right] \\  
&+\int^{\infty }_{E_{c_2}}{\frac{dE}{E}}\int^{D_2(\varphi)}_{D_1(\varphi)}{\frac{dD}{D}}({\rm same~integrand~as~above}\dots ) \\
\end{split}
\label{eq933}
\end{equation}
where we have defined the functions:
\begin{eqnarray}
&&\sigma \left({\rm E}\right)=\frac{{\Omega }^{{\rm 2}}_{{\rm R}}}{{{\rm 2}\hbar\Omega }^{{\rm 3}}_{{\rm G}}}{\rm =}\frac{{\Omega }^{{\rm 2}}_R}{{\rm 2}\hbar {\left({\Omega }^{{\rm 2}}_R{\rm +}{{\rm (}{\omega }_0-\omega {\rm )}}^{{\rm 2}}\right)}^{{\rm 3/2}}}  \nonumber \\
&&{\rm S}\left({\theta }_{{\rm 1}},{\theta }_{{\rm 2}}\right)=\sin \left(\Omega_G\tau_1\right) \sin^2\left(\Omega_G\tau_2/2\right) \\
&&{\rm C}\left(\theta_1,\theta_2\right)=-2\frac{\omega_0-\omega}{\Omega_R}\sin ^2\left(\Omega_G \tau_1/2\right)\sin ^2\left(\Omega_G \tau_2/2\right) \nonumber 
\label{eq934}
\end{eqnarray}
with $\theta_{1,2}=\Omega_G\tau_{1,2}$ the so-called pulse areas. $E_{c_{1,2}}$ 
are as in the previous Sections, whilst 
$D_{1,2}\left(\varphi\right)=\sqrt{E^2-D^2_{0max,min}\varphi^2}$ 
and $E={\hbar \omega }_0$.

In going from Eq.~(\ref{eq930}) to Eq.~(\ref{eq933}) we have tacitly made some 
assumption on the (fully averaged) spectral diffusion term  
$e^{-i\int^{2{\overline{\tau }}_{12}}_0{\Delta \omega \left(t'\right)s\left(t'\right)dt'}}$. 
The theory of spectral diffusion (SD) for the magnetic multi-welled ATS is a 
chapter still open, however we can safely assume that what was found by many 
Authors for NMR's spin-echoes and for the 2LS polarization echoes in glasses 
holds for the ATS as well. Namely, that there is a wide range of waiting times 
where the decay of the echo amplitude is a simple exponential in ${\tau }_{12}$ 
so that one can replace the SD term with 
$e^{-{2{\overline{\tau }}_{12}}/{{\tau }_\varphi}}$ , where 
${\tau }_\varphi(T)$ is a SD characteristic time depending only on 
temperature. There must be a SD time ${\tau }_{\varphi(3)}$ for the ATSs as 
well as a SD time ${\tau }_{\varphi(2)}$ for the standard 2LSs' ensemble. For 
the latter, theory shows \cite{Gal1988,Bla1977} that this parameter is 
independent of the energy gap $E$ and thus for the ATS we shall assume the same 
and, moreover, that (like for the phonon damping rate and Rabi frequency) its 
dependence on the magnetic field is weak or absent. This allows us to lump the 
SD problem together with phonon damping, yielding an overall exponential 
relaxation rate:
\begin{equation}
w(E,D)={{\tau }_\varphi}^{-1}+{\tau }^{-1}(E,D)
\label{eq935}
\end{equation}
in which the SD time is typically much shorter than the phonon-damping time 
$\tau $ and depends only on temperature through:
\begin{equation}
{{\tau }_\varphi(3)}^{-1}=c_{ATS}T
\end{equation}
with $c_{ATS}$ an appropriate constant. The assumption of an overall 
simple-exponential decay of the echo amplitude with ${\tau }_{12}$ is well 
verified experimentally \cite{Ens1996}.

\begin{figure}[!Htbp]
  \centering
  \includegraphics[scale=0.30] {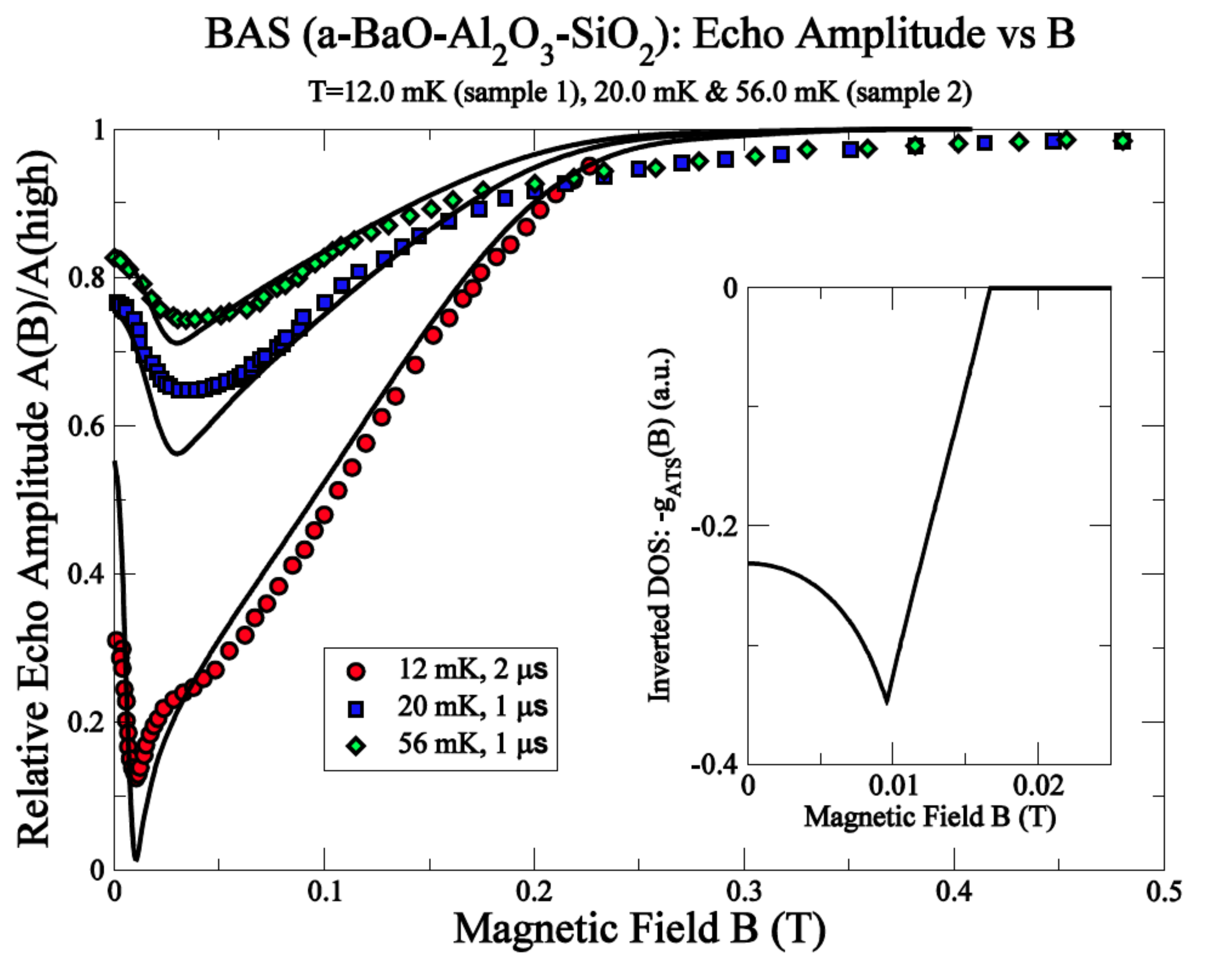}
\caption{(colour online) Magnetic field dependence of the polarization echo 
amplitude (relative to its value at ``high'' fields where saturation occurs) 
for the AlBaSiO glass \cite{Lud2002} (also referred to as BAS) at given 
experimental conditions. We believe two separate samples have been used. 
Continuous curves from our theory. Nominal frequency 1 GHz, 
$\tau_2=2\tau_1$=0.2 $\mu$s. Inset: behaviour of the ATS DOS for the same 
parameters (the physical origin of the effect).}
\label{image7e}
\end{figure}

\begin{figure}[!Htbp]
  \centering
  \includegraphics[scale=0.70] {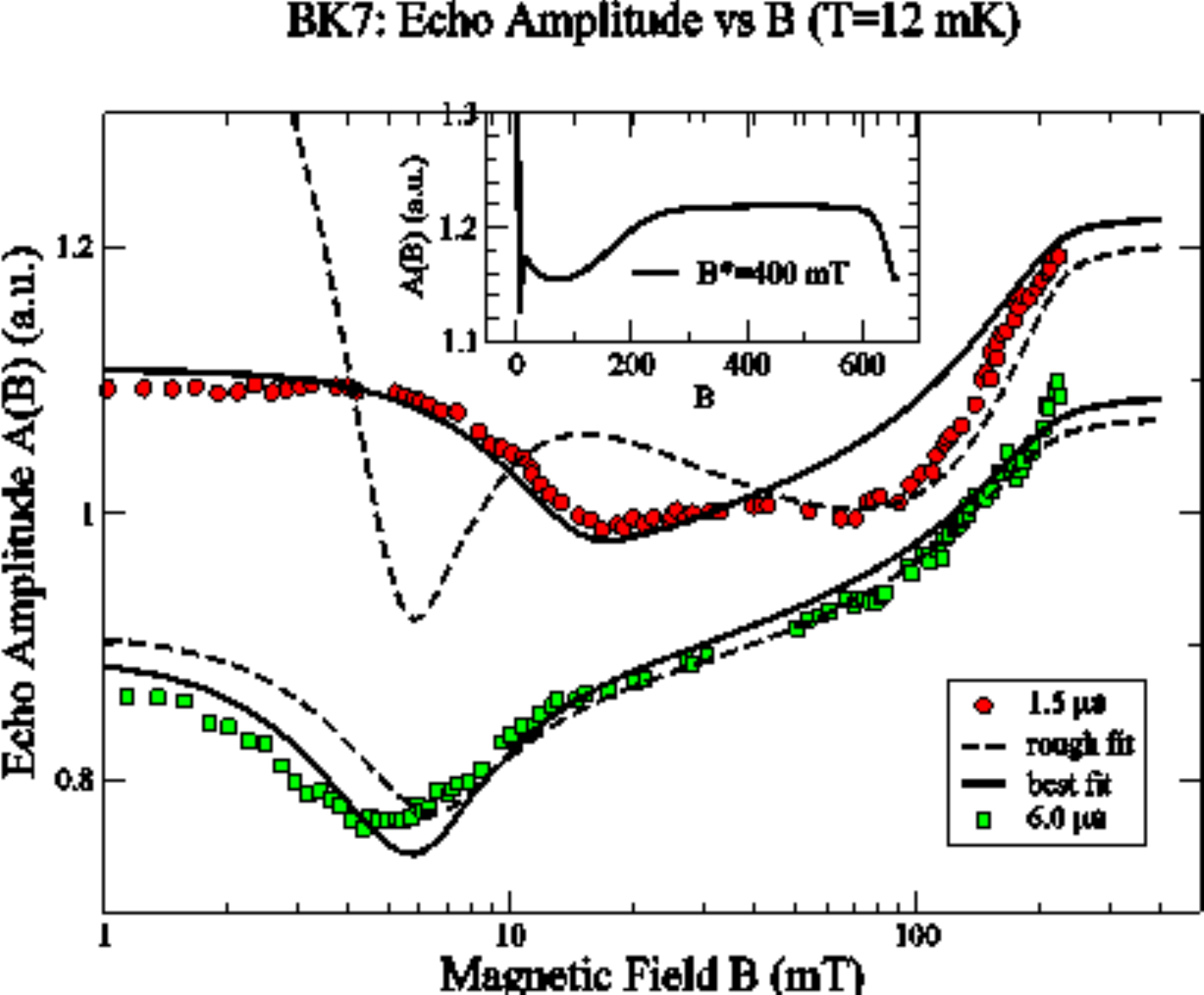}
\caption{(colour online) Magnetic field dependence of the polarization echo 
amplitude for the BK7 glass \cite{Lud2002} at given experimental conditions. 
Dashed curves (rough fit) and continuous curves from our theory; there are no 
more than two observable maxima or minima (no true oscillations). Nominal 
frequency 0.9 GHz, $\tau_2=2\tau_1$=0.2 $\mu$s. Inset: our prediction for the 
higher magnetic field regime ($B^{*}$ as defined in Section 5).}
\label{image8e}
\end{figure}

We now make use of Eq.~(\ref{eq933}) to fit the experimental data for the 
multi-silicates, the idea being that the total amplitude is 
$A\left(\varphi\right)=A_{2LS}+\Delta A\left(\varphi\right)$ (which must be 
averaged wrt the ATS magnetic orientation angle $\beta$).
Fig.~\ref{image7e} shows the experimental results for the relative echo 
amplitude in AlBaSiO (BAS glass) as a function of the magnetic field; values 
of \textit{B} up to 0.6 T have been explored and for three temperatures. The 
data are fitted with our theory with parameters as reported in 
Table~\ref{table91}. The agreement between theory and experiment is highly 
satisfactory, given the simplifications 
used in the theory. There is only one minimum in $A(B)$ and the inset in Fig. 
\ref{image7e} shows that again it is the ATS density of states (DOS) that is 
responsible for the magnetic effect (Section 3). Indeed, by enforcing the 
strict-resonance condition $\sigma (E)\to \delta (E-\hbar \omega )$ 
Eq.~(\ref{eq933}) collapses to a quantity very much like the DOS (convoluted 
with slow-varying corrections) and with the same behaviour, thus reproducing 
the main shape of $\Delta A(B)$. It is the non-resonant convolution of this 
quasi-DOS with other $E$-dependent functions that produces the rounding of the 
minimum and the $B^{-2}$ saturation. Interestingly, though $\tau_\phi\ll \tau$, 
the phonon-damping term plays a main role in the rounding of the high-$B$ tail 
to a $B^{-2}$ (as observed) saturation. The ATS approach predicts also a 
linear in $B$ intermediate decay regime of the echo amplitude, and this is 
often experimentally observed.
Next, in Fig.~\ref{image8e} we present the comparison of theory and experiment 
for data for the echo amplitude in BK7 (good optical glass, hence devoid of 
true microcrystals, bust nevertheless containing RERs) at two different values 
of the waiting time $\tau_{12}$. It is remarkable how our theory, despite the 
simplifications and the total absence of multi-level physics (as advocated by 
the NEQM approach), can reproduce all the features of the experimenatl data, 
including every change of curvature in $A(B)$ vs. $B$. A rough fit, not aiming 
at high ${\chi }^2$ agreement, reproduces the two maxima (and minima) that the 
NEQM approach takes as indication of the multiple (rapid) oscillations ensuing 
from the quantum beatings due to the Zeeman- and NEQM-splitting of the generic 
2LS \cite{Wur2002}. There are never more than two observed minima, in practice, 
and these can be reproduced by our simple ATS model.
Finally, in the inset of Fig.~\ref{image8e}, we show what the experimentalists 
missed by not exploring higher magnetic-field values. Using the simple-minded 
correction for the lower energy gap at higher fields, we plot the expected 
behaviour of $A(B)$ for intermediate fields. After the two minima, there is 
only an apparent saturation and new interesting features should characterise 
$A(B)$ at higher fields ($B>$600 mT), just like it happens for the dielectric 
constant (Section 5). A full description of the effect, however, requires a 
calculation involving all three ATS energy levels.
\begin{table}[!Htbp]
\begin{center}
\begin{tabular}{ |c|c|c|c|c|c|c|c| }
\hline
Glass type & $D_{min}$ & $D_{0min}\left|\frac{q}{e}\right|S_{\Delta }$ & $D_{0max}\left|\frac{q}{e}\right|S_{\Delta }$ & ${\Gamma }^{-1}$ & $c_{ATS}-c_{2LS}$ & ${{\rm p}}_{{\rm 1}}{{\rm F}}_0$ & ${\tan  \omega \Delta \tau }$ \\
 & ${\rm (mK)}$ & $({\rm K}\AA^2)$ & $({\rm K}\AA^2)$ & ${\left(\mu {\rm s}{{\rm K}}^5\right)}^{-1}$ & ${\left(\mu {\rm s} {\rm K}\right)}^{-1}$ & ${\rm D}~{\rm kV}~{{\rm m}}^{-1}$ &  \\
\hline
AlBaSiO & 17.74 & 0.95$\times$10${}^{3}$ & 2.13$\times$10${}^{4}$ & 9.22$\times$10${}^{6}$ & 5.008 & 0.461 & 0.247 \\
(sample 1) & & & & & & & \\
\hline
AlBaSiO & 27.20 & 1.14$\times$10${}^{3}$ & 8.96$\times$10${}^{3}$ & 2.57$\times$10${}^{5}$ & 3.825 & 0.450 & 0.245 \\
(sample 2) & & & & & & & \\
\hline
BK7\newline & 16.76 & 0.92$\times$10${}^{3}$ & 1.34$\times$10${}^{4}$ & 8.91$\times$10${}^{6}$ & 1.03 (*) & 0.60 & 0.207 \\ 
(1.5 $\mu $s) & & & & & & & \\
\hline
BK7\newline & 15.94 & 0.89$\times$10${}^{3}$ & 3.31$\times$10${}^{4}$ & 3.25$\times$10${}^{6}$ & 5.72 (*) & 0.98 & 0.204 \\
(6 $\mu $s) & & & & & & & \\
\hline
\end{tabular}
\caption{Fitting parameters for the echo amplitude's magnetic field dependence. (*) For BK7 (best-fit parameters only), $c_{ATS}$ only is involved.}
 \label{table91}
\end{center}
\end{table}

\subsection{Amorphous glycerol and the so-called isotope effect}
We now come to the astonishing case of amorphous glycerol 
(C${}_{3}$O${}_{3}$H${}_{8}$ vitrifies around $T_g\simeq$47 K), deuterated and 
natural. The first question is whether our model applies to this system, which 
nominally is single-component. We firmly believe it does and therefore that the 
\textit{a}-glycerol polarization echo experiments can also be explained by the 
presence of RERs/microphasing in the samples.

\begin{figure}[!Htbp]
\centering
   \subfigure[]{\includegraphics[scale=0.23] {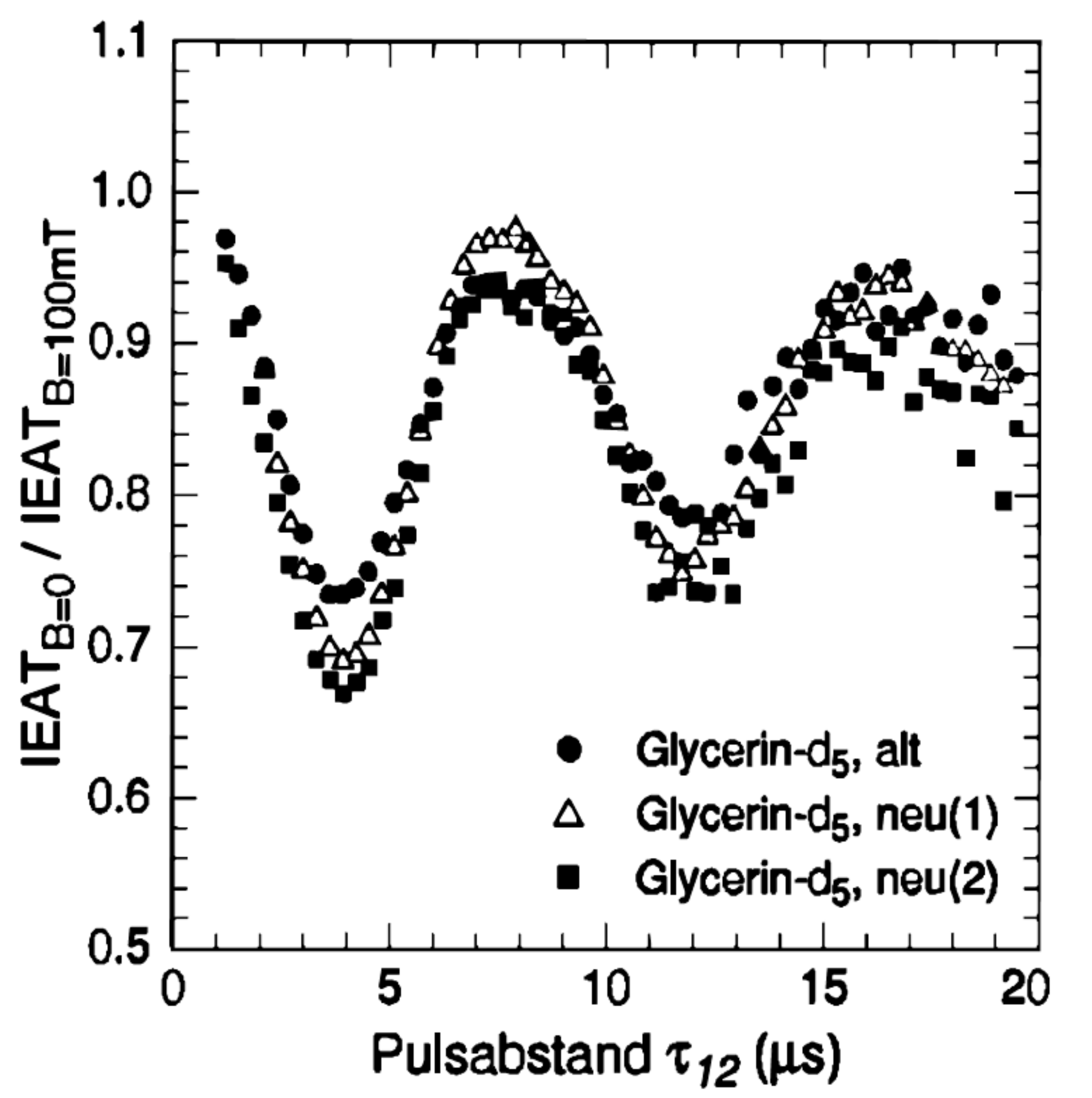}}
    \subfigure[]{\includegraphics[scale=0.23] {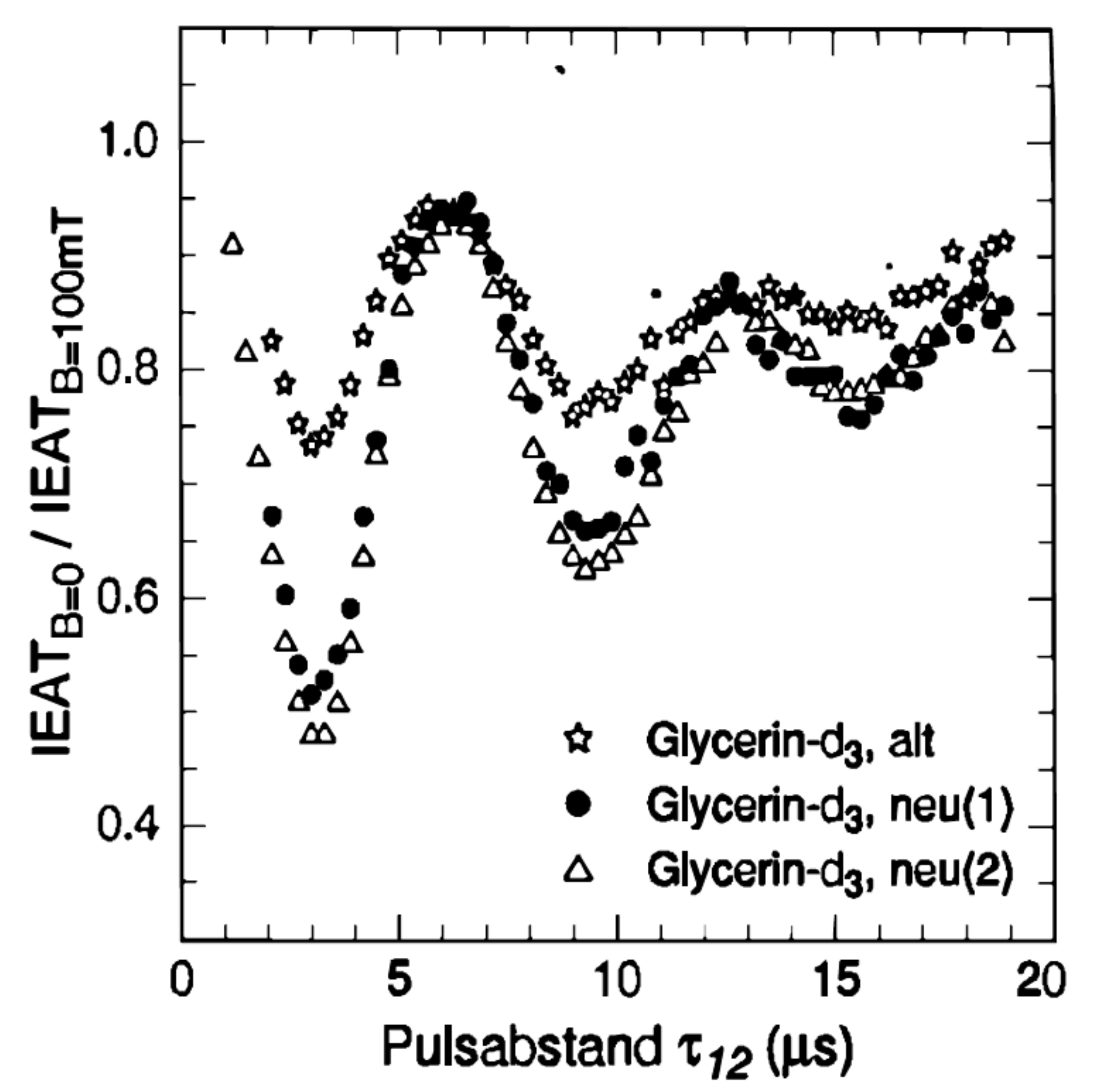}}
\caption{(colour online) Relative integrated echo amplitude vs. $\tau _{12}$ 
for various deuterated amorphous glycerol samples at $B=0$, nominal frequency 
0.85 GHz, $T=$13.5 mK. The samples termed ``old'' (alt) were left liquid 
several days in the air before cooling; two separate fresh preparations 
(neu(1) and neu(2)) were also employed, obtained from producer-sealed 
containers just before cooling. From \cite{Bra2004b}. }
\label{image9e}
\end{figure}

In Brandt's Ph.D. dissertation \cite{Bra2004b} it is reported that the liquid 
glycerol, from which the glass samples were made of, were contaminated by water 
(see Table~\ref{table92}). 
Moreover, experiments on samples with different deuterium molar content and 
different before-cooling open-air shelf-storage times gave definitely different 
results. The available experimental data are reported in Fig.~\ref{image9e}. 
One observes a much greater variation in the experimental data for the 
C${}_{3}$O${}_{3}$D${}_{3}$H${}_{5}$ samples left in the air before freezing 
than in the similiarly prepared C${}_{3}$O${}_{3}$D${}_{5}$H${}_{3}$ samples.
\begin{table}[!Htbp]
\begin{center}
\begin{tabular}{ |c|c|c| }
\hline
Sample & Chemical purity & Water content \\ 
\hline
glycerol-d${}_{0}$ & 99.9\% & {\rm ppm} H${}_2$O \\ \hline
glycerol-d${}_{3}$ & 99\% & = 1.5\% H${}_{2}$O \\ \hline
glycerol-d${}_{5}$ & 98\% & = 0.11\% H${}_{2}$O \\ \hline
glycerol-d${}_{8}$ & 98\% & no information \\ 
\hline
\end{tabular}
\caption{Purity and water-contents data for the studied glycerol samples.}
 \label{table92}
\end{center}
\end{table}

Glycerol is a highly hygroscopic substance, owing to the polarity (similar to 
H${}_{2}$O's) of its molecules which tend to shed one of the three hydrogens 
(or deuteria) attached to the oxygens, but none of those firmly attached to 
the carbon atoms. The equilibrium process is of the type, e.g., 
C${}_{3}$O${}_{3}$D${}_{3}$H${}_{5} \longleftrightarrow$ 
C$_{3}$O$_{3}$D$_{2}$H$_{5}^{-}$+D${}^{+}$ causing similar equilibrium 
concentrations as in the familiar dissociation process 
H$_{2}$O $\longleftrightarrow$ OH$^{-}$+H$^{+}$. Thus it is not unreasonable to 
imagine that the water contained in the air causes some hydrogens to substitute 
for some O-attached deuteria, in time, when absorbed in liquid glycerol-d$_3$, 
- but not in the case of glycerol-d$_5$. The glycerol-d$_3$ hygroscopic effect 
on the echo would support the NEQM theory, but the (albeit smaller) effect in 
glycerol-d$_5$ does not. Moreover, oddly enough the more hygroscopically 
contaminated samples (``alt'') yield the smaller effect on the echo's amplitude 
oscillations, whilst the fresher samples (``neu'' (1) and (2) give wider and 
different (neu(1) $\neq$ neu(2) amplitude oscillations.

The lack of reproducibility of the effect leads us to believe that the origin 
of the observed magnetic phenomena lies again in the formation of 
RERs/microcrystallites during the freezing process, which in glycerol are 
nucleated by the presence of dissociated H$_2$O. The concentration of micro- 
or nano-crystals is therefore very much sample-dependent. 

We remark from further tests \cite{Pal2011} that (Section 9) the number of 
coherently tunneling ions (most probably D$^{+}$ in glycerol-d$_{3}$) may be 
large and/or that the electric-field intensity (unknown for the glycerol 
experiments) may also be huge (thus renormalizing the value of $D_{min}$ way 
upwards).

The lack of reliable experimental dielectric data for amorphous glycerol makes 
the fitting of the polarization-echo data frustrating. All the more so, when 
considering that -- so far -- no evidence for a magnetocapacitance effect in 
a-glycerol has been found (albeit at the nominal frequency of 0.986 GHz) 
\cite{Bra2004b}. This last fact, however, could be explained by the suspected 
large value of $D_{min}$, for the order of magnitude of the magnetocapacitance 
is given, in our theory, by the combination 
${\pi P^*\overline{p^2_1}x_{ATS}}/{(2{\varepsilon }_0{\varepsilon }_rD_{min})}$ 
(Section 5) and for a-glycerol ${\varepsilon }_r=42.5$ and $D_{min}$ are much 
larger than for the multi-silicates. Low-frequency (kHz range) and weak 
electric-field intensity measurements of the dielectric constant of a-glycerol 
would be most useful. These lacking, we must limit our discussion of the 
magnetic dipole echo phenomena in a-glycerol to the qualitative level of 
understanding.

\begin{figure}[!Htbp]
\centering
   \subfigure[]{\includegraphics[scale=0.25] {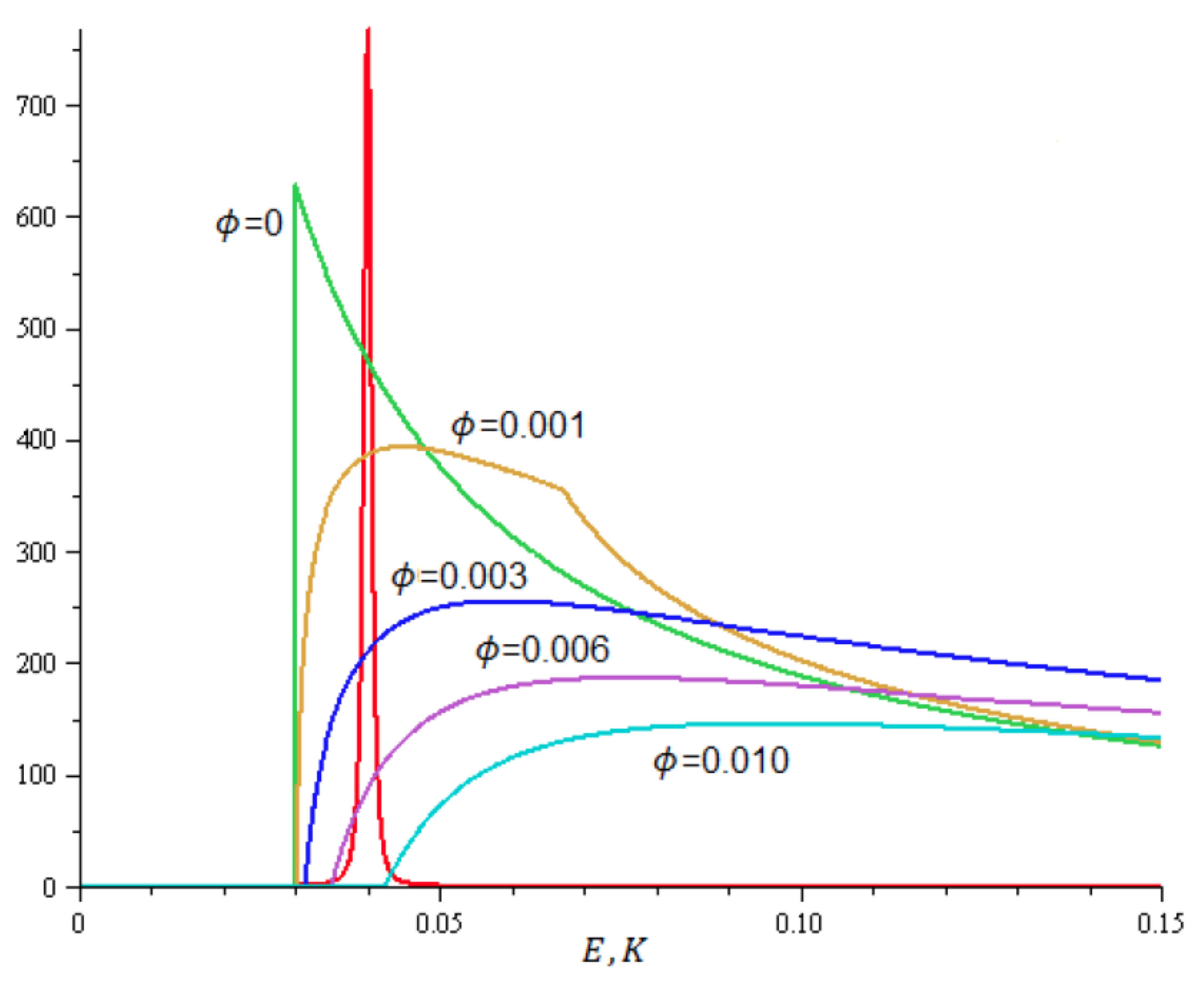}}
   \subfigure[]{\includegraphics[scale=0.25] {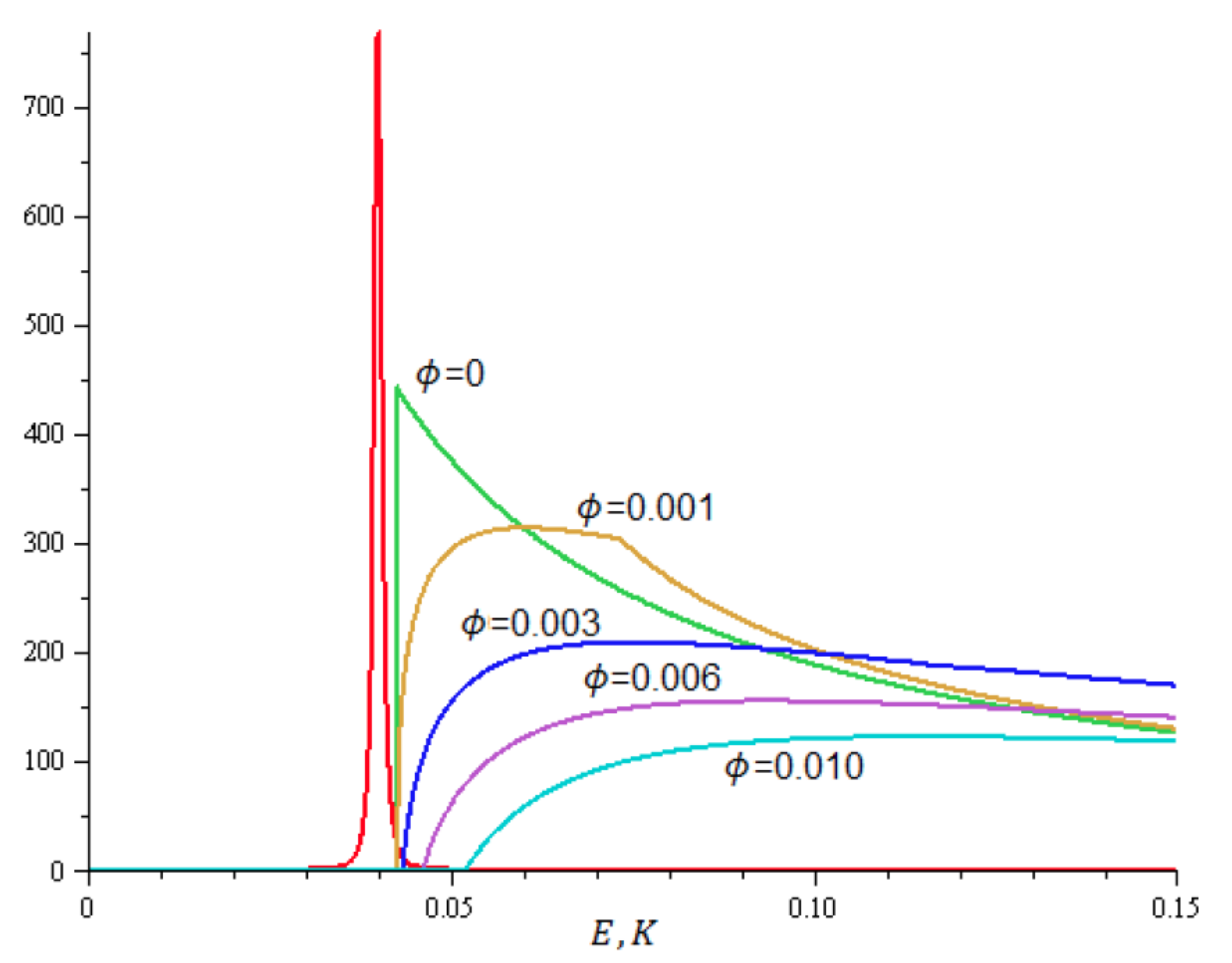}}     
\caption{(colour online) Convolution of the DOS $g_{ATS}(E,\varphi)$ and the 
spectral delimiter $\chi (E)$ in the integral giving the IEA. a) If 
$D_{min}<\hbar \omega $ there is a significant effect on the echo amplitude 
as $\varphi$ increases; b) In the opposite case $D_{min}>\hbar \omega $ the 
effect is much reduced and vanishes in the strict resonant case. This is our 
qualitative explanation for the ``isotope effect'' \cite{JugXXXX}. }
\label{image12-13e}
\end{figure}

The integrated echo amplitude, IEA, is readily obtained from Eq.~(\ref{eq930}), 
after the averaging procedure as in Eq.~(\ref{eq932}) and a subtraction of the 
pump-frequency mode $\omega $ are carried out (rotating reference frame 
description):
\begin{equation}
\begin{split}
\Delta IEA(\varphi)\cong & {-A}_0\frac{d}{{\varepsilon }_0{\varepsilon }_r}x_{ATS}\frac{4\pi {\hbar }^2P^*}{{{\rm F}}_0}\\
&\times \bigg\{\int^{E_{c_2}}_{E_{c_1}}{\frac{dE}{E}}\int^{D_2\left(\varphi\right)}_{D_{min}}\frac{dD}{D}{\tanh  \left(\frac{E}{2k_{B}T}\right)\ }\frac{E^2}{E^2-D^2}{\Omega }^2_R\sigma \left(E\right)\frac{2w}{w^2+{\left({\omega }_0-\omega \right)}^2} \\
&\times \left[S\left({\theta }_1,{\theta }_2\right){\cos  \left(2\left({\omega }_0-\omega \right){\tau }_{12}\right)\ }+C\left({\theta }_1,{\theta }_2\right){\sin  \left(2\left({\omega }_0-\omega \right){\tau }_{12}\right)}\right] \\
& +\int^{\infty }_{E_{c_2}}{\frac{dE}{E}}\int^{D_2\left(\varphi\right)}_{D_1(\varphi)}{\frac{dD}{D}}({\rm same~integrand~as~above}\dots )\bigg\} 
\end{split}
\end{equation}
with $\sigma (E),w,S$ and $C$ as in Eqs.~(\ref{eq934},\ref{eq935}) above. We 
are now in the presence of an even narrower spectral delimiter than 
$\sigma (E)$:
\begin{equation}
\chi \left(E\right)=\frac{w}{\pi \hbar \left(w^2+{\left({\omega }_0-\omega \right)}^2\right)}
\end{equation}
is a (normalized) very sharply-peaked function of $E=\hbar {\omega }_0$ since 
$w\ll {\Omega }_R\ll \omega $. We can then take the limit $p_1F_0\to \infty $ 
that seems to be appropriate for the glycerol experiments and replace 
$\sigma (E)$ by a constant, 
$\sigma (E)\to {\left(2\hbar {\Omega }_R\right)}^{-1}$, to get:
\begin{equation}
\begin{split}
\Delta IEA\left(\varphi\right)&\approx {-A}_0\frac{d}{{\varepsilon }_0{\varepsilon }_r}x_{ATS}\frac{2{\pi }^2{\hbar }^2P^*}{{{\rm F}}_0}\\
&\times \bigg\{\int^{E_{c_2}}_{E_{c_1}}{\frac{dE}{E}}\int^{D_2\left(\varphi\right)}_{D_{min}}\frac{dD}{D}{\tanh  \left(\frac{E}{2k_{B}T}\right)}\frac{E^2}{E^2-D^2}\\
& \cdot {\Omega }_R\chi \left(E\right)S\left({\theta }_1,{\theta }_2\right){\cos  \left(2\left({\omega }_0-\omega \right){\tau }_{12}\right)}\\
&+\int^{\infty }_{E_{c_2}}{\frac{dE}{E}}\int^{D_2\left(\varphi\right)}_{D_1(\varphi)}{\frac{dD}{D}} ({\rm same~integrand~as~above}\dots )\bigg\}
\end{split}
\label{eq939}
\end{equation}
For fixed and large $\tau_{12}$ the above is essentially the energy 
convolution of a function very much like the DOS $g_{ATS}(E)$ times some 
slowly-varying functions of $E$ and the sharply-peaked spectral delimiter 
$\chi \left(E\right)$. The result of this convolution is depicted in 
Fig. \ref{image12-13e} and very much depends on whether 
$D_{min}<\hbar \omega $ or $D_{min}>\hbar \omega $.
In the first case, the evolution of the $g_{ATS}(E)$ with $\varphi$ (or $B$) 
gives rise to a sizable magnetic effect on the echo amplitude (or IEA) even 
in the strict-resonant limit, whilst in the second case, 
$D_{min}>\hbar \omega $, the magnetic effect is drastically reduced and 
completely vanishes in the strict resonant limit, 
$\chi \left(E\right)\to \delta (E-\hbar \omega )$.

This is the explanation for the so-called isotope effect in amorphous glycerol, 
which actually is a mere mass-substitution effect. Indeed, we claim that for 
natural glycerol the situation $D_{min}>\hbar \omega $ applies and that for the 
deuterated samples, as for the multi-silicates, $D_{min}<\hbar \omega $ 
applies instead. Then, $D_{min}$ is roughly given by (Section 3):
\begin{equation}
D_{min}\approx \frac{1}{2}\hbar \sqrt{\frac{k}{M}} \begin{array}{ccc}
 & {\rm moreover:} & D_0\approx  \end{array}
\hbar \sqrt{\frac{k}{M}}\ e^{-\frac{d}{\hbar }\sqrt{2MV_0}}
\label{eq940}
\end{equation}
where $k$ is the bonding constant that depends only on the chemistry and not 
on the isotope's mass, $M$. Thus we immediately see that 
$D_{min}({\rm deuterium})=\ D_{min}({\rm hydrogen}){\mathbf /}\sqrt{2}$ and 
this is sufficient to make the transition from case b) to case a) in Fig. 
\ref{image12-13e} 
when a substantial mass change of the tunneling particles through isotopic 
substitution is made. The case of partial isotopic substitution requires 
separate considerations. A further consequence of mass substitution, as seen 
in Eq.~(\ref{eq940}), is that a larger mass $M$ will make -- the chemistry 
being unchanged -- the parameters $D_{0min,\ max}$ much smaller and thus will 
give rise to a much slower variation of the echo's amplitude with $B$, as is 
indeed verified in the experiments (Fig. \ref{image5e}). The highly 
non-uniform shape of the DOS for the ATS model thus qualitatively (and for 
the multi-silicates also quantitatively) explains all of the experimental 
findings.

\begin{figure}[!Htbp]
\centering
   \subfigure[] {\includegraphics[scale=1.0] {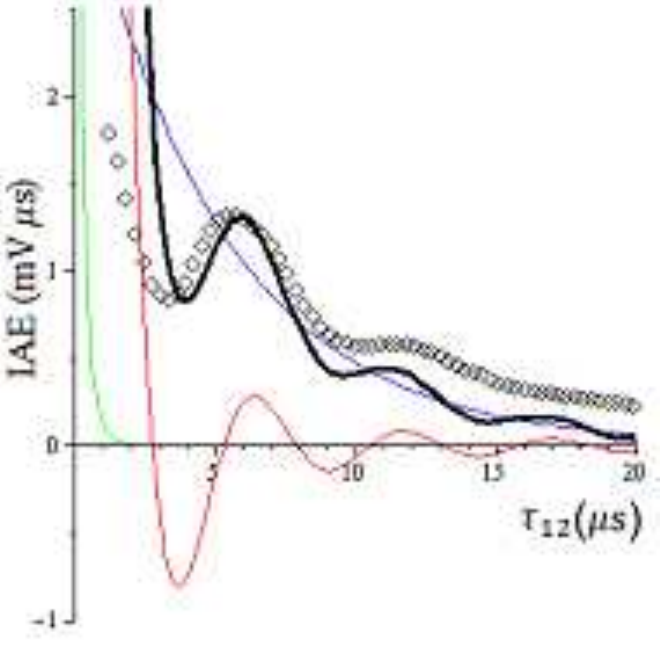}}
   \subfigure[] {\includegraphics[scale=1.0] {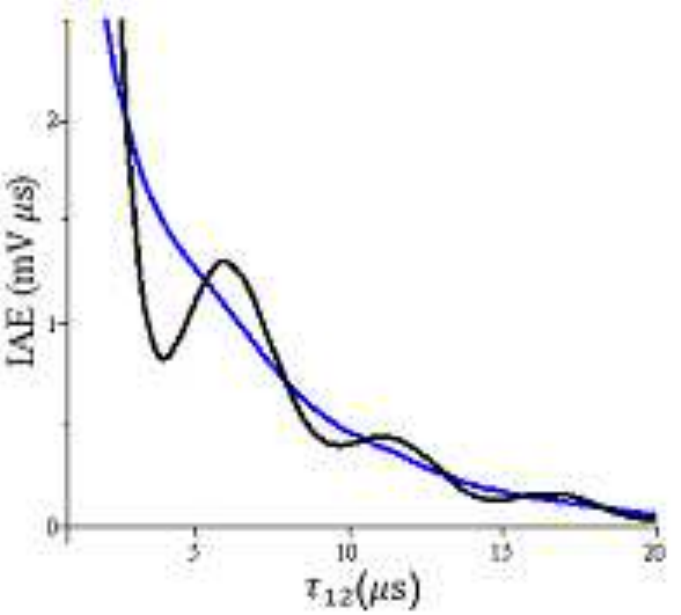}}
\caption[]{(colour online) The IEA from Eq.~(\ref{eq944}) plotted as a function 
of $\tau_{12}$ for parameters as in the text. a) The simplified theory almost 
reproduces (black curve) the experimental data (diamonds) for glycerol-d$_{3}$ at $B=0$, nominal 
frequency $\omega/2\pi$=0.887 GHz and $T=$13.5 mK \cite{Bra2004} (blue 
curve: -- 2LS contribution, green and red curves: -- ATS contributions, black 
curve: -- sum of 2LS and ATS contributions); b) Disappearance of the IAE 
oscillations when $\tau_{\phi(3)}$ decreases or $\Delta\omega $ increases (blue 
curve). From \cite{Pal2011}.}
\label{image14-15e}
\end{figure}

The last topic we discuss is our explanation for the dramatic effect for the 
echo's IEA dependence on the waiting time $\tau_{12}\ $near $B=0$ 
\cite{JugXXXX}. To explain this we take the $B\to 0$ limit of 
Eq.~(\ref{eq939}), when only the second term contributes and we get, 
evaluating the $D$-integral exactly and lumping all slowly-varying functions 
into an overall constant ${\cal A}^{*}$:
\begin{equation}
\begin{split}
{\lim  \Delta IEA\left(\varphi\right)}\approx & {-{\mathcal A}}^*\int^{\infty}_{D_{min}}{\frac{dE}{E}}{\tanh  \left(\frac{E}{2k_{B}T}\right)\ }{\ln  \left({\left(\frac{D_{0max}}{D_{0min}}\right)}^2\frac{E^2-D^2_{0min}\varphi^2}{E^2-D^2_{0max}\varphi^2}\right)}\\
&{\rm \times }\chi \left(E\right){\cos  (2\left({\omega }_0-\omega \right){\tau }_{12})}
\end{split}
\end{equation}
Inspection of this last expression shows, that the ATS contribution decreases
quadratically as the AB phase (or the magnetic field) increases; hence, the 
disappearance of the oscillations in ${\tau }_{12}$ is slow (as experimentally 
reported \cite{Lud2002}) for very weak but increasing $B$. To understand the 
oscillations themselves (and their absence for the silicates) we set $B=0$ and 
redefining the overall constant we arrive at (since $\tau_\phi$ is the shortest 
decay time involved):
\begin{equation}
{\lim  \Delta IEA\left(0\right)\ }\approx {-{\mathcal B}}^*\int^{\infty}_{D_{min}}{\frac{dE}{E}}\frac{{\tau }^{-1}_\varphi}{\pi \hbar {[\tau }^{-2}_\varphi+{({\omega }_0-\omega )}^2]}{\cos  (2\left({\omega }_0-\omega \right){\tau }_{12})}
\label{eq942}
\end{equation}
(${\cal B}^{*}$ being another overall constant). The remaining integral can be 
rewritten as an integral in the interval $[0,\infty ]$, which yields an 
exponentially decaying contribution in $\tau_{12}$, plus the integral in the 
energy interval $[0,\ \hbar \omega -D_{min}]$. The latter is responsible for 
the oscillations in $\tau_{12}$ since:
\begin{equation}
\int^b_0{dx}\frac{{\cos  x}}{x^2+a^2}=\frac{\pi }{2a}e^{-a}+{\rm Im}~\left\{\frac{e^{-a}}{2a}{\rm Ei}\left(-a+ib\right)-\frac{e^a}{2a}{\rm Ei}~\left(a+ib\right)\right\}
\end{equation}
where ${\rm Ei}(z)$ denotes the exponential-integral function. Rearranging 
Eq.~(\ref{eq942}), also adding the 2LS standard exponentially decaying 
contribution, we come to the following functional form for the total IEA of the 
sample:
\begin{equation}
\begin{split}
\lim  IEA\left(0\right) &\approx C_{2LS}e^{-\frac{2{\tau }_{12}}{{\tau }_{\varphi(2)}}}+C_{ATS}\frac{\tau_{\varphi(3)}}{\tau_{12}}e^{-\frac{2{\tau }_{12}}{{\tau }_{\varphi(3)}}}\\
&+K_{ATS}{\rm Im}\left\{\frac{e^{-a}}{2a}{\rm Ei}\left(-a+ib\right)-\frac{e^a}{2a}{\rm Ei}\left(a+ib\right)\right\} \\
\end{split}
\label{eq944}
\end{equation}
where $C_{2LS}$, $C_{ATS}$ and $K_{ATS}$ are appropriate constants and where:
\begin{equation}
a=2{\tau }_{12}{\tau }^{-1}_{\phi\left(3\right)} \begin{array}{ccc}
 &  & b= \end{array}
2{\tau }_{12}\left(\omega -\frac{D_{min}}{\hbar }\right)
\end{equation}
We plot expression~(\ref{eq944}) in Fig. \ref{image14-15e} as a function of 
$\tau_{12}$, assuming some reasonable value of $\tau_{\phi(3)}$=1.0 $\mu$s for 
the ATS spectral-diffusion time, $\tau_{\phi(2)}$=10.0 $\mu$s for the 2LS one, 
and of $\Delta\omega =\omega -\frac{D_{min}}{\hbar }={\rm 0.6\ MHz}$ for the 
frequency offset.
As is shown in Fig.~\ref{image14-15e}(a) (black curve), this (over)simplified 
theoretical treatment almost reproduces the data for glycerol-d$_{3}$ at $B=0$. 
Fig.~\ref{image14-15e}(b) also shows (blue curve) that reducing to 
$\tau_{\phi(3)}$=0.3 $\mu$s causes the oscillations to disappear; 
alternatively, this can be achieved by increasing $\Delta\omega $.

This treatment also explains the findings in Fig.~\ref{image9e}. Leaving the 
sample in the air causes a change in $D_{min}$ simply because more RERs or 
micro-crystals are nucleated (hence a larger $\Delta\omega$) by the absorbed 
water vapour, but also because of mass substitution through dissociation. The 
subtle role of the true value of the resonating frequency $\omega$ (the sample 
sits in a reentrant resonating cavity which is loaded with a reacting 
dielectric and thus develops its own resonance frequency) is as yet not 
completely understood.

Clearly, if $\Delta\omega <0$ (case of glycerol-d$_{0}$) there can only be 
exponential decay (from the 2LS contribution, mostly), but if $\Delta\omega >0$ 
is also too large the oscillations die out. We believe this is the case for the 
multi-silicates.

Qualitatively at least, our ATS model explains all experimental observations 
so far. In terms of the parameters used for the best fits, one cannot fail to 
notice (see Table \ref{table91}) that for the echo experiments the values of 
the cutoffs for the combination of parameters $D_0\frac{q}{e}S_{\triangle}$ are 
about one order of magnitude lower than for the other experiments, $C_p$ and 
$\epsilon$, carried out inside higher temperature ranges, in fact. This can be 
explained by a mechanism whereby the number $N$ of atomic tunneling systems 
within each ATS (hence within the interstices between the RERs or 
micro-crystals) picks up a temperature dependence $N(T)$.

\section{The Magnetic Field Dependent SQUID-Magnetization }
\subsection{Theory}
Having obtained qualitatively (and also quantitatively, barring the various 
approximations applied) good fits to the $C_p$, $\epsilon$ and $A_{\rm echo}$ 
experimental data with the ATS contributions added to the 2LS's (and when 
appropriate to the paramagnetic impurities'), we address the question of why
the values of $D_0\frac{q}{e}S_{\triangle}$ are so high. The first conclusion
is that the ATSs now appear to carry considerably high magnetic moments 
$\mu_{ATS}$ {\it per se}. Estimating from the definition ($T=0$)
$\mu_{ATS}= -\frac{\partial}{\partial B}\big(-\frac{1}{2}E\big)$,
where $E=\sqrt{D^2+D_0^2\varphi^2}$ is the ATS lower energy gap, we get for
not too small fields $B$ ($\mu_{ATS}$ vanishes linearly with $B$ when
$B\to 0$, but saturates at high enough $B$):
\begin{equation}
\mu_{ATS}\simeq\frac{\pi}{\Phi_0}SD_0=\frac{\pi}{\phi_0}\left(
\big\vert\frac{q}{e}\big\vert SD_0 \right).
\label{atsmoment}
\end{equation}
Thus the very same combination $\frac{q}{e} SD_0$ of parameters appears,
whilst $\phi_0\equiv h/e$ is the electronic magnetic flux quantum. Using the
values extracted from the $C_p$ best fit (e.g. Table~\ref{tab_d0_duran_extr})
we deduce from Eq.~(\ref{atsmoment}) that (for Duran) $\mu_{ATS}$ ranges from
about 3.8$\mu_B$ to 27.1$\mu_B$. This fact alone indicates that a large group
of correlated charged atomic particles is involved in each single ATS and that
an important ATS contribution to the sample's magnetization is to be expected
(Fe$^{2+}$ and Fe$^{3+}$ have magnetic moment $\mu_J=2\sqrt{6}\mu_B$ and 
$\sqrt{35}\mu_B$, respectively).

The magnetization~$M$ of a sample containing dilute paramagnetic impurities as
well as dilute magnetic-field sensitive ATSs is, like $C_p$, also given by the
sum of two different contributions:
\begin{enumerate}
\item Langevin's well-known paramagnetic impurities' contribution (Fe$^{2+}$
and Fe$^{3+}$, with $n_J$ concentration of one species having spin $J$), given
by the standard expression~\cite{AM}
\begin{equation}
M_J=n_J g \mu_B J B_J(z), ~~~\bigg(z=\frac{g\mu_B BJ}{k_{B}T}\bigg)
\label{magnet_impur}
\end{equation}
where the Brillouin function~$B_J$ is defined by:
\begin{equation}
B_J(z)=\frac{2J+1}{2J}\coth\bigg(\frac{(2J+1)}{2J}z\bigg)-\frac{1}{2J}\coth\bigg(\frac{1}{2J}z\bigg)
\end{equation}
and its low-field susceptibility is the known Curie law:
\begin{equation}
\frac{M}{B}\cong\frac{n_Jg^2\mu_B^2J(J+1)}{3k_{B}T}
\end{equation}
\item the ATS tunneling currents' contribution, given by the following novel
expression as the sum of contributions from ATSs of lowest gap $E$:
\begin{equation}
\begin{split}
M_{ATS}=\pi~P^{\ast} n_{ATS} \frac{1}{B} \bigg\lbrace &\int_{E_{c1}}^{E_{c2}} \mathrm{d}E \tanh\bigg(\frac{E}{2k_{B}T}\bigg)\ln \bigg(\frac{E^2-D_{0min}^2\varphi^2}{D_{min}^2}\bigg) \\
+&\int_{E_{c2}}^{\infty} \mathrm{d}E  \tanh\bigg(\frac{E}{2k_{B}T}\bigg)\ln \bigg(\frac{E^2-D_{0min}^2\varphi^2}{E^2-D_{0max}^2\varphi^2}\bigg) \bigg\rbrace
\end{split}
\end{equation}
and which can be also re-expressed (like in the case of $C_{ATS}$) using
$y=\frac{E}{2k_{B}T}$ in the following form:
\begin{equation}
\begin{split}
M_{ATS}=2\pi~P^{\ast} n_{ATS} k_B T \frac{1}{B} \bigg\lbrace &\int_{x_{c1}}^{x_{c2}} \mathrm{d}y \tanh y \ln \bigg(\frac{y^2-x_{0min}^2\varphi^2}{x_{min}^2}\bigg) \\
+&\int_{x_{c2}}^{\infty} \mathrm{d}y  \tanh y \ln \bigg(\frac{y^2-x_{0min}^2\varphi^2}{y^2-x_{0max}^2\varphi^2}\bigg) \bigg\rbrace
\end{split}
\label{magnet_adim}
\end{equation}
with, as before:
\begin{itemize}
\item $E_{c1}=\sqrt{D_{min}^2+D_{0min}^2 \varphi^2}$ and $E_{c2}=\sqrt{D_{min}^2+D_{0max}^2 \varphi^2}$;
\item $x_{c1,2}=\frac{E_{c1,2}}{2k_{B}T}$, $x_{min}=\frac{D_{min}}{2k_{B}T}$, etc.;
\end{itemize}
\end{enumerate}
We present this expression here for the first time, also motivated by the fact
that we expect a contribution to the measured magnetization $M$ from the ATSs
that is comparable to, or even greater than, Langevin's paramagnetism of the
diluted Fe impurities. The above expression follows from a straightforward
application of standard quantum statistical mechanics, with
$${\bf M}_{ATS}=n_{ATS}\langle -\frac{\partial {\cal H}_{3LS}}{\partial {\bf B}}
\rangle,$$
$n_{ATS}$ being the ATSs' concentration (a parameter always lumped together
with $P^{\ast}$) and with ${\cal H}_{3LS}$ given by Eq.~(\ref{3lsmagtunneling}).
The angular brackets $\langle \cdots \rangle$ denote quantum, statistical and
disorder averaging.

The above formula for $M_{ATS}$ is in fact correct for weak magnetic fields.
For higher fields the correction implied by Eq. (\ref{correction}) has to be
implemented and it corresponds to an improved analytic expression for the
lowest ATS gap:
\begin{equation}
E=\sqrt{D^2+D_0^2\varphi^2(1-\frac{1}{27}\varphi^2)}
\label{3lsgapimproved}
\end{equation}
which needs to be used for high fields \cite{Pal2011}. In practice, as seen
in Eq. (\ref{correction}), this corresponds -- where appropriate -- to the
replacement of $B$ with $B\sqrt{1-\frac{1}{45}(B/B^{\ast})^2}$. 
A full derivation and the study of the $B$- and $T$-dependence of $M_{ATS}$
will be presented elsewhere. Here, it suffices to say that temperature-dependent
fitting parameters are necessary to achieve good fits for $M(T)$ in such wide
temperature range (4$<T<$300 K). The underlying physical reason is that as 
temperature drops the RERs of the cellular structure of the glass (Section 2)
begin to fuse together and adsorb atoms/ions from the interstices between 
them, growing at their expense. As a result the number $N(T)$ of atomic 
tunnelers in each interstitial ATS decreases with decresing temperature: 
$T_g>T\to 0$ and a reasonable temperature-dependence is of the Arrhenius type:
\begin{equation}
N(T)=N_0\exp\left\{ -\frac{E_0}{k_{B}T} \right\},
\label{arrhenius}
\end{equation}
where $E_0$ is a suitable activation energy. Then, advocating coherent 
tunneling of the group of $N(T)$ atomic tunnelers making up each interstitial 
ATS \cite{Jug2013} we must use the temperature dependent parameters: 
\begin{eqnarray}
D_{min}&=&D_{min}^{(0)}\exp\left\{ -\frac{E_0}{k_{B}T}+\frac{E_0}{k_{B}T_0} 
\right\} \cr
D_{0min}\frac{q}{e}S_{\triangle}&=&\left[ D_{0min}\frac{q}{e}S_{\triangle}
\right]^{(0)} \exp\left\{ -\frac{3E_0}{k_{B}T}+\frac{3E_0}{k_{B}T_0} \right\} \cr
D_{0max}\frac{q}{e}S_{\triangle}&=&\left[ D_{0max}\frac{q}{e}S_{\triangle}
\right]^{(0)} \exp\left\{ -\frac{3E_0}{k_{B}T}+\frac{3E_0}{k_{B}T_0} \right\} 
\label{renormaliz}
\end{eqnarray}
and a similar one for $B^{\ast}$, containing two parameters proportional to
$N(T)$ in Eq. (\ref{arrhenius}). Here, $T_0$ is a temperature corresponding to
the parameters' combinations marked with a $(0)$-superscript and the values 
we extract from the best fits below correspond to parameters at that 
temperature (typically, $T_0$ is the average temperature of our $C_p$ fits in 
Section 4). 
\subsection{Comparison with available data}
\begin{figure}[!Htbp]
  \centering
  \includegraphics[scale=0.50] {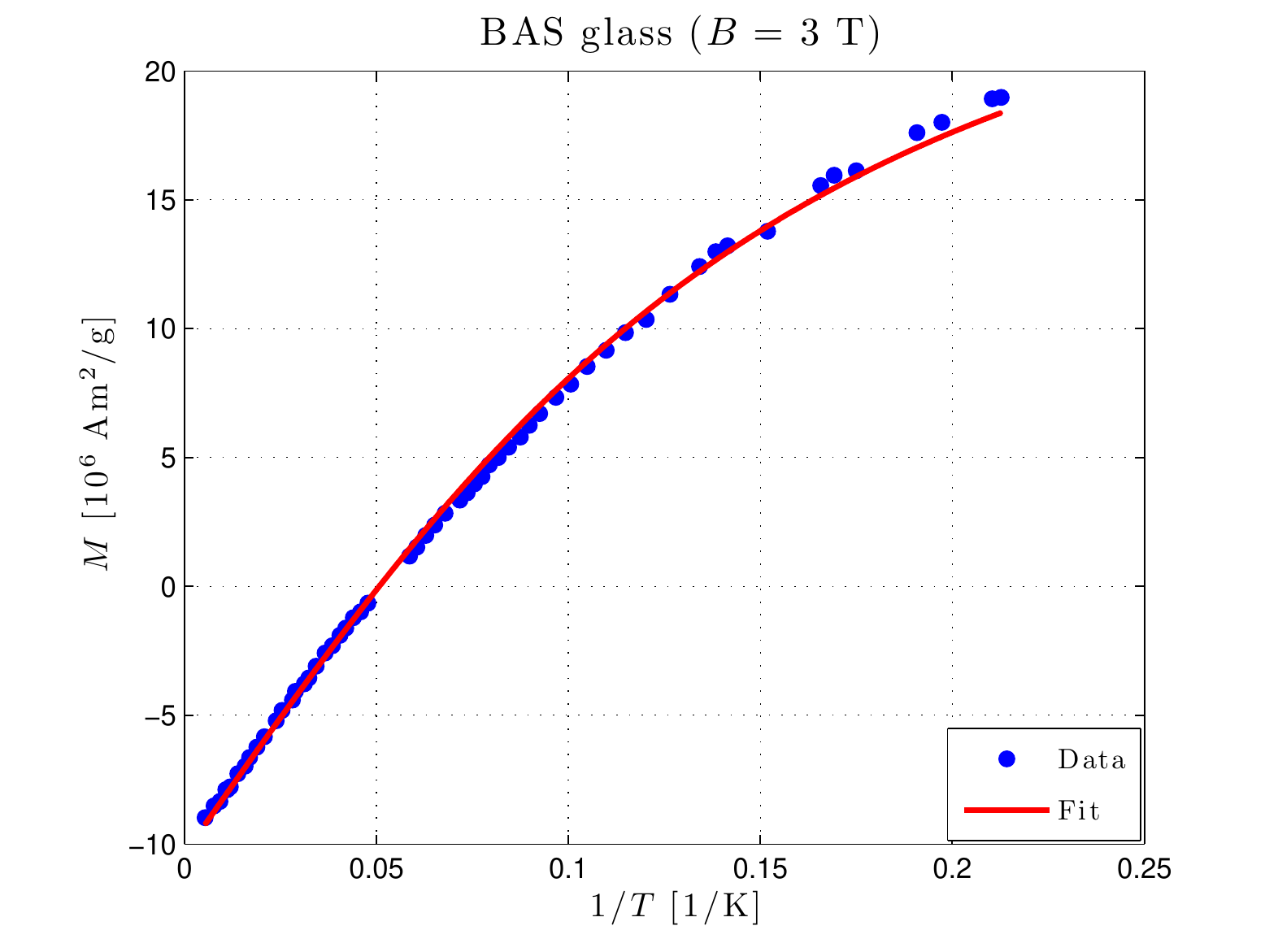}
\caption{(colour online) The best fit of the magnetization data~\cite{Sie2001} for the BAS glass, using Eq.~(\ref{magnet_impur}) (for the Fe$^{2+}$ and Fe$^{3+}$ impurities) and Eq.~(\ref{magnet_adim}) (for the ATSs). }
\label{magnetizat_albasi}
\end{figure}

The magnetization data~\cite{Sie2001} were best-fitted with
Eq.~(\ref{magnet_impur}) (for the Fe$^{2+}$ and Fe$^{3+}$ contributions) as
well as with Eq.~(\ref{magnet_adim}) (for the ATSs'), using the parameters
from the $C_p$-fits as input.
The best fit for the BAS glass is reported in Fig.~\ref{magnetizat_albasi} and
the extracted parameters in Table~\ref{tabular_all_albasi_magn}.

\begin{figure}[!Htbp]
  \centering
  \includegraphics[scale=0.50] {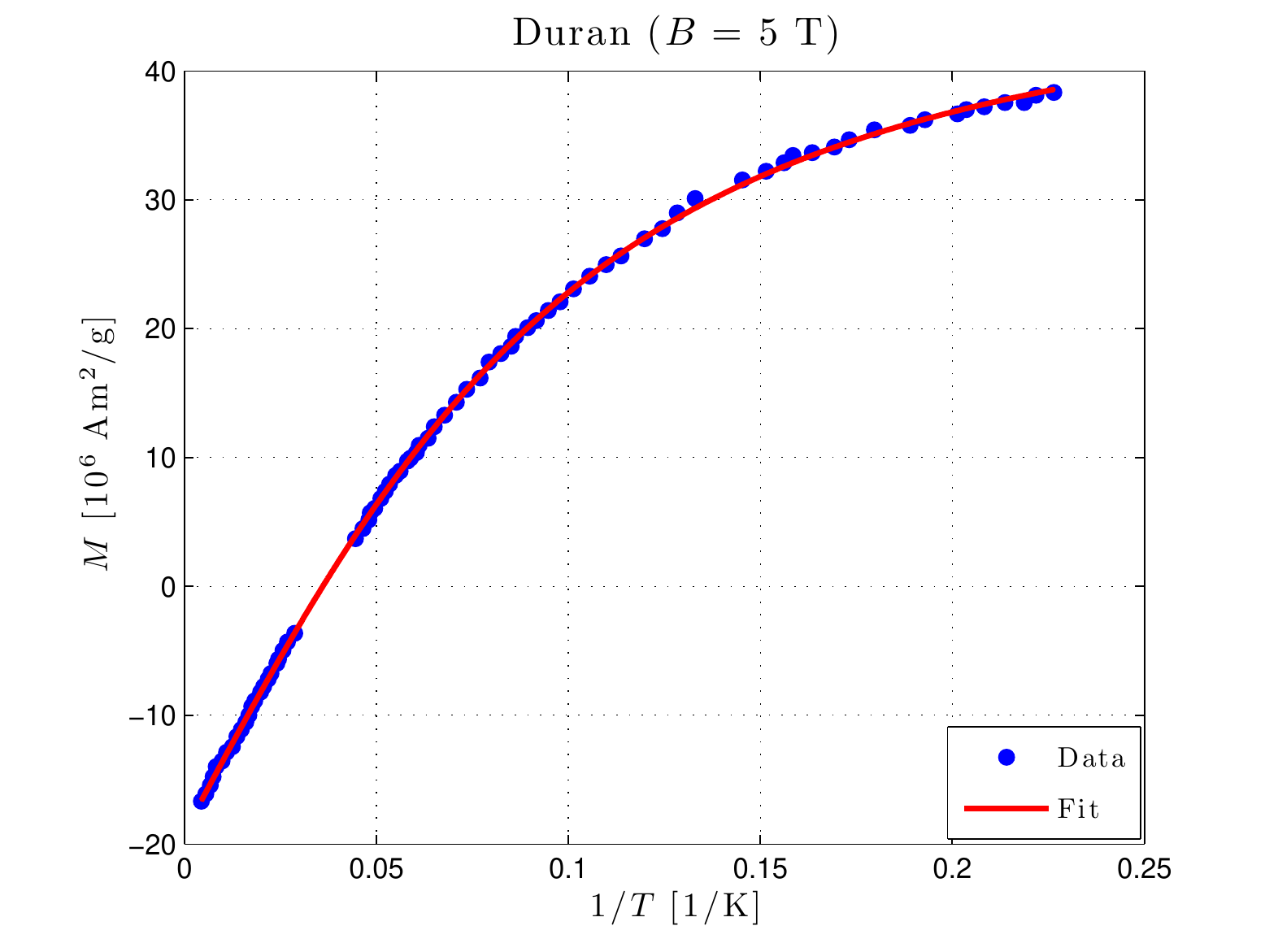}
\caption{(colour online) The best fit of the magnetization data~\cite{Sie2001} 
of Duran, using Eq.~(\ref{magnet_impur}) (for the Fe$^{2+}$ and Fe$^{3+}$ 
impurities) and Eq.~(\ref{magnet_adim}) (for the ATSs).}
  \label{magnetizat_duran}
\end{figure}

\begin{table}[!Htbp]
\begin{center}
\begin{tabular}{|c|c|}
\hline
Parameter & BAS glass \\
\hline
\hline
\textbf{$n_{Fe^{2+}}$} $\mathrm{[g^{-1}]}$ & 1.08$\times10^{17}$  \\
\textbf{$n_{Fe^{3+}}$} $\mathrm{[g^{-1}]}$ & 5.01$\times10^{16}$ \\
\textbf{$P^{\ast}n_{ATS}$} $\mathrm{[g^{-1}]}$ & 5.74$\times10^{16}$ \\
$D_{min}$ [K]  & 8.01$\times10^{-2}$\\
$D_{0min}\vert\frac{q}{e}\vert S$ [K$\AA^2$]  & 1.31$\times10^{5}$\\
$D_{0max}\vert\frac{q}{e}\vert S$ [K$\AA^2$] & 2.44$\times10^{5}$\\
$B^{\ast}$ [T] & 1.02 \\
$E_0$ [K] & 0.42 \\
$\mathrm{vert. offset}$ [Am$^2$g$^{-1}$] & -1.04$\times10^{-5}$\\
\hline
\end{tabular}
\caption{Extracted parameters (from the magnetization data of \cite{Sie2001})
for the concentration of ATS and Fe-impurities of the BAS glass. The vertical
offset represents the residual Larmor diamagnetic contribution.}
\label{tabular_all_albasi_magn}
\end{center}
\end{table}
The best fit for Duran is reported in Fig.~\ref{magnetizat_duran} and the
extracted parameters in Table~\ref{tabular_all_duran_magn}.
\begin{table}[!Htbp]
\begin{center}
\begin{tabular}{|c|c|}
\hline
Parameter & Duran \\
\hline
\hline
\textbf{$n_{Fe^{2+}}$} $\mathrm{[g^{-1}]}$ & 3.07$\times10^{17}$  \\
\textbf{$n_{Fe^{3+}}$} $\mathrm{[g^{-1}]}$ & 2.13$\times10^{17}$ \\
\textbf{$P^{\ast}n_{ATS}$} $\mathrm{[g^{-1}]}$ & 8.68$\times10^{16}$ \\
$D_{min}$ [K]  & 5.35$\times10^{-2}$\\
$D_{0min}\vert\frac{q}{e}\vert S$ [K$\AA^2$]  & 2.00 $\times10^{5}$\\
$D_{0max}\vert\frac{q}{e}\vert S$ [K$\AA^2$] & 2.81$\times10^{5}$\\
$B^{\ast}$ [T] & 1.63 \\
$E_0$ [K] & 0.34 \\
$\mathrm{vert. offset}$ [Am$^2$g$^{-1}$] & -1.97$\times10^{-5}$\\
\hline
\end{tabular}
\caption{Extracted parameters (from the magnetization data of \cite{Sie2001})
for the concentration of ATS and Fe impurities of Duran.}
\label{tabular_all_duran_magn}
\end{center}
\end{table}
We recall how we convert the Fe-concentrations thus obtained to atomic ppm 
concentrations (ppma), using the formula \cite{Bon2015b}:
\begin{equation}
n_J=\bar{n}_J\frac{N_A}{\sum_i \xi_i A_i}
\label{conversion}
\end{equation}
where $n_J=N_J/M$ is the mass density and $\bar{n}_J=N_J/N_{at}$ the atomic
concentration of the spin-$J$ Fe species (to be multiplid by 100 to get the
ppm) and where: $N_J$ is the number of Fe-ions in the sample with spin $J$ 
$\xi_i$ the molar fraction of the i-$th$ species, $A_i$ the molar mass of 
the i-$th$ species, $N_{at}$ the total number of atoms in the sample and
$N_A$ Avogadro's number (6.022$\times$10$^{23}$~mol$^{-1}$) for the Fe$^{2+}$ 
($J$=2) and Fe$^{3+}$ ($J$=5/2) impurities.

The nominal concentration of Fe$^{3+}$ for the BAS glass is (using 
Eq.~(\ref{conversion})):
\begin{equation}
\begin{split}
\bar{n}_{Fe^{3+}_{nom}}&=102~\mathrm{ppm} \\
n_{Fe^{3+}_{nom}}&=\frac{10^{-6}\cdot 102 \cdot 6.022\times10^{23}~\mathrm{mol}^{-1}}{80.530 \mathrm{\frac{g}{mol}}}=7.63\cdot 10^{17} \mathrm{g^{-1}}\\
\end{split}
\end{equation}
which is inadequate (as in Duran's case below) to explain the behaviour of the
heat capacity as a function of $B$, here presented (Fig.~\ref{heat_capa_fit})
and as a function of $T$ (studied in~\cite{Jug2004}).
Table~\ref{tab_imp_extr_alba_tot} summarizes the concentrations found from
our best fits of heat capacity and magnetization data, for the BAS glass.
The parameters found in~\cite{Jug2004} were
$P^{\ast}n_{ATS}$=6.39$\times10^{16}$~g$^{-1}$ and $\bar{n}_{Fe}$=20.44~ppm
where this latter was for the Fe$^{2+}$ concentration only.
The present study confirms that most of the Fe-impurities in these two glasses
are of the Fe$^{2+}$ type~\cite{Jug2004}.
\begin{table}[!Htbp]
\begin{center}
\begin{tabular}{ |l|l| }
  \hline
  \multicolumn{2}{|c|}{{BAS glass}} \\
  \hline
  \multicolumn{2}{|c|}{Heat Capacity fit} \\
  \hline
  \textbf{$n_{Fe^{2+}}$} & 1.06$\times10^{17}~\mathrm{g^{-1}}$ = 14.23~ppm \\
  \textbf{$n_{Fe^{3+}}$} & 5.00$\times10^{16}~\mathrm{g^{-1}}$ =  6.69~ppm \\
  \textbf{$P^{\ast}n_{ATS}$} & 5.19$\times10^{16}~\mathrm{g^{-1}}$  \\
  \hline
  \hline
  \multicolumn{2}{|c|}{Magnetization fit} \\
  \hline
  \textbf{$n_{Fe^{2+}}$} & 1.08$\times10^{17}~\mathrm{g^{-1}}$ = 14.38~ppm  \\
  \textbf{$n_{Fe^{3+}}$} & 5.01$\times10^{16}~\mathrm{g^{-1}}$ = 6.70~ppm\\
  \textbf{$P^{\ast}n_{ATS}$} & 5.74$\times10^{16}~\mathrm{g^{-1}}$ \\
   \hline
\end{tabular}
\caption{Comparison between the concentrations extracted from the two different
best-fitted experimental data sets for the BAS glass.}
 \label{tab_imp_extr_alba_tot}
\end{center}
\end{table}

\begin{figure}[!Htp]
  \centering
  \includegraphics[scale=0.50] {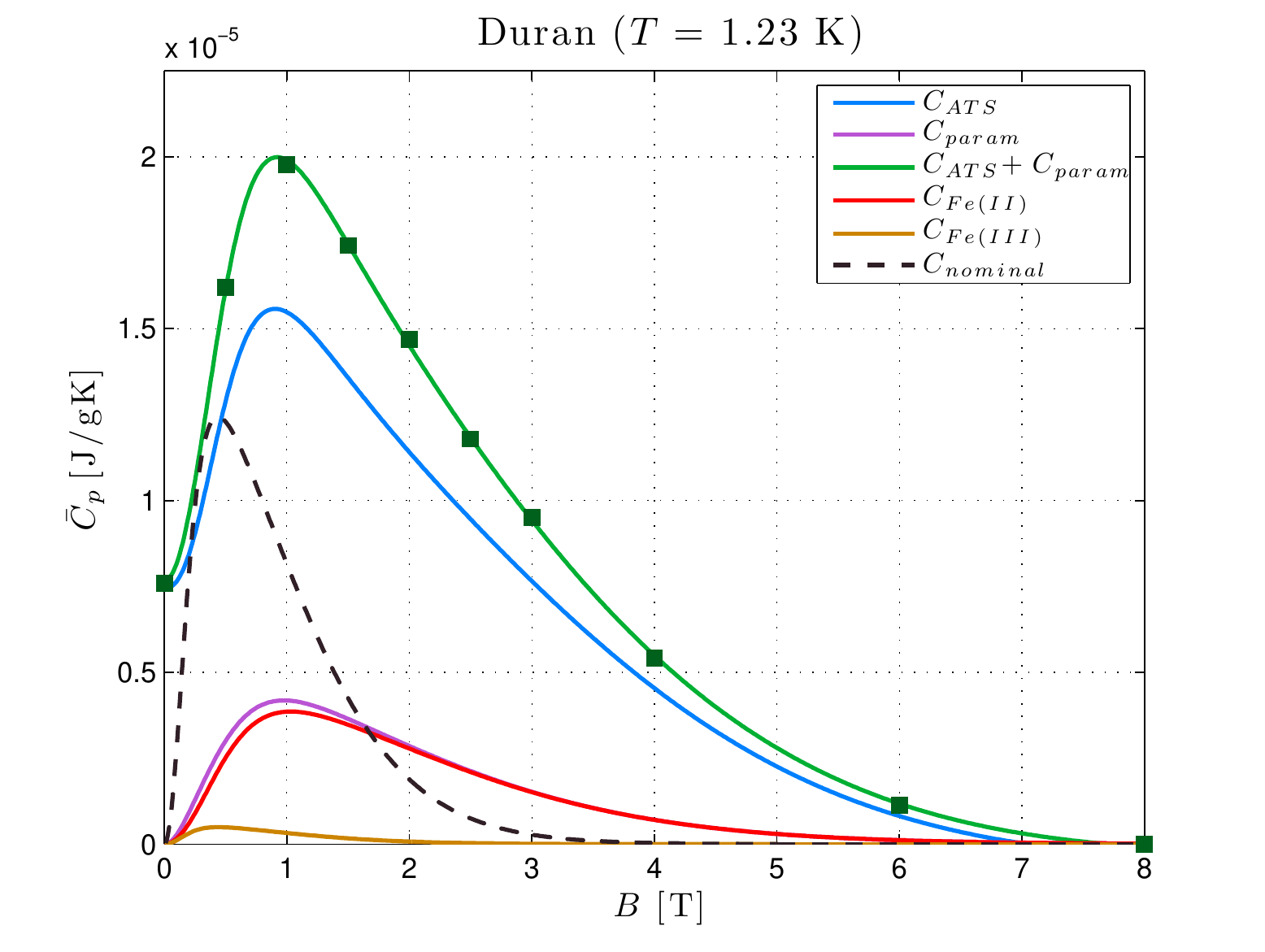}
\caption{(colour online) The curves represent the different terms that 
contribute to the heat
capacity of Duran in our best fit of the data from \cite{Sie2001}. The dashed
curve is for Langevin's contribution only, but with the nominal concentration
of $\bar{n}_{Fe^{3+}}$=126 ppm (no ATS). }
  \label{spec_heat_contri_duran}
\end{figure}

The nominal concentration of Fe$^{3+}$ for Duran is (using 
Eq.~(\ref{conversion})):
\begin{equation}
\begin{split}
\bar{n}_{Fe^{3+}_{nom}}&=126~\mathrm{ppm} \\
n_{Fe^{3+}_{nom}}&=\frac{10^{-6}\cdot 126 \cdot 6.022\times10^{23}~\mathrm{mol}^{-1}}{61.873 \mathrm{\frac{g}{mol}}}=1.23\cdot 10^{18} \mathrm{g^{-1}}\\
\end{split}
\end{equation}
which again is inadequate to explain the behaviour of the heat capacity as a
function of $B$ (see Fig.~\ref{spec_heat_contri_duran}).
Table~\ref{tab_imp_extr_duran_tot} summarizes the concentrations found from
our best fits of heat capacity and magnetization data, for Duran. The
parameters found in~\cite{Jug2004} were
$P^{\ast}n_{ATS}$=6.92$\times10^{16}$~g$^{-1}$ and $\bar{n}_{Fe}$=47.62~ppm
where this latter was the Fe$^{2+}$ concentration only.

\begin{figure}[!Htp]
  \centering
  \includegraphics[scale=0.50] {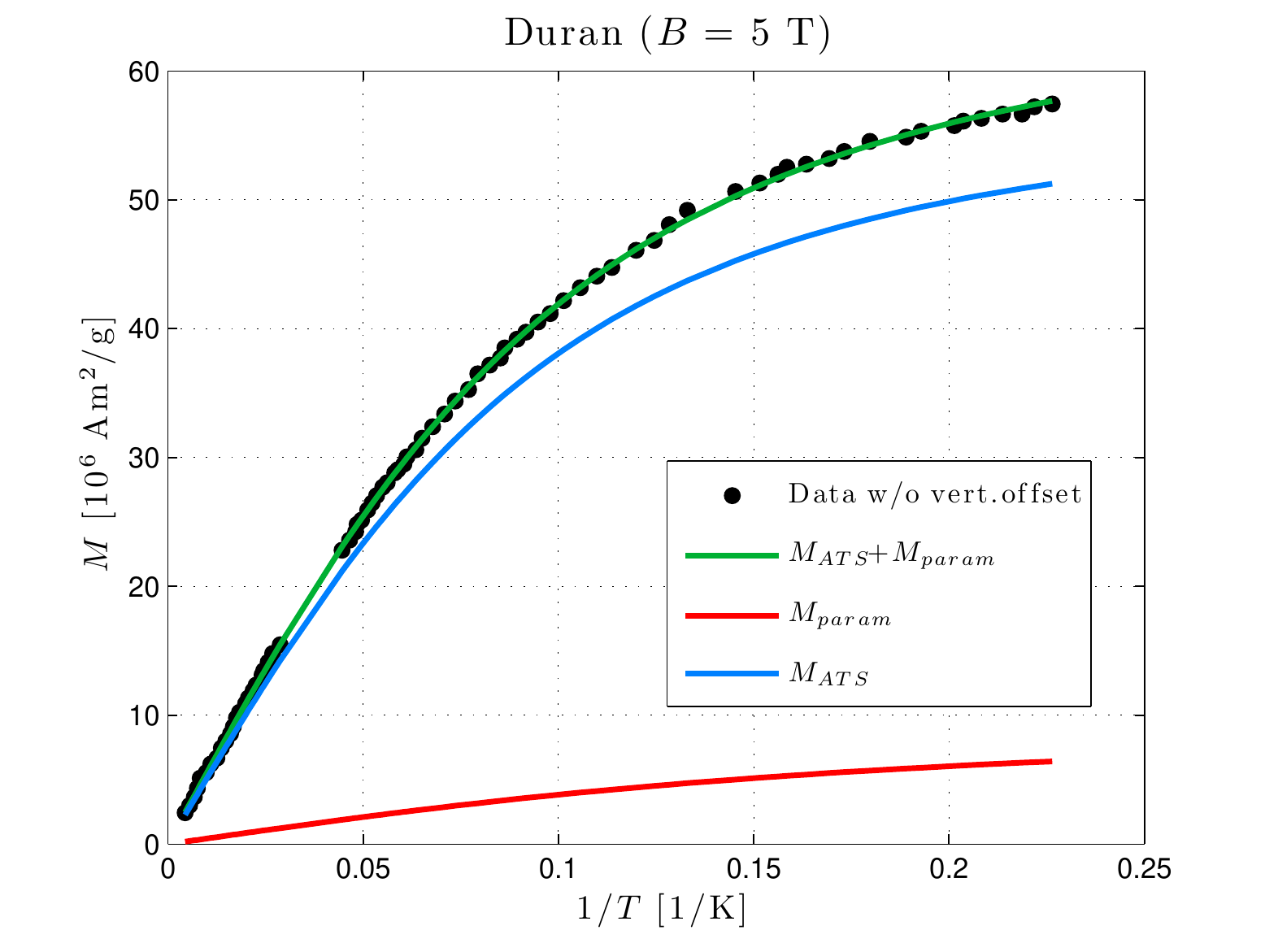}
\caption{(colour online) The curves represent the different contributions to the magnetization
of Duran in our best fit of the data from \cite{Sie2001}.}
  \label{magnetiz_contri_duran}
\end{figure}

\begin{figure}[!Hbp]
  \centering
  \includegraphics[scale=0.50] {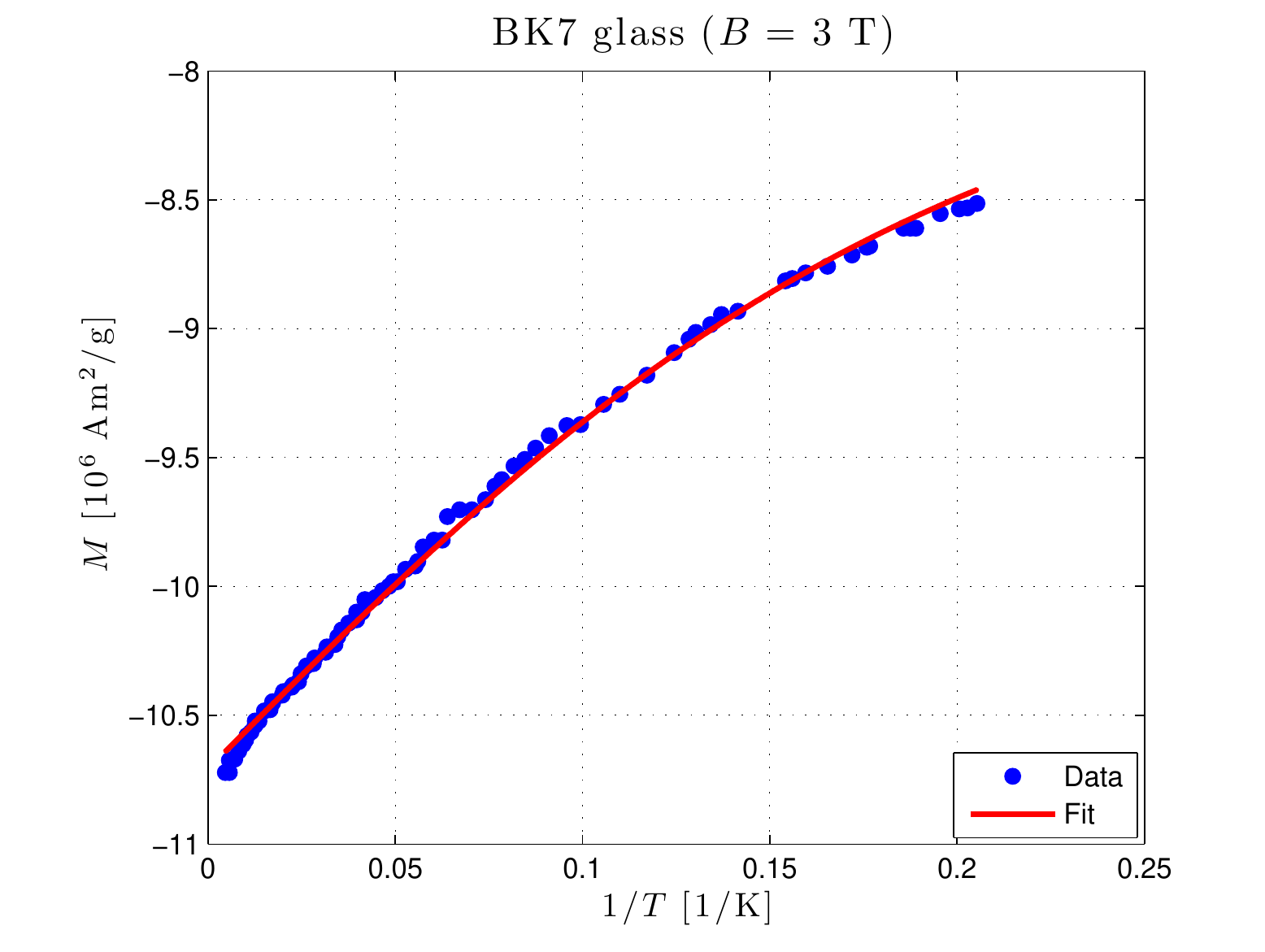}
\caption{(colour online) The best fit of the magnetization data~\cite{Sie2001} 
for BK7, using
Eq.~(\ref{magnet_impur}) (for the Fe$^{2+}$ and Fe$^{3+}$ impurities) and
Eq.~(\ref{magnet_adim}) (for the ATSs). Data from \cite{Sie2001}.}
  \label{magnetizat_BK7}
\end{figure}

Fig.~\ref{spec_heat_contri_duran} presents the behaviour of the different
contributions to the heat capacity as a function of $B$ for Duran;
$C_{Fe^{2+}}$ and $C_{Fe^{3+}}$ are given by Eq.~(\ref{param_impu_formula_cp}),
respectively with the Fe$^{2+}$ and Fe$^{3+}$ parameters, $C_{param}$ is the
sum of these latter two contributions, C$_{ATS}$ is given by
Eq.~(\ref{heat_capa_tot_formula}) and the green line represents the result of
the best fit. The dashed line corresponds to the $\bar{C}_p(B)$ one would get
from the nominal concentration $\bar{n}_{Fe}$ of 126 ppm~\cite{Sie2001} as
extracted from the SQUID magnetization measurements fitted with the Langevin
contribution only (no ATS contribution). Likewise
Fig.~\ref{magnetiz_contri_duran} presents the behaviour of the different
contributions (Eq.(\ref{magnet_impur}) and Eq.(\ref{magnet_adim})) to the
magnetization as a function of $B$, also for Duran. It can be seen that the ATS
contribution is in both cases dominant, also (in the case of the magnetization)
at the higher temperatures.
\begin{table}[!Htbp]
\begin{center}
\begin{tabular}{ |l|l| }
  \hline
  \multicolumn{2}{|c|}{{Duran}} \\
  \hline
  \multicolumn{2}{|c|}{Heat Capacity fit} \\
  \hline
  \textbf{$n_{Fe^{2+}}$} & 3.21$\times10^{17}~\mathrm{g^{-1}}$ = 33.01~ppm \\
  \textbf{$n_{Fe^{3+}}$} & 2.11$\times10^{17}~\mathrm{g^{-1}}$ =  21.63~ppm \\
  \textbf{$P^{\ast}n_{ATS}$} & 8.88$\times10^{16}~\mathrm{g^{-1}}$  \\
  \hline
  \hline
  \multicolumn{2}{|c|}{Magnetization fit} \\
  \hline
  \textbf{$n_{Fe^{2+}}$} & 3.07$\times10^{17}~\mathrm{g^{-1}}$ = 31.58~ppm  \\
  \textbf{$n_{Fe^{3+}}$} & 2.13$\times10^{17}~\mathrm{g^{-1}}$ = 21.86~ppm\\
  \textbf{$P^{\ast}n_{ATS}$} & 8.68$\times10^{16}~\mathrm{g^{-1}}$ \\
   \hline
\end{tabular}
\caption{Comparison between the concentrations extracted from the two different
best-fitted experimental data sets for Duran.}
 \label{tab_imp_extr_duran_tot}
\end{center}
\end{table}
We finally present our preliminary study of the SQUID magnetization data (also
available from \cite{Sie2001}) for the borosilicate glass BK7, for which
however no substantial magnetic effect in the heat capacity $C_p$ has been
reported~\cite{Sie2001}. This glass has a nominal Fe-impurity concentration
of $\bar{n}_{Fe^{3+}}$=6 ppm~\cite{Woh2001,Sie2001,Lud2003}, yet our best fit
in Fig.~\ref{magnetizat_BK7} with both Langevin (Eq.~(\ref{magnet_impur})) and
ATS (Eq.~(\ref{magnet_adim})) contributions produces the concentrations and
parameters given in Table~\ref{tabular_all_BK7_magn}. The best fit was carried
out with knowledge of ATS parameters from our own theory~\cite{Jug2014} for
the magnetic effect in the polarization-echo experiments at mK
temperatures~\cite{Lud2003}. We conclude that our main contention is once
more confirmed, in that the concentration of Fe in BK7 we extract in this way
is only about 1.1 ppm and the bulk of the SQUID magnetization is due to the
ATSs. Table~\ref{tabular_all_BK7_magn} reports our very first estimate of
$n_{ATS}P^{\ast}$ for BK7. Assuming $P^{\ast}$ to be of order 1 and about
the same for all glasses, we conclude that the concentration $n_{ATS}$ of the
ATSs nesting in the RERs is very similar for all of the multi-silicate glasses
by us studied for their remarkable magnetic effects. From the present
SQUID-magnetization best fits we have obtained 5.74$\times10^{16}$ g$^{-1}$
(BAS glass), 8.68$\times10^{16}$ g$^{-1}$ (Duran) and 1.40$\times10^{16}$
g$^{-1}$ (BK7). The almost negligible magnetic effect in $C_p$ for BK7 is
due, in our approach, to the low values of the cutoffs $D_{0min}$ and
$D_{0max}$ for this system (these parameters appearing in the prefactor and
in the integrals' bounds determining the ATS contribution to
$C_p$ ~\cite{Jug2013}).
\begin{table}[!Htbp]
\begin{center}
\begin{tabular}{|c|c|}
\hline
Parameter & BK7 \\
\hline
\hline
\textbf{$n_{Fe^{2+}}$} $\mathrm{[g^{-1}]}$ & 6.69$\times10^{15}$ = 0,71 ppm  \\
\textbf{$n_{Fe^{3+}}$} $\mathrm{[g^{-1}]}$ & 3.43$\times10^{15}$ = 0.36 ppm\\
\textbf{$P^{\ast}n_{ATS}$} $\mathrm{[g^{-1}]}$ & 1.40$\times10^{16}$ \\
$D_{min}$ [K]  & 5.99$\times10^{-2}$\\
$D_{0min}\vert\frac{q}{e}\vert S$ [K$\AA^2$]  & 8.87$\times10^{4}$\\
$D_{0max}\vert\frac{q}{e}\vert S$ [K$\AA^2$] & 1.20$\times10^{5}$\\
$B^{\ast}$ [T] & 1.30 \\
$E_0$ [K] & 0.48 \\
$\mathrm{vert. offset}$ [Am$^2$g$^{-1}$] & -1.08$\times10^{-5}$\\
\hline
\end{tabular}
\caption{Extracted parameters (from the magnetization data of \cite{Sie2001})
for the concentration of ATSs and Fe impurities of the BK7 ($\sum_i\xi_iA_i$=
63.530 g~mol$^{-1}$~\cite{Sie2001}). The vertical offset represents the
residual Larmor diamagnetic contribution.}
\label{tabular_all_BK7_magn}
\end{center}
\end{table}
We find values of the combinations of parameters $D_{0min}\vert\frac{q}{e}S$ 
and $D_{0max}\vert\frac{q}{e}S$ again as high as those extracted from $C_p$,
$\epsilon$ and $A_{echo}$ experimental data. We believe \cite{Jug2013} this
is to be ascribed to the coherent tunneling of an average number $N(T)$ of
atomic tunnelers trapped -- and strongly interacting -- in each interstitial 
ATSs (Fig.~\ref{cellstructure}).

\section{Nature of the Tunneling Systems and the Potential Energy Landscape}
So far we have described how the anomalous behaviour of glasses in the
low temperature regime is related to localized hoppings between adjacent
local configurations, through qm tunneling. The STM postulates that in the
configuration space each TS can be represented by a double well potential
(DWP) and makes certain assumptions regarding the distribution of its
asymmetry and barrier height. Although from a qualitative point of view
this picture is certainly reasonable and appealing, in practice one has to
see whether in a real glass-forming system such DWPs with the right
characteristics are indeed present. Through experiments, it is very
difficult to obtain microscopic information about the properties of the TSs
since they do not give {\it direct} access to the relevant information
(distribution of barrier heights, asymmetry, nature of the motion of the
participating particles, and so on (but see the attempts in
\cite{Fle2007,Bar2013})).

Thus at present the only promising approach to make progress in this
field are atomistic computer simulations. In the past such simulations
have already helped to clarify many properies of the potential energy
landscape (PEL) of glass-forming systems and allowed to understand their
relaxation dynamics~\cite{Sas1998,Deb2001,Sci2005,Heu2008}. (We recall that
the PEL is the surface of the potential energy as a function of the $3N_{at}$
coordinates of the system, where $N_{at}$ is the number of particles. Note
that this PEL does not depend on temperature, but the regions that the system
explores with a significant probablity does.) One can therefore hope that
such simulations will also permit to gain insight into the properties of
glasses at intermediate, low, and very low temperatures.

The question that has to be addressed in such simulations is whether
or not one does indeed find in the PEL local DWPs that have the right
features. Note that one expects that the number of local minima is
exponentially large in $N_{at}$, i.e. it is a huge number and therefore it
will be completely impossible to find all of the DWPs. Furthermore one faces
the conceptual problem of how this purely classical concept of the PEL can
be used to describe the tunneling processes, i.e. a quantum phenomenon. The
simplest approach to address these issues is the introduction of a reaction
coordinate that parametrizes the hopping form one local minimum to a
neighboring one and from this to calculate the effective barrier height
and thus the tunneling parameters.  Subsequently this information can be
fed into theoretical approaches like those discussed in the previous
Sections to calculate quantities like the heat capacity $C_p$ and so on.

\begin{figure}[!h]
\centering \includegraphics[scale=0.45] {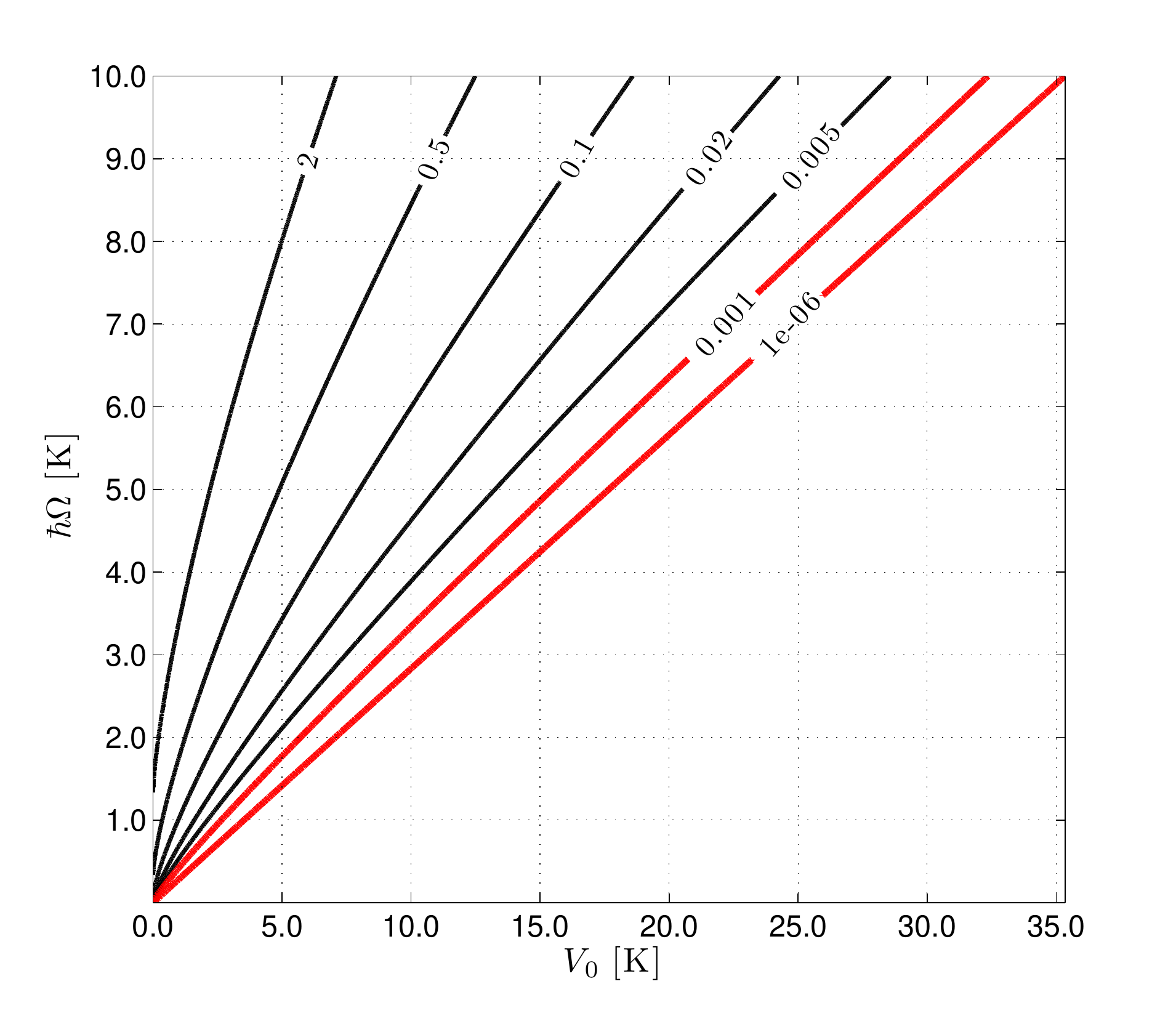}
\caption{Iso-lines of the tunneling parameter $\Delta_0$ as a function of
the barrier height~ $V_0$ and the single-well energy $\hbar\Omega$. The
physically relevant range for $\Delta_0$ is delimited by the two bold red
lines.}
\label{2dplot}
\end{figure}

First of all we have to discuss the order of magnitude of the main
parameters that characterize the TSs, i.e. (for the 2LSs) the tunneling
matrix element $\Delta_0$, the energy of the single well $\hbar\Omega$,
and the barrier height $V_0$. The tunneling matrix element $\Delta_0$ is
given, for example, by Eq.~(\ref{overlap}), i.e.
$\Delta_0=\frac{1}{2}\hbar\Omega \bigg(3-\sqrt{\frac{8V_0}{\pi \hbar\Omega}}
\bigg) e^{-2\frac{V_0}{\hbar \Omega}}$. From the literature one can
infer the typical range of values for the first two quantities for the
most studied glasses (the multi-silicates):
\begin{itemize}
\item
$10^{-6}~\text{K}<\Delta_0<~10^{-3}~\text{K}$ for the tunneling parameter
\cite{Pal2011},
\item
$10^{-5}~\text{K}<\hbar\Omega<10~\text{K}$ for the well energy
\cite{Phi1987,Pal2011}
\end{itemize}
(it is thought that $\hbar\Omega$ should be the largest energy in the
problem, hence of the order of the gap energy's upper cutoff $E_{max}$).
This limits have been verified also in the context of the present ETM
theory's various applications and allow us to deduce the range of variation
of the barrier height~$V_0$. Figure~\ref{2dplot} shows some iso-lines of
$\Delta_0$ as a function of the barrier height $V_0$ and the well energy
$\hbar\Omega$.  The range mentioned above for $\Delta_0$ corresponds to the
region delimited by the two red bold lines. This plot shows that the typical
barrier height range goes up to a maximum of 20~K or so and therefore this
value represents a discriminant threshold for the relevant parameters of the
TSs (2LSs, specifically).
In the past most simulation studies of the properties of the PEL have
focused on the so-called ``inherent structures'' (ISs), i.e. the local
minima of the PEL~\cite{Sas1998,Deb2001,Sci2005,Heu2008}. Following the
pioneering work of Goldstein who was the first to point out that the
corrugated nature of the PEL is important to understand the slow relaxation
dynamics of glass-forming liquids~\cite{Gol1969}, researchers started to
study the properties of this landscape for simple glass-forming
systems~\cite{Sas1998,Web1985,Ang2003,Bro2000}. Weber and
Stillinger~~\cite{Web1985} were the first to analyze the PEL by studying
how the system evolves in this complex landscape and explores the various
minima. For this they periodically quenched the configurations
of molecular dynamics (MD) simulations to zero temperature and identified
nearby minima as well as the connecting transition states, i.e. the saddle
points (SPs). Subsequently Heuer and coworkers carried out detailed studies
in an attempt to determine the properties of the DWPs that are relevant for
the low temperature properties of glasses~\cite{Heu1993,Rei2006}. This was
done by searching for the DWPs in the PEL in a wide range of the energy
asymmetry~$\Delta$, distance between the wells~$d$, and barrier
height~$V_0$.  After parametrizing these quantities with a polynomial
function and subsequently applying an interpolation procedure, they
inferred the properties of TSs from the collected sets of DWPs. However,
their conclusions regarding the nature of the DWPs in terms of
configurations and parameters (asymmetry,...) depend strongly on the
manner they have scanned the PEL and as a consequence one does not have, in
fact, at present a real good understanding on the nature of the DWPs, or
more generally, of the possible atomic configurations that are mutually
connected via a tunneling process.

As discussed above, the barrier height of the TSs that are relevant at
low $T$ is rather small. However, most previous studies have found that
the overwhelming majority of the barriers between two adjacent ISs is not
that small, i.e. it is a sizable fraction of the interaction energy
between the particles, thus on the order of 100~K~\cite{Ang2003}. Hence
this indicates that in fact the TSs are {\it not} related to a process in
which one particle changes its nearest neighborhood and thus moves a good
fraction of the typical nearest neighbor distance, i.e. 0.3-2 \AA. So,
what other options remain for excitations that give rise to DWPs that
have a small barrier height and thus a reasonable tunnelling parameter?
One possibility is that these excitation occur in the {\it interior} of
the basin of attraction of an IS. In other words, it could be that the
basin of attraction is not just a simple slightly deformed bowl, but is
instead corrugated and characterized internally by small channels and
valleys. These kind of local deformations in the structure of the PEL can
give rise to \textit{effective} DWPs and therefore we believe that
both the TSs and the ATSs are associated not with jumps of the system from
one IS to a neighboring one, but rather to the motion
inside the basin of attraction of a given IS. The TSs could be therefore
associated with the motion from one of these valleys to a nearby one and
the tunneling can occur among different channels.

\begin{figure}[!Htbp]
  \centering \includegraphics[scale=0.95] {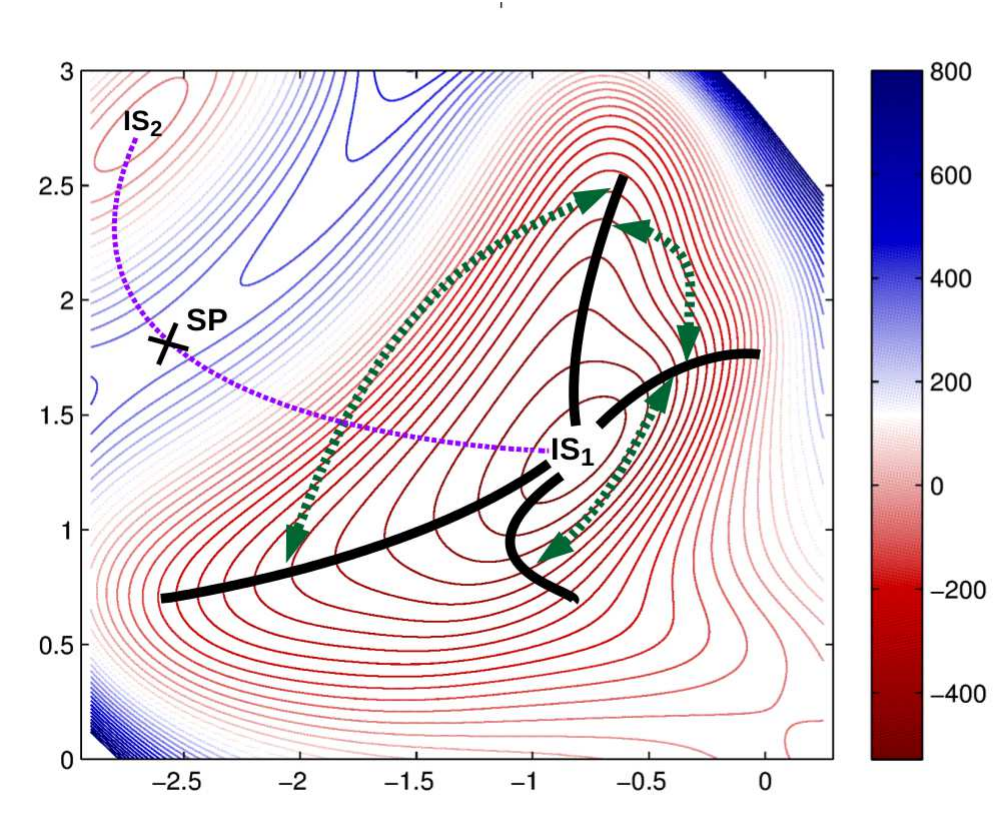}
\caption{(colour online) Simplified view of the glass potential energy
landscape (PEL) of a glass-forming system. We show two local minima, i.e.
inherent structures (ISs), that are connected by a first-order saddle
point (SP). Also shown are some of the valleys (bold black solid lines)
that emanate from the IS and often lead to a SP. The bold green dotted
lines represent possible transition pathways from one valley to another.
Tunneling takes place across the ``reliefs'' of the PEL.  Note that the
pathis of these valleys do not necessarily lead to a saddle.}
  \label{valley}
\end{figure}

To make this idea more specific we show in Fig.~\ref{valley} what is
intended to be an oversimplified representation of the real PEL of a glass,
that we computed in order to clarify the concepts; as can be noted, the
immediate surrounding of the IS is still harmonic, but as soon as one moves
away (always remaining inside the same basin of attraction) different
valleys (ending roughly at the N, E, S and W cardinal points in the drawing)
develop (as indicated by solid lines), separated by regions in which the
potential is higher and where the tunneling processes could occur (a
tunneling path is marked by a bold green dashed line). We have two ISs
(IS$_1$ and IS$_2$) that are connected via a reaction path (thin dashed
line) that goes through a SP. Note that this type of transition usually
involves only
a small number of degrees of freedom, i.e. on the order of 4-10 particles,
since the excitation is localized in the three dimensional space. Thus the
vast majority of the other particles change their position only marginally
in this transition event, since they basically react only in an elastic
manner, i.e. without changing their neighborhood. As mentioned above,
the typical energy of the SPs is too large to permit a tunneling process
with a reasonable probability and thus such processes might be irrelevant
for the properties of glasses at low temperatures. We therefore conclude
that the TSs should not be thought as a restricted subset of ISs forming
DWPs, but rather as small irregular valleys that exist inside a single
IS's basin and characterize the very bottom of the landscape. As is
exemplified in Fig. \ref{valley}, the 2LSs are tunneling processes of the
N-W dashed bold green-path type, whilst the much shallower N-E and S-E
tunneling bold green-paths are probably good examples for the more elusive
ATSs.
We note, however, that to each IS are connected on the order of
$N_{at}$ valleys (bold solid lines), many of which lead to one of the
$O(N_{at})$ SPs. These valleys are local minima {\em within} the basin of
attraction of a given IS. Since these valleys start all at the IS, their
energy difference is small, as long as they are close to the IS.
Furthermore one can expect that also the barrier that the system has to
cross when moving from one valley to another is not very large. (Such
processes are indicated by the bold green dotted lines with arrows.) Thus
such an inter-valley transition can approximately be described by a DWP
in which the two wells have almost the same energy and the barrier height
is not very large, i.e. the DWPs do indeed have the features that we need
in order to describe the TSs. Thus at this stage we can speculate that the
TSs are related not to the IS-IS transitions, how is usually done, but to
valley-valley transitions {\em inside} the basin of attraction of a single
IS. For the moment we can of course only say that the valley-valley
transitions will give rise to DWPs that have the right {\em
semi-quantitative} features. It will thus be important to check by means
of numerical simulations to what extent these DWPs can indeed be used to
describe the low-temperature anomalies of glasses on a {\em quantitative}
basis. In addition it will be important to understand how the specific
macroscopic properties of the glass former (composition, local structure
etc.) influences the features of the PEL {\em within} the basin of
attraction of an IS, and hence to establich a link between the macroscopic
properties of the glass and its behaviour at low temperatures.

From the numerical-analysis point of view, in order to understand
the low temperature anomalies found experimentally, there is a need
to perform local and systematic analysis of the bottom of the IS
basin. Since it has been found~\cite{Ang2003} that, around the glass
transition temperature, the system moves quite close to the IS and
the majority of the saddles found is of order~1, the new approach would
be the exploration of the IS's surroundings starting from configurations
obtained with standard MD simulations at the lowest
temperature that can be equilibrated. The mere location of the SPs is
not enough to characterize the PEL, in fact, as we can imagine, a given
IS is associated to more than one nearby saddle and furthermore there
is no guarantee that following the direction of the softest eigenvalue
necessarily leads to the saddle, as depicted in Figure~\ref{valley}: the
branches that develop rather tend to vanish in the complexity of the PEL.

We therefore intend to analyze the properties of simple glass-forming
systems in order to systematically locate all of the TSs and verify
whether they can represent a probe of the very local structure of the PEL
and, therefore in this sense, contribute also to the properties of the
glass transition. Such numerical simulations should not be too difficult
and are currently on the way.

\section{Conclusions and Outlook}
In conclusion, the ATS tunneling model for the magnetic effects in 
multi-component glasses (the multi-silicates BAS (or AlBaSiO), Duran and BK7) 
and contaminated mono-component glass glycerol has been fully justified in 
terms of a (not entirely) new vision for the intermediate-range structure of 
real glasses. In this scheme the particles are organized in regions of enhanced
regularity (RERs) and more mobile {\em charged} particles trapped in the 
interstices between the (on average) close-packed RERs. These are coherent
(owing to proximity and strong Coulomb forces) atomic tunnelers that can be
modeled in terms of a single quasi-particles with highly renormalized 
parameters. This model explains a large number of experimental data and facts 
with remarkable consistency also in terms of cross-checks like the 
determination of the concentration of trace paramagnetic impurities. 

The fact that pure amorphous silica (Spectrosil-I, for example \cite{Lud2003})
shows no detectable magnetic effects is a consequence of the extremely small
size of the RERs for a-SiO${}_2$, deprived of almost any nucleation centres
for both micro-crystal and polycluster (RER) formation. These RERs will 
therefore trap a very small number $N$ of dangling-bond ionic particles, or
none at all given the covalent nature of the Si-O bonds. Hence, no magnetic
effects are observable in the purest, single-component glasses. 

The present theory shows that the magnetic (and compositional) effects are a
mere manifestation of the inhomogeneous, cellular-like intermediate atomic
structure of real glasses. A cellular-type structure that has been advocated 
by scientists, especially (but not only) in the ex-USSR, now for almost a 
century. It is not impossible that with this vision the TSs could become in 
the near future the right probes with which to study the structure of glasses
in the laboratory.

We have also commented on the possible atomistic computational checks for this
theory and it is hoped that in the near future the MD simulations will provide
strong evidence for the nature, here proposed, of the tunneling states.

\section*{Acknowledgements}
One of us (SB) acknowledges support from the Italian Ministry of Education,
University and Research (MIUR) through a Ph.D. Grant of the Progetto Giovani
(ambito indagine n.7: materiali avanzati (in particolare ceramici) per
applicazioni strutturali), as well as from the Bando VINCI-2014 of the
Universit\`a Italo-Francese. We are very grateful to Maksym Paliienko for his 
help with data fitting and manuscripts preparation. GJ gratefully acknowledges
stimulating discussions with A.S. Bakai.

\end{document}